%% file: MeCN_v4_07.tex
 \documentclass[final,5p,times,twocolumn]{elsarticle}

 \usepackage{graphics}
 \usepackage{graphicx}

\usepackage{amssymb}
\usepackage{amsmath}

 \usepackage{lineno}

\journal{Journal of Molecular Spectroscopy}

\begin{document}

\begin{frontmatter}



\title{Toward a global model of the interactions in low-lying states of methyl cyanide: 
       rotational and rovibrational spectroscopy of the $\varv _4 = 1$ state and 
       tentative interstellar detection of the $\varv _4 = \varv _8 = 1$ state in Sgr~B2(N)}


\author[Koeln]{Holger S.P.~M\"uller\corref{cor}} 
\ead{hspm@ph1.uni-koeln.de} 
\cortext[cor]{Corresponding author.} 
\author[Bonn]{Arnaud Belloche} 
\author[Koeln]{Frank Lewen} 
\author[JPL]{Brian J. Drouin} 
\author[JPL]{Keeyoon Sung} 
\author[UoVA]{Robin T. Garrod} 
\author[Bonn]{Karl M. Menten}

\address[Koeln]{I.~Physikalisches Institut, Universit{\"a}t zu K{\"o}ln, 
  Z{\"u}lpicher Str. 77, 50937 K{\"o}ln, Germany}
\address[Bonn]{Max-Planck-Institut f{\"u}r Radioastronomie, Auf dem H{\"u}gel 69, 
  53121 Bonn, Germany}
\address[JPL]{Jet Propulsion Laboratory, California Institute of Technology, 
  Pasadena, CA 91109-8099, USA}
\address[UoVA]{Departments of Chemistry and Astronomy, University of Virginia, 
  Charlottesville, VA 22904, USA}

\begin{abstract}

Rotational spectra of methyl cyanide were recorded newly and were analyzed together with 
existing spectra to extend the global model of low-lying vibrational states and their 
interactions to $\varv _4 = 1$ at 920~cm$^{-1}$. The rotational spectra cover large 
portions of the 36$-$1439~GHz region and reach quantum numbers $J$ and $K$ of 79 and 16, 
respectively. Information on the $K$ level structure of CH$_3$CN is obtained from IR spectra. 
A spectrum of $2\nu_8$ around 717~cm$^{-1}$, analyzed in our previous study, covered also 
the $\nu _4$ band. The assignments in this band cover 880$-$952~cm$^{-1}$, attaining 
quantum numbers $J$ and $K$ of 61 and 13, respectively.

The most important interaction of $\varv _4 = 1$ appears to be with $\varv _8 = 3$, 
$\Delta K = 0$, $\Delta l = +3$, a previously characterized anharmonic resonance. 
We report new analyses of interactions with $\Delta K = -2$ and $\Delta l = +1$, with 
$\Delta K = -4$ and $\Delta l = -1$, and with $\Delta K = -6$ and $\Delta l = -3$; 
these four types of interactions connect all $l$ substates of $\varv _8 = 3$ in energy 
to $\varv _4 = 1$. A known $\Delta K = -2$, $\Delta l = +1$ interaction with 
$\varv _7 = 1$ was also analyzed, and investigations of the $\Delta K = +1$, 
$\Delta l = -2$ and $\Delta K = +3$, $\Delta l = 0$ resonances with $\varv _8 = 2$ 
were improved, as were interactions between successive states with $\varv _8 \le 3$, 
mainly through new $\varv _8 \le 2$ rotational data.

A preliminary single state analysis of the $\varv _4 = \varv _8 = 1$ state was carried out 
based on rotational transition frequencies and on $\nu _4 + \nu _8 - \nu _8$ hot band data. 
A considerable fraction of the $K$ levels was reproduced within uncertainties in its entirety 
or in part, despite obvious widespread perturbations in $\varv _4 = \varv _8 = 1$.

In addition to the interstellar detection of rotational transitions of methyl cyanide from 
within all vibrational states up to $\varv _4 = 1$, we report the tentative detection of 
$\varv _4 = \varv _8 = 1$ toward the main hot molecular core of the protocluster Sagittarius B2(N) 
employing the Atacama Large Millimeter/submillimeter Array.

\end{abstract}


\end{frontmatter}




\section{Introduction}
\label{introduction}

\begin{figure}
 \begin{center}
  \includegraphics[width=8.0cm]{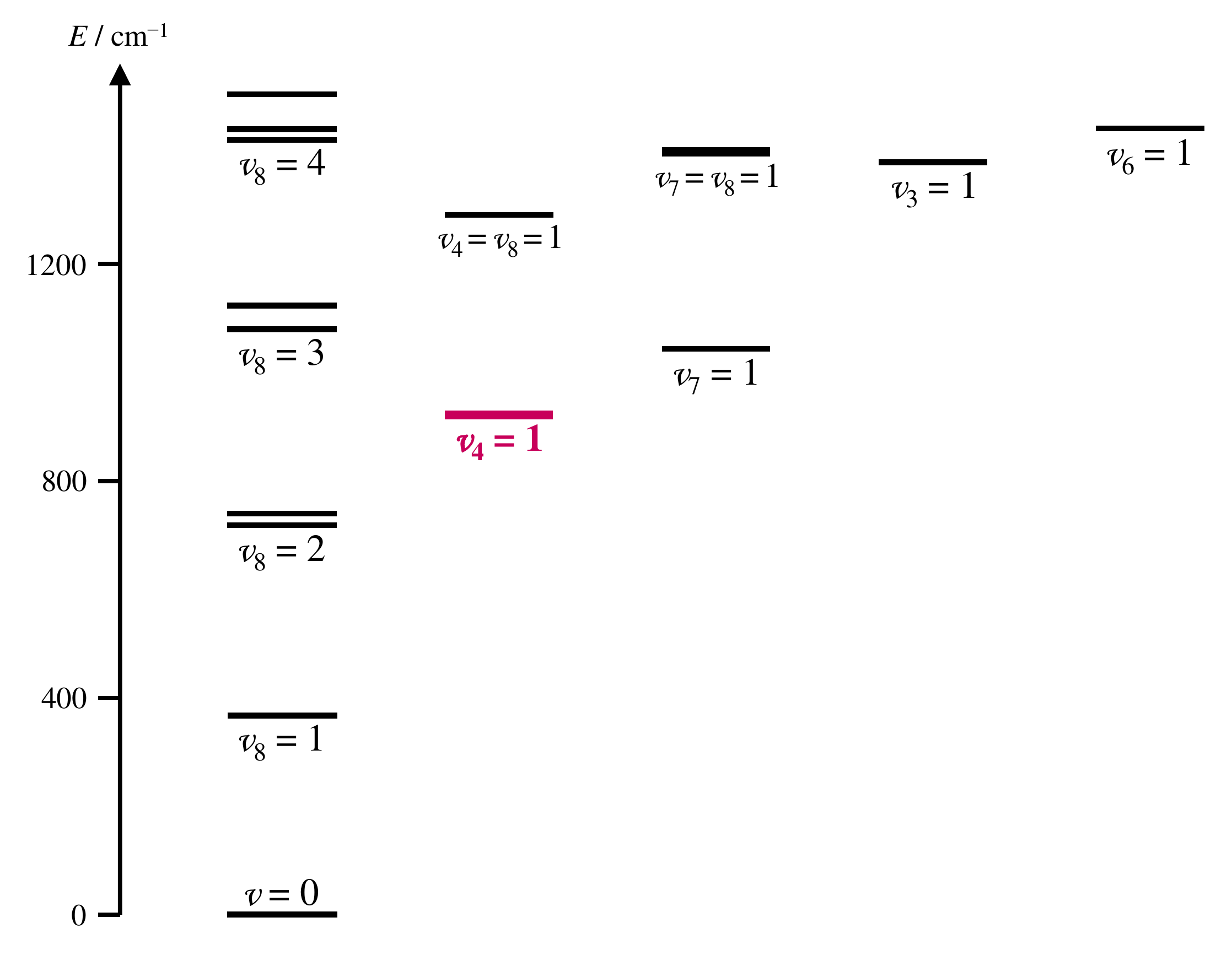}
 \end{center}
  \caption{Schematic representation of the energies of low-lying vibrational states 
           of CH$_3$CN with origins of the $l$ substates.}
  \label{fig_vib_energies}
 \end{figure}

Methyl cyanide was detected in Sagittarius (Sgr) A and B almost 50 years ago as one of the first 
molecules observed by radio-astronomical means \cite{MeCN_det_T_1971}. Since then, the molecule 
has been found in very diverse astronomical sources, a fairly detailed overview was given in our 
previous work on vibrational states $\varv _8 \le 2$ of CH$_3$CN \cite{MeCN_v8le2_2015}. 
We point out that numerous rare isotopologs have been detected as well, which include 
$^{13}$CH$_3$$^{13}$CN \cite{EMoCA_2016} and CHD$_2$CN \cite{CHD2CN_det_2018}. More important 
for the present study is the detection of excited state transitions of CH$_3$CN up to 
$\varv _4 = 1$ at 920~cm$^{-1}$ \cite{EMoCA_2016}; see \textbf{Fig.~\ref{fig_vib_energies}} 
for an overview of the low-lying vibrational states of methyl cyanide and 
\textbf{Table~\ref{vib_energies}} for a summary of the vibrational energies, including 
those of the $l$ substates.


\begin{table}
\begin{center}
\caption{Energies (cm$^{-1}$) and symmetries Sym of low-lying vibrational states of methyl cyanide.}
\label{vib_energies}
{\footnotesize
\begin{tabular}[t]{lcr@{}ll}
\hline 
State                     & Sym. & \multicolumn{2}{c}{Energy} & Reference                                     \\
\hline
$\varv = 0$               & $A$  &          0&.0              & per definitionem                              \\
$\varv_8 = 1$             & $E$  &        365&.024            & this work, Ref.~\cite{MeCN_v8le2_2015}        \\
$\varv_8 = 2^0$           & $A$  &        716&.750            & this work, Ref.~\cite{MeCN_v8le2_2015}        \\
$\varv_8 = 2^2$           & $E$  &        739&.148            & this work, Ref.~\cite{MeCN_v8le2_2015}        \\
$\varv_4 = 1$             & $A$  &        920&.290            & this work, Ref.~\cite{MeCN_nu4_nu7_3nu8_1993} \\
$\varv_7 = 1$             & $E$  &       1041&.855            & Ref.~\cite{MeCN_nu4_nu7_3nu8_1993}            \\
$\varv_8 = 3^1$           & $E$  &       1077&.79             & this work, Ref.~\cite{MeCN_nu4_nu7_3nu8_1993} \\
$\varv_8 = 3^3$           & $A$  &       1122&.35             & this work$^a$                                 \\
$\varv_4 = \varv_8 = 1$   & $E$  &       1290&.05             & this work                                     \\
$\varv_3 = 1$             & $A$  &       1385&.2              & Ref.~\cite{pentade_1994}                      \\
$\varv_7 = \varv_8 = 1^0$ & $A$  &       1401&.7              & Ref.~\cite{pentade_1994}                      \\
$\varv_7 = \varv_8 = 1^2$ & $E$  &       1408&.9              & Ref.~\cite{pentade_1994}                      \\
$\varv_8 = 4^0$           & $A$  & $\sim$1426&.               & Ref.~\cite{MeCN_v8le2_2015}$^b$               \\
$\varv_8 = 4^2$           & $E$  &       1447&.9              & Ref.~\cite{pentade_1994}                      \\
$\varv_6 = 1$             & $E$  &       1449&.7              & Ref.~\cite{pentade_1994}                      \\
$\varv_8 = 4^4$           & $E$  & $\sim$1514&.               & Ref.~\cite{MeCN_v8le2_2015}$^b$               \\
\hline \hline
\end{tabular}\\[2pt]
}
\end{center}
{\footnotesize
$^a$ 1122.15~cm$^{-1}$ in Ref.~\cite{MeCN_nu4_nu7_3nu8_1993}.\\
$^b$ Estimated value; see also section~\ref{spec-discussion}. 
}
\end{table}


The identification of gaseous methyl cyanide relies mostly on laboratory spectroscopic 
information; in astronomical sources this is almost exclusively done with rotational 
spectroscopy from the microwave to the submillimeter region. The first study of the 
rotational spectrum of CH$_3$CN, and of its isomer CH$_3$NC, dates back to the early 
days of microwave spectroscopy \cite{MeCN_rot_1947}. A detailed account on previous work 
involving vibrational states up to $\varv _8 = 2$ was given in our investigation 
of these states \cite{MeCN_v8le2_2015}. A Fermi resonance between $\varv _8 = 1^{-1}$ 
and $2^{+2}$ ($\Delta l = 3$) was identified at $K = 13$ and 14 and analyzed by means 
of rotational spectroscopy. Such resonances occur also between $\varv _8 = 2^{-2}$ 
and $3^{+1}$ at $K = 12$ and 13 and between $\varv _8 = 2^0$ and $3^{+3}$ at $K = 15$. 
Transitions up to $K = 11$ and 13 have been accessed for $\varv _8 = 2^{-2}$ and $2^0$, 
respectively. This particular type of resonance was reported, to the best of our 
knowledge, for the first time in studies involving the corresponding bending states 
of propyne \cite{MeCCH_nu10+Dyade_2002,CH3CCH_v39_2004,MeCCH_10mue_2009}. 
Rotational spectroscopy was instrumental in untangling analogous resonances. 
A study of such resonances was also reported for CH$_3$NC in its $\varv _8 \le 2$ 
states \cite{MeNC_v8le2_2011}. In the case of CH$_3$CN, additional resonances of 
the type $\Delta \varv_8 = \pm 1$, $\Delta K = \mp 2$, $\Delta l = \pm 1$ 
were identified and analyzed in detail for $\varv _8 = 1^{-1}$ and $2^{0}$ 
($K = 13$ and 11) and $\varv _8 = 1^{+1}$ and $2^{+2}$ ($K = 15$ and 13). 
Whereas these resonances caused pronounced perturbations, an analogous resonance 
between $\varv = 0$ and $\varv _8 = 1^{+1}$ at $K = 14$ and 12, respectively, 
displayed only small perturbations. However, these were strong enough to cause 
observable cross-ladder transitions between the states, thus connecting 
strongly these two vibrational states in energy.


 \begin{figure}
 \begin{center}
  \includegraphics[width=8.0cm]{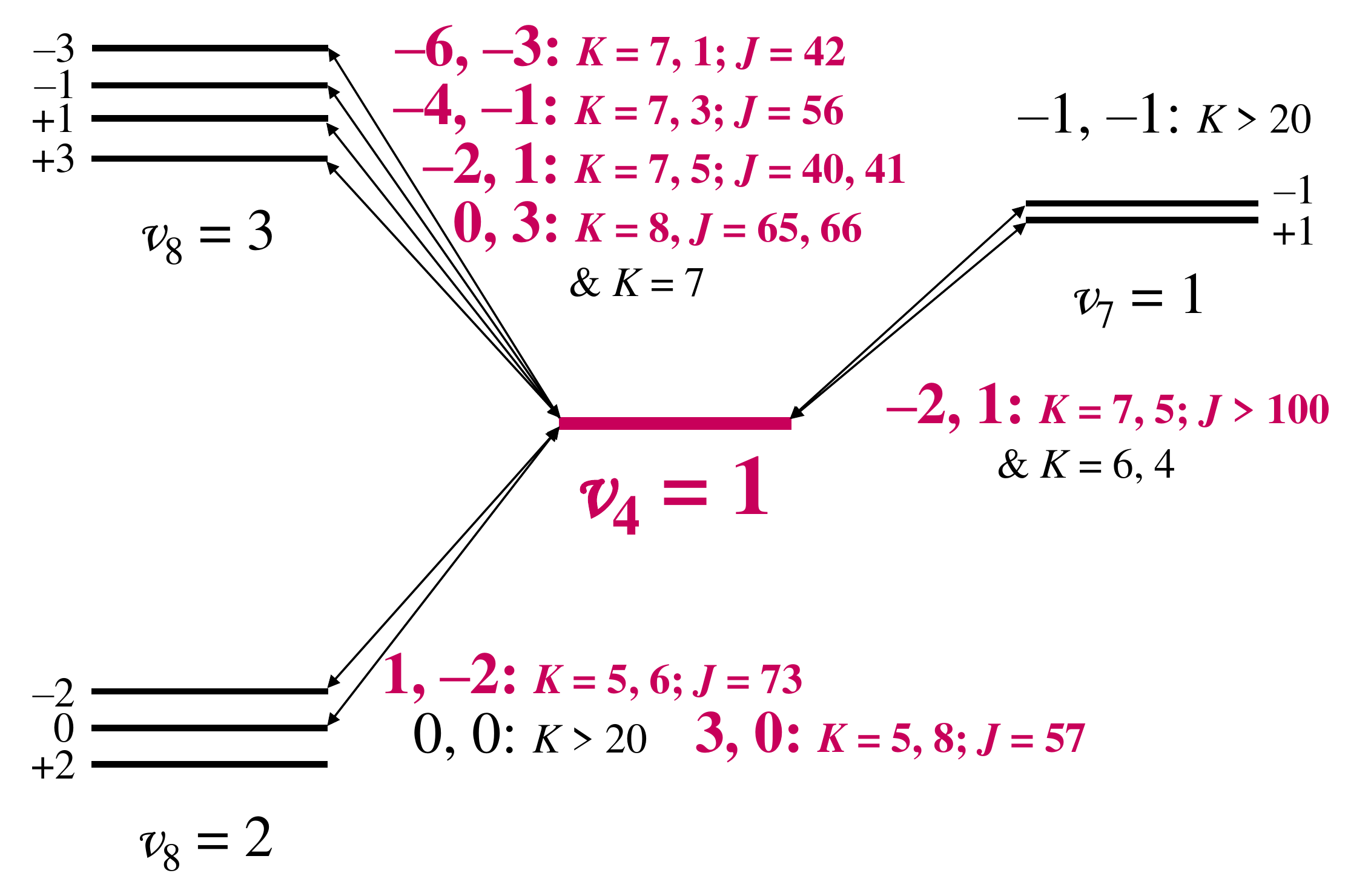}
 \end{center}
  \caption{Representation of the rovibrational interactions involving $\varv _4 = 1$. 
           The $l$ substates of each vibrational state are ordered as they appear at 
           intermediate and higher $K$. Arrows indicate interacting $l$ substates. 
           $\Delta K$ and $\Delta l$ are given with respect to $\varv _4 = 1$. 
           The first $K$ refers to $\varv _4 = 1$, the second to that of the 
           interacting $l$ substate. Distant resonances or resonances with no 
           crossing in $J$ are colored black, resonances with crossing in $J$ in 
           bold and dark red. The most perturbed $J$ is also given.}
  \label{v4-resos}
 \end{figure}


The analyses of the next three higher-energy states, $\varv _4 = 1$ at 920~cm$^{-1}$, 
$\varv _7 = 1$ at 1042~cm$^{-1}$, and $\varv _8 = 3$ at 1078 and 1122~cm$^{-1}$, and 
their interactions took many years until a fairly comprehensive and sufficiently 
accurate level was achieved. Interactions involving $\varv _4 = 1$ are shown 
schematically in \textbf{Fig.~\ref{v4-resos}}.

Kondo and Person evaluated the strength of the Coriolis interaction between 
$\nu _4$ and $\nu _7$ through intensity perturbations of $\nu _4$ in a 
low-resolution ($\sim$1~cm$^{-1}$) IR spectrum \cite{low-res_nu4-nu7_etc_IR_1974}. 
Bauer \cite{Bauer_thesis_1970} studied the rotational spectra of CH$_3$CN and 
CH$_3$C$^{15}$N up to $\varv _4 = 1$ (here and in the following, unlabeled atoms 
refer to $^{12}$C and $^{14}$N). The $\varv _4 = 1$ data with $6 \le J'' \le 9$ 
and $K \le 6$, and with frequencies up to 184~GHz, were published in a journal 
later \cite{MeCN-14-15_v4=1_rot_1975}. Duncan et al. carried out a comparative 
study of the IR spectra of several methyl cyanide isotopologs along with a force 
field calculation \cite{FF_Duncan_1978}. They proposed a strong Fermi interaction 
between $\nu _4$ and $2\nu _8^0$, a strong anharmonic resonance between $\nu _4$ 
and $3\nu _8^3$, and a moderate anharmonic resonance between $\nu _7$ and $3\nu _8^1$. 
Rackley et al. performed a laser Stark investigation of $\nu _4$ and $\nu _7$ 
and determined in particular a $\nu _4$/$\nu _7$ interaction parameter, even 
though they point out that the resonance in $\nu _7^{-1}$ occurs at $K \approx 23$ 
\cite{Triade_1982}. In addition, they examined the $\nu _7^{+1}$/$3\nu _8^{+1}$ 
anharmonic resonance, which is most strong at $K = 7$ and 8 \cite{Triade_1982}. 
Mori et al. \cite{MeCN_nu7_etc_Laser-Stark_1984} carried out more extensive 
analyses of CH$_3$CN IR bands. In addition to the interactions analyzed earlier, 
they proposed a Fermi resonance between $2\nu _8 ^{-2}$ and $\nu _7^{+1}$ at 
$K = 13$ and 14. Mito et al. \cite{Triade_15N_1984} performed a laser Stark 
investigation of the $\nu _4$ band of CH$_3$C$^{15}$N in order to analyze 
the resonance with $3\nu _8^{+3}$, which they found to be much weaker than 
proposed by Duncan et al. \cite{FF_Duncan_1978}. They found a crossing between 
$K = 7$ and 8, with the latter farther apart at low $J$. Wallraff et al. 
\cite{nu4_1985} extended the $K$ range of $\nu _4$ for CH$_3$CN and obtained 
essentially the same results concerning the corresponding resonance. 
Bocquet et al. \cite{MeCN-vib_le_v4_J=19_1988} recorded submillimeter 
transitions of methyl cyanide, $J'' = 19$ up to $K = 12$ in the case of 
$\varv _4 = 1$. Cosleou et al. \cite{v4=1_v8=3_1991} extended the $J$ range 
of rotational transitions in $\varv _4 = 1$ up to 24. 
They also analyzed the resonance between $\varv _4 = 1$ and $\varv _8 = 3 ^{+3}$, 
located a $\Delta K = -2$, $\Delta l = +1$ interaction between $\varv _4 = 1$ and 
$\varv _7 = 1 ^{+1}$ at $K = 6$ and 4, respectively, and proposed a $\Delta K = +3$, 
$\Delta l = 0$ interaction between $\varv _4 = 1$ and $\varv _8 = 2 ^{0}$ at $K = 5$ 
and 8, respectively. The most comprehensive and accurate analysis of the $\nu _4$, 
$\nu _7$, and $3\nu _8$ band system of CH$_3$CN was presented by Tolonen et al. 
\cite{MeCN_nu4_nu7_3nu8_1993}. They included most of the resonances mentioned 
before, with the exceptions of the $\Delta K = 0$ and 3, $\Delta l = 0$ interactions 
between $\varv _4 = 1$ and $\varv _8 = 2^{0}$. They introduced a $\Delta K = +1$, 
$\Delta l = -2$ resonance between $K = 5$ and 6 of $\varv _4 = 1$ and 
$\varv _8 = 2^{-2}$, respectively.

The next five vibrational states are $\varv _3 = 1$, $\varv _6 = 1$, 
$\varv _4 = \varv _8 = 1$, $\varv _7 = \varv _8 = 1$, and $\varv _8 = 4$, see 
\textbf{Fig.~\ref{fig_vib_energies}} and \textbf{Table~\ref{vib_energies}}. 
Investigations of the interactions between these states began 50 years ago. 
Matsuura analyzed the Fermi resonance between $\nu _6^{\pm1}$ and 
$(\nu _7 + \nu _8)^{\mp2}$ \cite{nu6_1971}. Duncan et al. \cite{Pentade_1971} 
and later Matsuura et al. \cite{Pentade_1982} included the Coriolis resonance 
between $\nu _6$ and $\nu _3$ in their analyses. Paso et al. \cite{pentade_1994} 
presented the latest, fairly comprehensive and quite accurate analysis of these 
bands. Their assignments covered extensive parts of $\nu _6$, a fair fraction 
of $(\nu _7 + \nu _8)^{\mp2}$, and some transitions in $\nu _3$ ($K = 5$ and 6), 
which gain intensity through the Coriolis resonance with $\nu _6$. 
Information on $\nu _4 + \nu _8$, $(\nu _7 + \nu _8)^{\pm0}$, and 
$4\nu _8^{\pm2}$ were obtained through various resonances; no information 
was presented for $4\nu _8^0$ and for $4\nu _8^{\pm4}$. A moderately weak 
anharmonic resonance between $(\nu _4 + \nu _8)^{-1}$ and 
$(\nu _7 + \nu _8)^{+2}$, mainly at $K = 5$, and a $\Delta K = -1$, 
$\Delta l = +2$ interaction between $(\nu _4 + \nu _8)^{-1}$ and $\nu _6^{+1}$, 
mainly at $K = 10$ and 9, respectively, were treated in their analysis. 
Mito et al. \cite{Pentade_15N_1984} studied the $\nu _4 + \nu _8 - \nu _8$ 
hot band of CH$_3$C$^{15}$N and analyzed an anharmonic resonance between 
$\varv _4 = \varv _8 = 1^{+1}$ and $\varv _8 = 4^{+4}$ with largest effect 
at $K = 8$ and a lesser one at $K = 9$ and anharmonic resonances between 
$\varv _4 = \varv _8 = 1^{-1}$, $\varv _8 = 4^{+2}$, and 
$\varv _7 = \varv _8 = 1^{+1}$ most strongly perturbed at $K = 6$ and 
less so at $K = 5$. Judging from Fig.~6 of Paso et al. \cite{pentade_1994}, 
the last three resonances occur at the same $K$ values in CH$_3$CN, with 
a possible difference in the resonance between $\varv _4 = \varv _8 = 1^{-1}$ 
and $\varv _7 = \varv _8 = 1^{+1}$, which may be strongest in $K = 5$.

Approximately 15 years ago, we started our project devoted to recording and analyzing 
low-lying vibrational states of methyl cyanide. The aims were providing predictions 
of rotational and rovibrational spectra for radio-astronomical observations and for 
studies of the atmospheres of Earth and Titan, among others. An additional, but also 
necessary aim were thorough investigations of perturbations within and between these 
vibrational states.

In the course of a line-broadening and -shifting study in the $\nu _4$ band region of 
CH$_3$CN \cite{MeCN_nu4_int_2008}, a preliminary analysis of $\varv _4 = 1$ and its 
interactions with other vibrational states was carried out. In addition, assignments 
were made for the $\nu _4 + \nu _8 - \nu _8$ hot band and for rotational transitions 
in $\varv _4 = \varv _8 = 1$. Subsequently, extended assignments for the ground state 
rotational spectra of six methyl cyanide isotopologs were based on measurements on 
a sample of natural isotopic composition \cite{MeCN_rot_2009}. Some time later, 
we carried out a similar study of three minor isotopologs, CH$_3$$^{13}$CN, 
$^{13}$CH$_3$CN, and CH$_3$C$^{15}$N, in their $\varv _8 = 1$ excited vibrational 
states \cite{MeCN_isos_v8_rot_2016}. The analysis of CH$_3$CN vibrational states 
up to $\varv _4 = 1$ was quite advanced about ten years ago \cite{v4=1_OSU_2010}. 
Attempts to introduce $\varv _7 = 1$ data into the fit were quite successful, 
but the inclusion of $\varv _8 = 3$ data proved to be more difficult. 
Data for both states were omitted for the present fit because both states are heavily 
interacting. Ultimately, our previous account on rotational and rovibrational data 
of CH$_3$CN \cite{MeCN_v8le2_2015} was limited to states with $\varv _8 \le 2$. 
The omission of $\varv _4 = 1$ data at that time was based on large residuals 
in the $\Delta K = 3$ ground state loops from Ref.~\cite{MeCN_DeltaK=3_1993} and 
concomitant changes in the purely axial ground state parameters ($A - B$, $D_K$, $H_K$). 
In addition, there were small, but systematic residuals in some $K$ series of the 
$\varv _4 = 1$ rotational data.

In our present study, we have reanalyzed carefully our $\varv _4 = 1$ data, recorded 
additional rotational transitions pertaining to $\varv _4 = 1$ and to lower vibrational 
states as well as to $\varv _4 = \varv _8 = 1$. We improved the analyses of all known 
resonances involving $\varv _4 = 1$ and those involving states differing in one quantum 
of $\varv _8$. Three higher-$\Delta K$ resonances between $\varv _4 = 1$ and $\varv _8 = 3$ 
were also investigated. These findings improve the parameters  for $\varv _4 = 1$ considerably 
and for some of the lower states to a lesser extent. We also report a preliminary analysis 
of $\varv _4 = \varv _8 = 1$. 
We use the spectroscopic results obtained for $\varv_4=1$ and $\varv_4=\varv_8=1$ in 
this study to investigate the vibrationally excited methyl cyanide emission in the main 
hot molecular core embedded in the high-mass star forming protocluster Sagittarius B2(N) 
observed with the Atacama Large Millimeter/submillimeter Array (ALMA) in the 
frame of the ReMoCA project \cite{Belloche19}.

The remainder of this article is outlined as follows: experimental details of the 
rotational and rovibrational spectra are given in Section~\ref{exptl_details}; 
Section~\ref{results} contains our results with details on the spectroscopy and 
interactions in low-lying vibrational states of CH$_3$CN, descriptions of the analyses 
carried out in the present study, a summary of the data obtained newly as well as 
those from previous investigations, and the determination of spectroscopic parameters. 
The astronomical results are described in Section~\ref{s:astro}; a discussion of the 
spectroscopic findings is presented in Section~\ref{spec-discussion}; 
Section~\ref{Conclusions} finally presents conclusions and an outlook from our study.

\section{Experimental details}
\label{exptl_details}
\subsection{Rotational spectra at the Universit{\"at} zu K{\"o}ln}

All measurements at the Universit{\"a}t zu K{\"o}ln were recorded at room temperature 
in static mode employing different Pyrex glass cells having an inner diameter of 
100~mm with pressures in the range of 0.5$-$1.0~Pa below 368~GHz, around 1.0~Pa 
between 1130 and 1439~GHz, and mostly 2~Pa up to 4~Pa between 748 and 1086~GHz. 
The measurements covered transitions pertaining to one $J$ of one or more 
vibrations in many cases, sometimes smaller groups of lines, and in many cases 
individual lines. 
The window material was Teflon at lower frequencies, whereas high-density 
polyethylene was used at higher frequencies. Frequency modulation was used 
throughout. The demodulation at $2f$ causes an isolated line to appear close 
to a second derivative of a Gaussian.

The $J = 4 - 3$ transitions of $\varv _4 = 1$ near 73~GHz were recorded using a 3~m long 
single pass cell, a backward-wave oscillator (BWO) based 4~mm synthesizer AMC~MSP1 
(Analytik \& Me{\ss}technik GmbH, Chemnitz, Germany) as source, and a Schottky-diode as detector. 
A small number of methyl cyanide rotational transitions were investigated around 875~GHz 
and around 892~GHz with the Cologne Terahertz Spectrometer \cite{THz-BWO_1994} using 
a BWO as source and a liquid helium cooled InSb hot-electron bolometer (QMC) as detector.

Transitions of $J = 2 - 1$ and $3 - 2$ around 37 and 55~GHz, respectively, were recorded 
with an Agilent E8257D microwave synthesizer as source and a home-built Schottky diode 
detector. A 7~m long double pass absorption cell was used for these measurements. 
Transitions of $J = 4 - 3$ to $6 - 5$ were measured in two coupled 7~m long double pass 
absorption cells. Source frequencies were generated using a Virginia Diodes, Inc. (VDI) 
tripler driven initially by an Agilent E8257D microwave synthesizer, later by a Rohde 
\& Schwarz SMF~100A synthesizer, and a Schottky diode detector was employed again. 
Additional information on the spectrometer is available elsewhere \cite{n-BuCN_rot_2012}.

Further transitions were covered in the 164$-$368~GHz region employing a 5~m long 
double pass absorption cell, VDI frequency multipliers driven by a Rohde \& Schwarz 
SMF~100A synthesizer, and Schottky diode detectors. Ref.~\cite{OSSO_rot_2015} 
contains more information on this spectrometer. Uncertainties down to 3~kHz were 
assigned for very symmetric lines with very good signal-to-noise ratio recorded 
with these two frequency multiplier based spectrometers. Almost as small uncertainties 
($\ge 5$~kHz) were assigned in a study of 2-cyanobutane \cite{2-CAB_rot_2017}, 
which has a much richer rotational spectrum. Lines of average quality were assigned 
10$-$20~kHz uncertainties, up to 50~kHz for weaker lines or lines close to stronger 
ones.

Measurements were also carried out in parts of the 1130$-$1439~GHz region employing 
a 3~m long single pass cell, two VDI frequency multipliers driven by an Agilent E8257D 
microwave synthesizer as source, and a liquid He-cooled InSb bolometer (QMC) as detector. 
This spectrometer was described in somewhat greater detail in the investigation of 
CH$_3$SH \cite{MeSH_rot_2012}. Our latest CH$_3$CN study \cite{MeCN_v8le2_2015} or 
an investigation of isotopic thioformaldehyde \cite{H2CS_rot_2019} demonstrate that 
accuracies of 10~kHz can be reached readily routinely for very symmetric lines with 
good signal-to-noise ratios (S/N). Lines of average quality have uncertainties of 
30 to 80~kHz, weaker lines, less symmetric lines or close to stronger lines were 
assigned uncertainties of 100 to 200~kHz, sometimes up to 300~kHz.

A similar setup with a 5~m long single pass cell, a VDI frequency multiplier driven 
by a Rohde \& Schwarz SMF~100A microwave synthesizer as source, and a closed cycle 
liquid He-cooled InSb bolometer (QMC) as detector was used to cover transitions 
in the 748$-$1086~GHz region.

\subsection{Rotational spectra at the Jet Propulsion Laboratory}

The CH$_3$CN rotational spectra taken with the JPL cascaded multiplier spectrometer 
\cite{JPL_multiplier_spectrometer_2005} are the same as employed for our latest two 
studies \cite{MeCN_v8le2_2015,MeCN_isos_v8_rot_2016}. Generally, the output of a 
multiplier chain source is passed through a 1$-$2 meter pathlength flow cell and is 
detected by a silicon bolometer cooled to near 1.7~K. The cell is filled with a 
steady flow of reagent grade acetonitrile at room temperature, and the pressure and 
modulation are optimized to enable good S/N with narrow lineshapes. 
The S/N ratio was optimized for a higher-$K$ transition (e.g. $K = 12$) because 
of the very strong ground state transitions of the main isotopolog with lower $K$, 
which frequently exhibit saturated line profiles. 
This procedure enables better dynamic ranges for the extraction of line positions for rare 
isotopologs and highly excited vibrational satellites. The frequency ranges covered were 
440$-$540, 530$-$595, 619$-$631, 638$-$648, 770$-$855, 780$-$835, 875$-$930, 967$-$1050, 
1083$-$1093, 1100$-$1159, 1168$-$1198, 1576$-$1591, and 1614$-$1627~GHz. 
Most of the employed multiplier sources were previously described 
\cite{MeCN_v8le2_2015,JPL_multiplier_spectrometer_2005}. In addition, recording 
conditions and sensitivities of detectors can have strong influences on the quality 
of the spectra. Particularly good S/N were reached around 600, 800, 900 and at 
1100$-$1200~GHz. The S/N changed considerably within each scan and was usually 
lower towards the edges. The uncertainties were judged exactly as the submillimter 
lines measured in Cologne, however, the distribution of the uncertainties differed 
somewhat with fewer lines having very small uncertainties and more lines 
in the 50$-$100~kHz range.

\subsection{Infrared spectrum}
\label{exptl_IR}

The lower wavenumber part of the infrared spectrum of CH$_3$CN, recorded between 600 
and 989~cm$^{-1}$, was already used in our previous study \cite{MeCN_v8le2_2015}. 
It was recorded at the Pacific Northwest National Laboratory (PNNL) with a Bruker 125~HR 
Fourier transform spectrometer at 0.0016~cm$^{-1}$ resolution using an MCT detector. 
A multi-pass absorption cell, set to an optical path length of 19.20~m, was filled 
to 0.226~Torr (30.1~Pa) of CH$_3$CN at 293.0~K. Small amounts of OCS ($\sim$3\%) 
and CO$_2$ ($\sim$0.4\%) were added to the sample for frequency calibration. 
Comparison of 63 well-isolated lines of the $\nu_3$ fundamental of OCS at 860~cm$^{-1}$ 
\cite{OCS_ref_1992} produced a calibration factor 1.000000868~(21) with an rms of 
0.0000182~cm$^{-1}$. Nearly 9000 line positions and relative intensities between 
880 and 952~cm$^{-1}$ were retrieved using non-linear least-squares curve-fitting 
\cite{curve-fitting_IR-positions_1983}. The accuracy in $\nu_4$ was 0.0001~cm$^{-1}$, 
about a factor five worse than the OCS calibration because of the congestion of 
the CH$_3$CN spectrum. Uncertainties in $\nu_4 + \nu_8 - \nu_8$ were mostly 
0.0002 or 0.0004~cm$^{-1}$ because of the greater congestion compared with $\nu_4$.


\section{Results}
\label{results}

Pickett's SPCAT and SPFIT programs \cite{spfit_1991} were used for calculations of 
the CH$_3$CN spectra and for fitting of the measured data. The programs were intended 
to be rather general, thus being able to fit asymmetric top rotors with spin and 
vibration-rotation interaction. They have evolved considerably with time because many 
features were not available initially \cite{editorial_Herb-Ed,intro_JPL-catalog}, 
in particular special considerations for symmetric or linear molecules or for higher 
symmetry cases. One of the latest additions is the option to define $l$-doubled 
states having $l \equiv 0$~mod~3.

We determined rotational, centrifugal, and hyperfine structure (HFS) parameters of 
the ground state as common for all vibrational states. Some of the data were measured 
or reported with partial or fully resolved HFS, but the majority of the data, in 
particular the IR data, were not affected by HFS. Therefore, all states were defined 
twice, with and without HFS. Vibrational changes $\Delta X = X_{\rm i} - X_0$ to 
the ground vibrational state were fit for excited vibrational states, where $X$ 
represents a parameter and $X_{\rm i}$ and $X_0$ the parameter in an excited and 
ground vibrational state, respectively. This is very similar to several previous 
studies on CH$_3$CN, for example \cite{MeCN_nu4_nu7_3nu8_1993,pentade_1994} and 
rather convenient because vibrational corrections $\Delta X$ are usually small with 
respect to $X$, especially for lower order parameters $X$. Moreover, this offers 
the opportunity to constrain vibrational corrections to $\varv_8 = 2$ to twice those 
of $\varv_8 = 1$ wherever appropriate, thus reducing the amount of independent 
spectroscopic parameters further.
New parameters in the fit were chosen carefully by searching for the parameter that 
reduces the rms error of the fit the most. We tried to assess if the value of a new 
parameter is reasonable in the context of related parameters and tried to omit or 
constrain parameters whose values changed considerably in a fit or had relatively 
large uncertainties. Care was also taken that a new parameter is reasonable with 
respect to quantum numbers of newly added transition frequencies or that it can 
account for systematic residuals.

The spectroscopic parameters used in the present analyses are standard symmetric rotor 
parameters defined and designated in a systematic way. The designation of the interaction 
parameters in particular may differ considerably with respect to other publications, and 
there may be small changes in the details of their definitions. Therefore, we give a 
summary of the interaction parameters in the following. Fermi and other anharmonic 
interaction parameters are designated with a plain $F$ and are used in the same way 
irrespective of a $\Delta l = 0$ or $\Delta l = 3$ interaction because the SPFIT and SPCAT 
programs use only $l = 0$ and $\pm1$. The parameters $G_b$ and $F_{ac}$ are first and 
second order Coriolis-type parameters, respectively, of $b$-symmetry, i.e., they are 
coefficients of $iJ_b$ and $(J_aJ_c + J_cJ_a)/2$, respectively. The parameters $G_a$ and 
$F_{bc}$ are defined equivalently. The interacting states are given in parentheses 
separarated by a comma; the degree of excitation of a fundamental and the $l$ quantum number 
are given as superscripts separated by a comma if necessary. Rotational corrections to these 
three types of parameters are designated with $J$ and $K$ subscripts, respectively, as is 
usually the case. There may also be $\Delta K = \Delta l = 2$ corrections (i.e. $J_+^2 + J_-^2$; 
where $J_{\pm} = J_a \pm iJ_b$) to these parameters; they are indicated by a subscript 2. 
Higher order corrections with $\Delta K = \Delta l = 4$ etc. are defined and indicated 
equivalently. Additional aspects relevant to the spectroscopy of CH$_3$CN were detailed 
earlier \cite{MeCN_v8le2_2015}. Further, and more general information on SPFIT and SPCAT 
is available in Refs.~\cite{spfit_Novick_2016} and \cite{spfit_Drouin_2017} and in the 
Fitting Spectra section\footnote{See https://cdms.astro.uni-koeln.de/classic/pickett} 
of the Cologne Database for Molecular Spectroscopy, CDMS \cite{CDMS_2005,CDMS_2016}.

After a brief description of the spectroscopy of CH$_3$CN and interactions in low-lying 
vibrational states, we summarize the previous data situation for individual states, 
the new data, and important aspects of the added spectroscopic parameters, starting 
with $\varv _4 = 1$ because it is the most extensive addition to our global fit of 
low-lying CH$_3$CN vibrational states. After presentation of the global fit parameters, 
we summarize our findings for $\varv _4 = \varv _8 = 1$.

\subsection{Overview of the spectroscopy and interactions in low-lying vibrational 
states of CH$_3$CN}
\label{intro-spec}


 \begin{figure}
 \begin{center}
  \includegraphics[width=5.0cm]{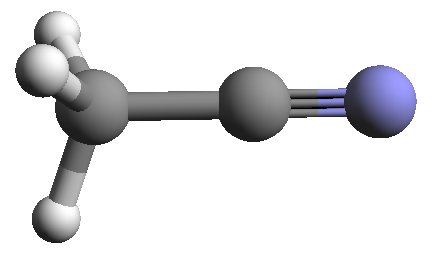}
 \end{center}
  \caption{Model of the methyl cyanide molecule. The C atoms are shown in gray, 
           the H atoms in light gray, and the N atom is shown in blue. The $a$-axis is 
           along the CCN atoms and is also the symmetry axis.}
  \label{CH3CN_molecule}
 \end{figure}


The six atoms in methyl cyanide lead to 12 vibrational degrees of freedom, four totally 
symmetric and four doubly degenerate fundamentals. 
The three lowest energy fundamentals are the doubly degenerate CCN bending mode $\nu_8$ 
at 365.024~cm$^{-1}$ \cite{MeCN_v8le2_2015}, the totally symmetric CC stretching mode 
$\nu_4$ at 920.290~cm$^{-1}$ \cite{MeCN_nu4_nu7_3nu8_1993}, and the doubly degenerate 
CH$_3$ rocking mode $\nu_7$ at 1041.855~cm$^{-1}$ \cite{MeCN_nu4_nu7_3nu8_1993}. 
These bands are comparatively weak with integrated room temperature cross sections 
in the range 2 to $5 \times 10^{-19}$~cm/molecule.

Methyl cyanide is a strongly prolate molecule ($A \gg B$) because the light H atoms 
are the only ones not on the symmetry axis of CH$_3$CN, as indicated in 
\textbf{Fig.~\ref{CH3CN_molecule}}. Rotational transitions obey the $\Delta K = 0$ 
selection rules. The large dipole moment of 3.92197~(13)~D \cite{MeCN-dipole} 
causes the transitions to be particularly strong. $\Delta K = 3$ transitions only 
gain intensity through centrifugal distortion effects and are usually too weak to 
be observed. The purely axial parameters $A$ (or $A - B$), $D_K$, etc. cannot be 
determined by rotational spectroscopy, unless perturbations are present. 
Rovibrational spectroscopy yields, strictly speaking, the differences $\Delta A$ 
(or $\Delta (A - B)$), $\Delta D_K$, etc. from single state analyses. Thus, the 
ground state axial parameters cannot be determined from such fits either. 
In the case of CH$_3$CN, they were determined through analyses of three IR bands 
involving two doubly degenerate vibrational modes, $\nu_8$, $\nu_7 + \nu_8$, and 
$\nu_7 + \nu_8 - \nu_8$ \cite{MeCN_DeltaK=3_1993} and improved through perturbations 
\cite{MeCN_v8le2_2015}.

The value of $A$, $\sim$5.27~cm$^{-1}$, leads to a rapid increase in rotational 
energy with $K$. The $J = K = 9$ level is at 429~cm$^{-1}$, higher than 
the vibrational energy of $\varv _8 = 1$. The highest $K$ levels observed 
involve $K = 21$, and the $J = K = 21$ energy is at 2313~cm$^{-1}$.

Low-lying degenerate bending modes commonly display strong Coriolis interaction 
between the $l$ components. The Coriolis parameter $\zeta$ in $\varv _8 = 1$ of 
CH$_3$CN is 0.8775, close to the limiting case of 1. The $K$ levels with $l = +1$ 
are pushed down in energy, and those with $l = -1$ are pushed up with the result 
that levels differing in $K$ by $\pm2$ and in $l$ by $\pm2$ are close in energy. 
These levels have the same symmetry and can thus repel each other through the 
$q_{22}$ interaction, which causes widespread effects in $\varv _8 = 1$ and its 
overtone states. The $\varv = 0$ $K$ levels rise in energy faster than those of 
$\varv _8 = 1^{+1}$ because of the shift of those $K$ levels to lower energies, 
see for example Fig.~4 in Ref.~\cite{MeCN_v8le2_2015}. The $K = 14$ levels of 
$\varv = 0$ and those of $K = 12$ of $\varv _8 = 1^{+1}$ cross between $J =42$ and 43. 
The effects are small ($\le 25$~MHz) and rather localized to $\sim$12 transitions in 
each vibrational state, of which several shifts are below 1~MHz. Nevertheless, the 
observation of the most perturbed transitions within each state and the much weaker 
transitions between the vibrational states introduced a very accurate local 
constraint on the CH$_3$CN energy level structure.

There are three $l$ components in the case of $\varv _8 = 2$, $l = 0$ with 
$E_{\rm vib} = 716.750$~cm$^{-1}$ and $l = \pm2$ with $E_{\rm vib} = 739.148$~cm$^{-1}$ 
\cite{MeCN_v8le2_2015}. The effective strength of the Coriolis interaction between 
the $l = +2$ and $l = -2$ levels is two times that betwen the $l$ components in 
$\varv _8 = 1$, causing not only levels with $\Delta K = \pm2$ and $\Delta l = \pm2$ 
to be close in energy, but also those with $\Delta K = \pm4$ and $\Delta l = \pm4$. 
There is a $q_{22}$ resonance between $K = 2$ of $\varv _8 = 2^{-2}$ and $K = 4$ of 
$\varv _8 = 2^{0}$. Consequently, transitions of the nominally forbidden $2\nu _8^{\pm2}$ 
band gain substantial intensity and are almost as strong as $2\nu _8^{0}$ for levels 
close to the resonance. This facilitated an accurate determination of the origin 
of $2\nu _8^{\pm2}$. The shifts in the $K$ levels in $\varv _8 = 1$ and 2 produce 
a level crossing between $K = 13$ and 14 for $\varv _8 = 1^{-1}$ and $\varv _8 = 2^{+2}$, 
a Fermi resonance with $\Delta l = 3$. Large effects of resonances such as this are 
fairly localized in $K$, but usually widespread in $J$.

Interactions with $\Delta \varv_8 = \pm 1$, $\Delta K = \mp 2$, $\Delta l = \pm 1$, such 
as that between $\varv _8 = 0$ and 1, occur also between $\varv _8 = 1$ and 2 and have been 
analyzed in our previous study \cite{MeCN_v8le2_2015}, as have been $\Delta l = 3$ Fermi 
resonances between $\varv _8 = 2$ and 3 through rotational transitions in $\varv _8 = 2$. 
Additional information will be given in Section~\ref{results_v8=2_rot}.

Adding the $K$ levels of $\varv _4 = 1$ and $\varv _7 = 1$ between $\varv _8 = 2$ and 3 
to the picture creates many more opportunities for resonant or near-resonant interactions 
as mentioned in the introduction and to some extent in the following section. 
Moving up in energy increases the number of vibrational states and potential 
resonances rapidly.

\subsection{The $\varv _4 = 1$ state}
\label{results_v4=1}

Around 2014, when our fit of low-lying vibrational states of CH$_3$CN was restricted to 
$\varv _8 \le 2$, the $\varv _4 = 1$ data set contained more than two thirds of the final 
$\nu _4$ lines; mostly lines with high $J$ or $K$ and $Q$-branch lines were added later. 
The rotational data at that time consisted of earlier published data with $J'' = 6$ to 9 
\cite{MeCN-14-15_v4=1_rot_1975}, $J'' = 19$ \cite{MeCN-vib_le_v4_J=19_1988}, and 
$J'' = 18$, 20, and 23 \cite{v4=1_v8=3_1991}, of which the $J'' = 6$, 7, and 23 lines 
were retained in the final line list. Transitions with $J'' = 24$ to 64 between 455~GHz 
and almost 1.2~THz were extracted from the spectra taken at JPL; transitions with 
$J'' = 86$ at almost 1.6~THz were too noisy and were omitted from the final line list. 
Spectra in the fit taken in Cologne covered at that time $J'' = 3$, high-$K$ lines 
of $J'' = 47$ and 48 and several transitions with $J'' = 61$ to 78 at 1.11 to 1.44~THz 
to varying degrees. The lines were straightforward to assign, except few very perturbed 
lines.


 \begin{figure}
 \begin{center}
  \includegraphics[width=8.5cm]{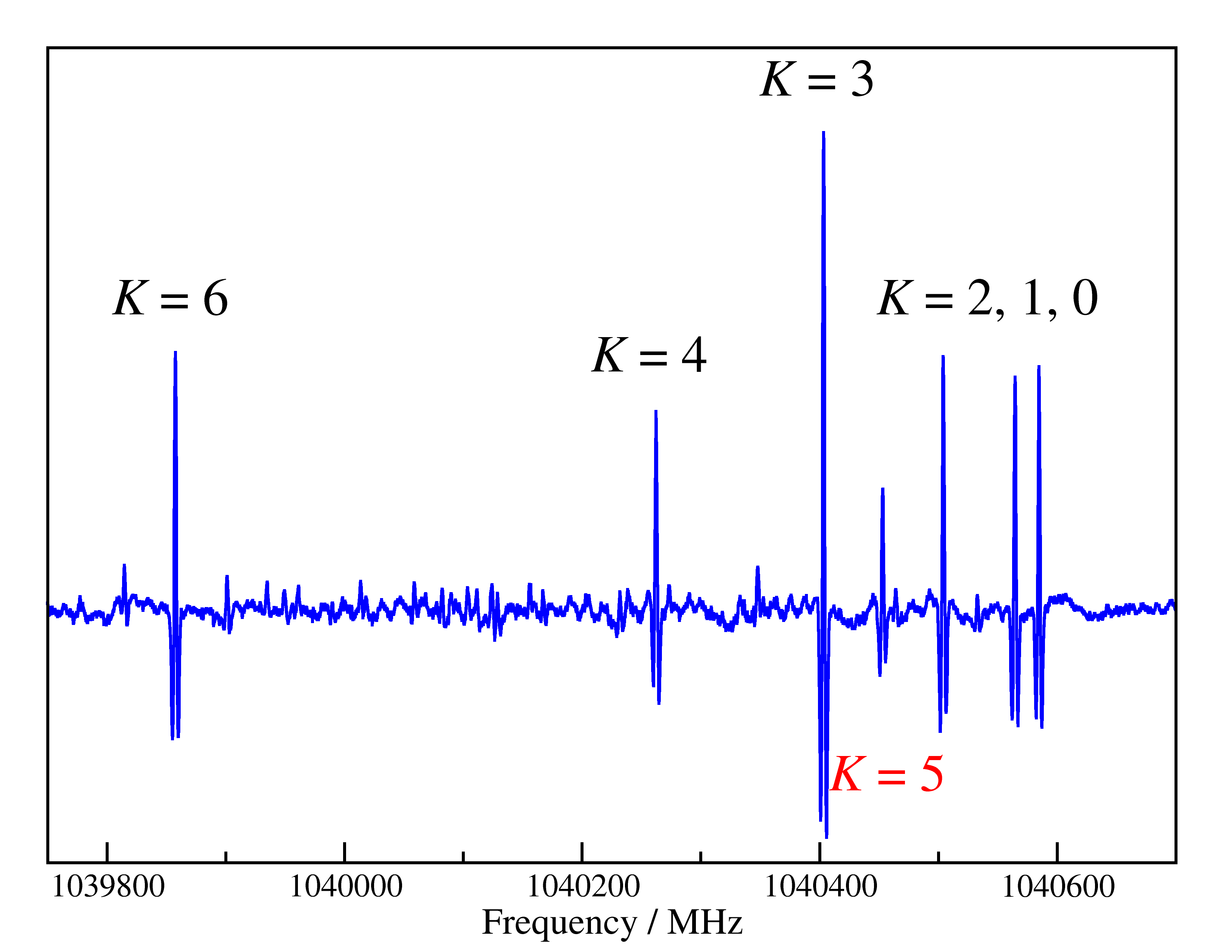}
 \end{center}
  \caption{Section of the  $\varv _4 = 1$ rotational spectrum of CH$_3$CN displaying 
           part of the $J = 57 - 56$ transitions. The $K = 5$ transition has been  
           shifted higher in frequency because of the resonance with $K = 8$ of  
           $\varv _8 = 2^0$.}
  \label{reso_v4_K5_v8x2_K8}
 \end{figure}


Fitting the lines was much more challenging, as a large number of distortion parameters 
appeared to be needed and some, mostly small, but systematic residuals remained. 
Moreover, all attempts fitting the $\varv _4 = 1$ data led to large residuals in the 
ground state $\Delta K = 3$ loops and changes in the purely axial ground state parameters 
well outside the uncertainties. In order to see the effect and the quality of newly added 
$\varv _4 = 1$ data or the influence of added parameters on the fit more easily, 
we performed a refit of $\varv _4 = 1$ initially constrained to only the rotational data 
in this state. In order to account for the effects caused by interactions with other 
vibrational states, we took the entire spectroscopic parameter set from our previous 
study \cite{MeCN_v8le2_2015} and kept all parameters fixed. Parameters were floated 
or new parameters added only if this resulted in a substantial improvement of the 
rms error as a measure of the quality of the fit. At each step, care was taken to 
float or add only the parameter that resulted in the greatest improvement of the 
rms error.

\subsubsection{Rotational data and analysis}
\label{results_v4=1_rot}

The refit was started with initial rotational data up to 440~GHz which encompassed 
the previously published data 
\cite{MeCN-14-15_v4=1_rot_1975,MeCN-vib_le_v4_J=19_1988,v4=1_v8=3_1991} and the $J'' = 3$ 
lines from Cologne. Most of the transition frequencies were calculated quite well by 
the initial parameters with the exception of the $K = 12$ lines, which were reported 
almost 1~MHz lower. The rms error before the fit of almost 2.0 was reduced to slightly 
below 1.0 with small to modest changes in the $\varv _4 = 1$ parameters.

Addition of $J'' = 1$, 2, 4, and 5 data caused minor changes. Many of these and of 
the $J'' = 3$ transitions displayed partially or fully resolved $^{14}$N hyperfine 
splitting. In addition, we observed some of the weaker $\Delta F = 0$ components. These 
data established that the excitation of $\varv _4$ affects the hyperfine structure 
negligibly as we obtained $\Delta eQq = -1.0 \pm 4.1$~kHz. This parameter was tested 
later again with essentially the same result and therefore not retained in the final fit.

The subsequent increase in $J$, $J'' = 24$ to 27 were added next to the line list, 
but more importantly the increase in $K$ to 15 required octic distortion parameters 
to be included, in particular $\Delta L_{KKJ}$ together with $\Delta L_{JK}$. 
Addition of $J'' = 28$ to 34 up to 640~GHz did not yet require further parameters, 
but perturbations in the $K = 7$ lines could not be accounted by the initial 
interaction parameters. However, the effects were small, less than 0.36~MHz, and 
the lines were given reduced weights preliminarily, effectively ignoring the perturbation 
at this stage of the analysis.

The addition of series of transitions with higher $J$ ($\ge 42$) did not yield satisfactory 
results with one or two additional $\varv _4 = 1$ distortion parameters. This situation did not 
change by adding newly recorded transitions in the 2 and 1~mm regions with $8 \le J'' \le 19$ 
and $K$ up to 16.

In order to test if the difficulties fitting the higher-$J$ $\varv _4 = 1$ data were caused 
by interactions with other vibrational states, we combined the $\varv _4 = 1$ rotational data 
up to $J'' = 34$ with the $\nu _4$ assignments existing at that time and the $\varv _8 \le 2$ 
data from Ref.~\cite{MeCN_v8le2_2015} supplemented by data recorded at 2 and 1~mm in 
the course of the present investigation as detailed in Sections~\ref{results_v8=2_rot} 
and \ref{results_v8=1_rot}. The spectroscopic parameters were essentially those of 
the previous study, with the exception for $\varv _4 = 1$, for which we took the latest set 
of parameters. The change in parameter values was small in most cases, and often within 
the present uncertainties. Notable exceptions were $\Delta (A-B)$, $\Delta D_K$, and 
$\Delta H_K$, which took values quite close to the now final ones, as well as 
$F_{ac}(8^{2,\pm2},4^1)$ and $F_{2ac}(8^{2,0},4^1)$, which describe the interactions between 
$\varv _8 = 2$ and $\varv _4 = 1$. Even though the ground state purely axial parameters 
changed little, they were kept fixed in the fit for now.

The addition of $K = 0$ to 3 lines with $J'' \le 79$ caused a pronounced change in $\Delta D_J$ and 
a relatively larger one in $\Delta H_J$. The parameter $\Delta L_{JJK}$ was added to the fit later. 
In the course of fitting these data, additional, limited measurements in the 748$-$1077~GHz 
region were made for $K = 3$, 4, and 10 to 15, and more extensive measurements were made for 
$K = 5$ to 9 in order to improve the analyses of the interactions between $\varv _4 = 1$ with 
other vibrational states. The impact of including $K = 4$ lines up to high $J$ in the fit was 
small. Subsequently, lines with $K = 5$, 6, 7, 9, and 8 were added one after the other.

There is a crossing in energy between $K = 5$ and 6 of $\varv _4 = 1$ and $K = 6$ and 7 of 
$\varv _8 = 2^{-2}$ on one hand and $K = 8$ and 9 of $\varv _8 = 2^{0}$ on the other hand, 
see \textbf{Fig.~\ref{v4-resos}} and also the $K$-level diagram in Fig.~2 of 
Ref.~\cite{MeCN_nu4_nu7_3nu8_1993}. Resonances occur in $K = 5$ of $\varv _4 = 1$ at $J = 73$ 
in the first case and at $J = 57$ and $K = 8$ in the second case. The previous line list 
contained the $J'' = 74$ transition of $\varv _8 = 2^{-2}$, $K = 6$ in the vicinity 
of the first resonance. 
We added $J'' = 71$ and 75$-$77 of $\varv _4 = 1$, $K = 5$ in the present study and, at 
the resonance, the two slightly stronger $J = 73 - 72$ transitions between the states and 
the weaker transition within $\varv _4 = 1$. In the case of the second resonance involving 
$K = 8$ of $\varv _8 = 2^{0}$ ($\Delta K = 3$), our previous line list already contained 
$J'' = 55$ and 56. In the course of the present investigation, we recorded the $J'' = 57$ 
and 58 transitions as well as the weaker $J = 57 - 56$ and $58 - 57$ transitions between 
both vibrational states. The displacement of the $J = 57 - 56$, $K = 5$ transition 
is shown in \textbf{Fig.~\ref{reso_v4_K5_v8x2_K8}}. The transitions associated with these 
resonances in $K = 5$ of $\varv _4 = 1$ affected not only the value of 
$F_{2ac}(8^{2,0},4^1)$, but also the $\varv _8 = 2^{\pm2}$ parameters; for example, 
$\eta_K$ of $\varv _8 = 2^{\pm2}$ was now essentially equal to $\eta_K$ of $\varv _8 = 1$ 
and $\Delta D_K$ of $\varv _8 = 2^{\pm2}$ was nearly identical to $\Delta D_K$ 
of $\varv _8 = 2^{0}$, as one would expect.


 \begin{figure}
 \begin{center}
  \includegraphics[width=8.5cm]{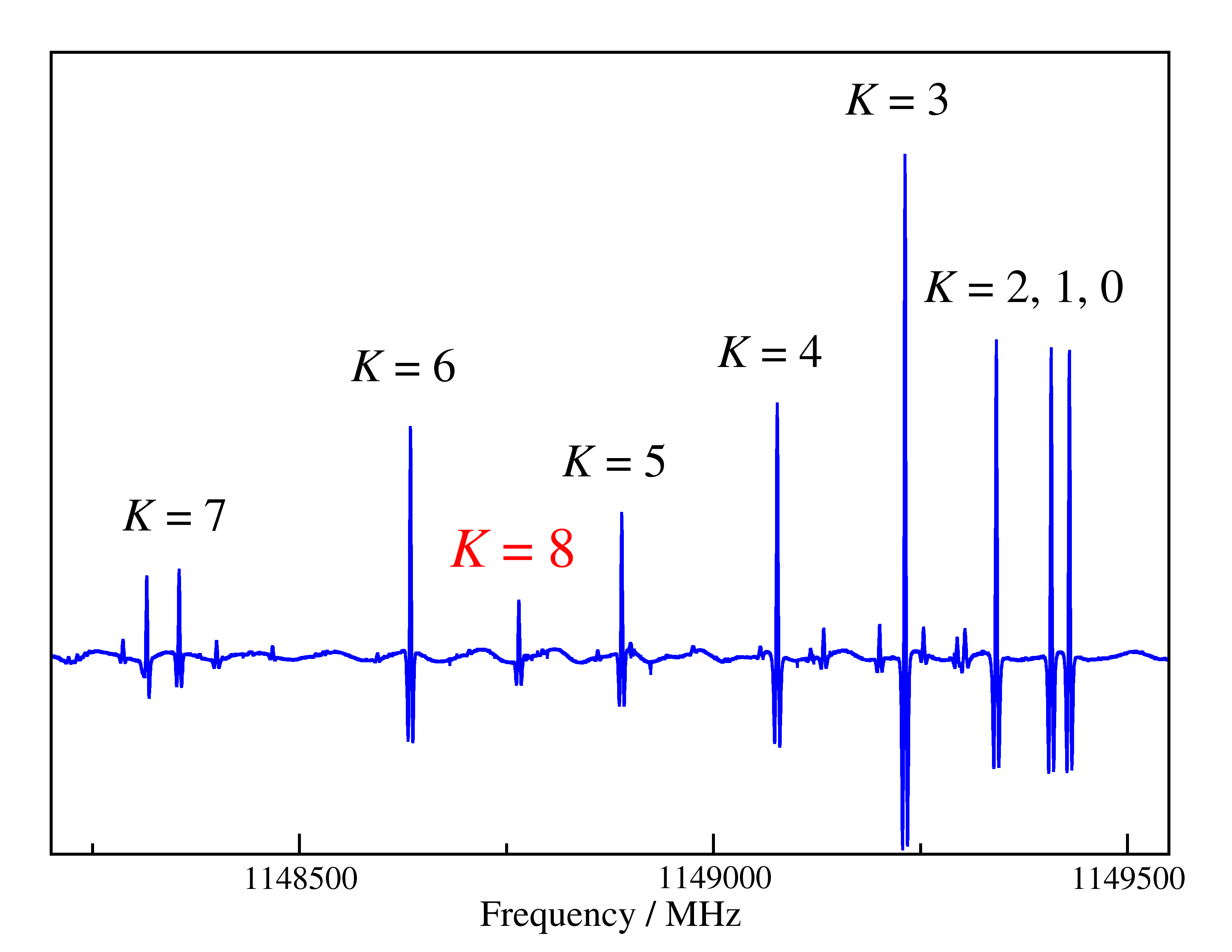}
 \end{center}
  \caption{Section of the $\varv _4 = 1$ rotational spectrum of CH$_3$CN displaying 
           part of the $J = 63 - 62$ transitions. The $K = 8$ transition has been  
           shifted higher in frequency because of the resonance with $K = 8$ of  
           $\varv _8 = 3^{+3}$. The largest perturbation occurs for $J = 66 - 65$, 
           see also \textbf{Fig.~\ref{fortrat_v4_K8_v8x3_K8}}.}
  \label{reso_v4_K8_v8x3_K8}
 \end{figure}


 \begin{figure}
 \begin{center}
  \includegraphics[width=8.5cm]{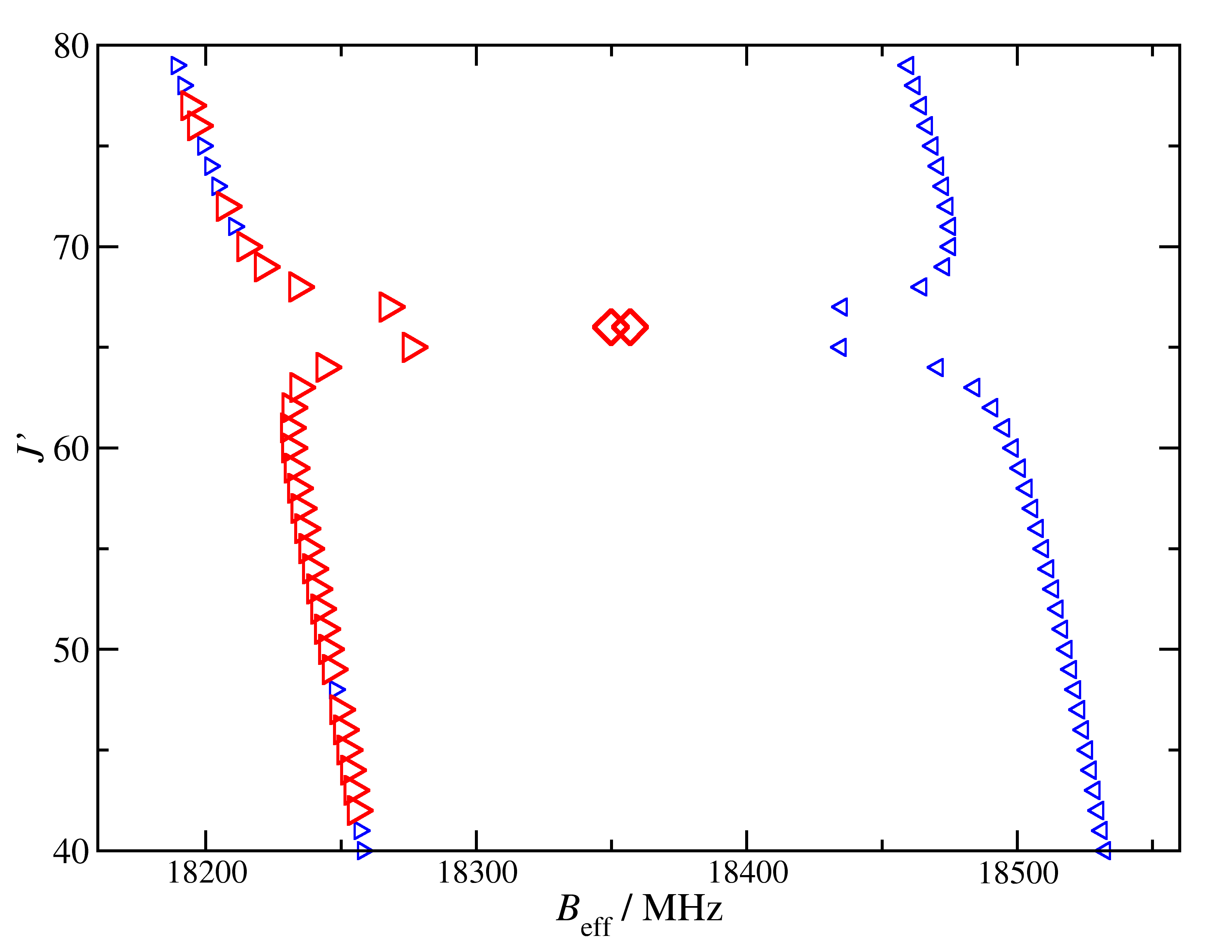}
 \end{center}
  \caption{Section of the Fortrat diagram of the rotational spectrum of CH$_3$CN 
           displaying the anharmonic resonance between $K = 8$ of $\varv _4 = 1$ 
           (left) and $\varv _8 = 3^{+3}$ (right). Measured transitions in the final fit 
           are shown in large triangles (online red), calculated transitions in smaller 
           ones (online blue). The two transitions at $J' = 66$ are nominally between 
           the states and shown as diamonds; see also \textbf{Fig.~\ref{reso_v4_K8_v8x3_K8}}.}
  \label{fortrat_v4_K8_v8x3_K8}
 \end{figure}


A crossing in energy occurs also between $K = 6$ and 7 of $\varv _4 = 1$ and $K = 4$ and 5 of 
$\varv _7 = 1^{+1}$. The energy difference at $J = 6$ of $K = 6$ and 4 is 6.2~cm$^{-1}$, 
increasing to 14.3~cm$^{-1}$ at $J = 78$, the highest value in our line list. In contrast, 
the energy difference at $J = 7$ of $K = 7$ and 5 is 18.7~cm$^{-1}$, decreasing to 9.4~cm$^{-1}$ 
at $J = 78$. Even though a crossing is predicted to occur at $J$ well above 100, it appears 
as if the perturbation in $K = 6$ of $\varv _4 = 1$ contributes more to the interaction parameter 
than the perturbation in $K = 7$.

Particularly challenging was fitting the $K = 7$ and 8 transitions. First, there is a crossing 
between $K = 7$ and 8 of $\varv _4 = 1$ with $K = 7$ and 8 of $\varv _8 = 3^{+3}$. At low $J$, 
the effect is larger in $K = 7$, as the energy difference is 8.0~cm$^{-1}$, while it is 
19.6~cm$^{-1}$ in $K = 8$. But the energy difference in $K = 7$ increases to 35.7~cm$^{-1}$ 
at $J = 78$, the highest value in our present data set, whereas there is a resonance at 
$J = 65$ and 66 in $K = 8$. The perturbations are pronounced already a few $J$ below 
the resonance, as can be seen in \textbf{Fig.~\ref{reso_v4_K8_v8x3_K8}}. The Fortrat diagram 
in \textbf{Fig.~\ref{fortrat_v4_K8_v8x3_K8}} demonstrates the good coverage of transitions in 
the fit on the $\varv _4 = 1$ side which include in particular the most strongly perturbed 
$J = 66 - 65$ transitions nominally between the two vibrational states. There are, however, 
additional resonances in $K = 7$ of $\varv _4 = 1$, first with $K = 5$ of $\varv _8 = 3^{+1}$ 
at $J = 40$ and 41, second with $K = 3$ of $\varv _8 = 3^{-1}$ at $J = 56$, and finally with 
$K = 1$ of $\varv _8 = 3^{-3}$ at $J = 42$. In other words, $\varv _8 = 3$ is linked 
with all of its $l$-components to $\varv _4 = 1$ through rovibrational interactions.

The addition of high-$J$ lines with $K = 7$ improved $F(4,8^{3,\pm3})$ considerably and 
$F_2(4,8^{3,\pm1})$ even more so. The parameters $\Delta P_{KJ}$ and $\Delta P_{JK}$ 
were added to the fit with the $K = 9$ lines.


\begin{table*}
\begin{center}
\caption{Maximum $J$ value and $K$ values and number of rotational and IR lines in 
         vibrational substates of methyl cyanide used in the present / previous study$^a$, 
         with numbers retained in the present fit from previous data in parentheses and 
         rms error for subsets of data.}
\label{statistics}
{\footnotesize
\begin{tabular}[t]{lccccccc}
\hline 
                          & $\varv = 0$       & $\varv _8 = 1^{+1}$ & $\varv _8 = 1^{-1}$ & $\varv _8 = 2^0$  & $\varv _8 = 2^{+2}$ & $\varv _8 = 2^{-2}$ & $\varv _4 = 1$    \\
\hline
$J_{\rm max}$(rot)        & 89/89             & 88/88               & 88/88               & 88/88             & 88/88               & 88/88               & 79/24             \\
$K_{\rm max}$(rot)        & 21/21             & 20/20$^b$           & 17/17               & 13/13$^c$         & 19/19$^c$           & 11/11               & 16/12             \\
no. of rot lines$^{d}$    & 316/316           & 605/596             & 549/527             & 488/424           & 607/546             & 351/299             & 705/86            \\
rms$^e$ / kHz             & 21.9/22.2         & 43.3/43.7           & 40.6/39.9           & 52.6/60.9         & 54.5/58.1           & 53.9/78.2           & 51.2/$-$          \\
rms error$^e$             & 0.862/0.859       & 0.803/0.807         & 0.797/0.796         & 0.965/0.982       & 0.906/0.915         & 0.971/0.957         & 1.021/$-$         \\
$J_{\rm max}$(IR)         & $-$/$-$$^f$       & 71/71               & 74/74               & 67/66             & 57/53               & 66/66               & 61/73             \\
$K_{\rm max}$(IR)         & 7/7               & 11/11               & 11/11               & 13/12             &  2/2                &  5/7                & 13/12             \\
no. of  IR lines$^{d}$    & 5/5$^f$           & 836/836             & 861/861             & 963/935           & 57/40               & 195/197             & 1290/1173         \\
rms / $10^{-4}$~cm$^{-1}$ & 0.07/0.04         & 0.22/0.22           & 0.19/0.19           & 0.17/0.19         & 0.26/0.38           & 0.20/0.24           & 0.09/0.13         \\
rms error$^e$             & 1.328/0.716       & 0.847/0.826         & 0.797/0.797         & 0.634/0.771       & 0.836/849           & 0.703/0.727         & 0.920/$-$         \\
\hline \hline
\end{tabular}\\[2pt]
}
\end{center}
{\footnotesize
$^a$ Previous data from Ref.~\cite{MeCN_v8le2_2015} and references therein for the $\varv _8 \le 2$ data. Previous $\nu_4$ data from Ref.~\cite{MeCN_nu4_nu7_3nu8_1993}.\\
$^b$ Initially incorrectly given as 19 \cite{MeCN_v8le2_2015} because one line with $K = 20$ was overlooked.\\ 
$^c$ These numbers were initially \cite{MeCN_v8le2_2015} interchanged.\\
$^d$ Each blend of lines counted as one line. Hyperfine splitting in present and previous data was considered in the present fit. 
     Transitions with $K = 0$ of $l$-doubled states associated with $\varv _8 = 1^{-1}$ and $\varv _8 = 2^{+2}$, respectively, in present data set. 
     Lines, which were weighted out, are not counted. Transitions between vibrational states counted for higher vibrational state.\\
$^e$ No rms error was given in Refs.~\cite{MeCN_nu4_nu7_3nu8_1993}; we assume parameter uncertainties are based on standard errors, 
     i.e. the rms error is 1.0 by definition. No rms value was given in that work for the rotational data.\\
$^f$ No individual $\Delta K = 3$ loops were given in Ref.~\cite{MeCN_DeltaK=3_1993}. We used the five $\Delta K = 3$ splittings from Table~II 
     in that work with the reported uncertainties.
}
\end{table*}


The inclusion of high-$J$ lines with $K = 8$ required in particular the adjustment of some 
$\varv _8 = 3^{\pm3}$ parameters. The origin was floated first, followed by $\Delta B$. 
Floating $A\zeta$ also improved the fit, but the minimum was much flatter than suggested 
by its uncertainty. The value of 138720~MHz was quite different from the initial 138527.8~MHz, 
but fairly close to 138656.0 and 138655.3~MHz determined in our present fits for $A\zeta$ 
of $\varv _8 = 1$ and 2, respectively. We estimated for $\varv _8 = 3^{\pm3}$ a value of 
138654.8~MHz and kept this value fixed. Subsequently, we also floated the origin of 
$\varv _8 = 3^{\pm1}$. Additional $\varv _4 = 1$/$\varv _8 = 3$ interaction parameters 
included in late fits are $F_{J}(4,8^{3,\pm3})$, $F_{2J}(4,8^{3,\pm1})$, 
$F_4(4,8^{3,\mp1})$, and $F_6(4,8^{3,\mp3})$ with which almost all of the $K = 7$ 
transitions of $\varv _4 = 1$ were calculated well. They cover for example $J'' = 43$ 
to 62 with the exception of $J'' = 50$ and 59. Unfortunately, it was difficult to 
fit either of the two potential $J = 43 - 42$ transition frequencies at 785142.02 
and 785181.83~MHz. The former line matches better the expected intensity whereas the 
latter is slightly closer to the presently calculated frequency of 785163.61~MHz. 
The line list contains three potential $J = 42 - 41$ transition frequencies; two also 
$\sim$20~MHz to either side of the present calculation and one with more than 50~MHz 
possibly too far away. The late observation of the $J = 41 - 40$ transition was 
instrumental for solving the puzzle. One line of correct intensity for this transition 
was found only $\sim$5~MHz higher than from a very late calculation. This line was easily 
accomodated by the fit. The value of $F_2$ was changed slightly, its uncertainty 
improved considerably. In contrast, the values of $F_4$ and $F_6$ changed relatively 
more, and their uncertainties were less improved. The calculated transition frequencies 
of $J'' = 41$ and 42 changed only by slightly more than 1~MHz. The over- or 
underestimation of the strength of the $\Delta K = -6$, $\Delta l = -3$ resonance 
between $\varv _4 = 1$ and $\varv _8 = 3$ most likely explains the presently poor 
fitting of the $K = 7$, $J'' = 41$ and 42 transitions. 
The authors of Ref.~\cite{MeCN_nu4_nu7_3nu8_1993} define $\Delta X$ with opposite sign 
than we do. Their experimentally determined $\Delta A$ of $\varv _8 = 3^3$ is 266~MHz, 
whereas their extrapolated value is 376~MHz, a difference of 110~MHz. This large 
discrepancy prompted us to try to increase our $\Delta (A - B)$ in magnitude in order 
to fit these two transitions better. A value of $-$392.7~MHz reproduced the $J'' = 42$ 
candidate line with the correct intensity and one of the $J'' = 41$ lines within 
uncertainties. 
Two $\nu _4$ transitions with $K = 7$ and $J' = 42$ and with the correct intensities, 
however, were observed $\sim$40~MHz lower than calculated, suggesting two other candidate 
transitions to be the correct ones. In contrast to expectations, $\Delta (A - B)$ had to 
be enlarged even further in magnitude to $-$442.06~MHz, almost to the extrapolated value 
\cite{MeCN_nu4_nu7_3nu8_1993}. Noting that several other $\varv _8 = 3$ parameters differ 
from what we would extrapolate from our $\varv _8 = 1$ and 2 parameters, we adjusted 
$\Delta D_K$ of $\varv _8 = 3$ and $A\zeta$ of $\varv _8 = 3^1$ in order to test 
the effects on $\Delta (A - B)$ of $\varv _8 = 3^3$. 
The resulting changes on $\Delta (A - B)$ of $\varv _8 = 3^3$ were $\sim$0.5~MHz in 
magnitude, but canceled due to their opposite signs. Changes to all other spectroscopic 
parameters were very small and affected the weighted rms of the fit very little, such that 
we refrained from adjusting further $\varv _8 = 3$ parameters. Inspection of the energy 
levels of $J = 42$ revealed that $K = 1$ of $\varv _8 = 3^{-3}$ has changed from 
3335~MHz or 0.11~cm$^{-1}$ below $K = 7$ of $\varv _4 = 1$ before the fit to 1979~MHz 
or 0.066~cm$^{-1}$ above $K = 7$ of $\varv _4 = 1$ after the fit. It is thus the 
perturbation between these levels which accounts for the initial deviations and for 
the subsequent constraining of $\Delta (A - B)$ of $\varv _8 = 3^3$. 
In this context it is remarkable that the $\Delta K = 6$ term $F_6(4,8^{3,\mp3})$ 
could be omitted from the final fit. The proximity of $K = 5$ of $\varv _8 = 3^{+1}$ 
to $K = 1$ of $\varv _8 = 3^{-3}$, less than \textbf{0.6}~cm$^{-1}$ for $J$ up to 45, 
may affect these results somewhat in later analyses with extensive $\varv _8 = 3$ data.

Adding the high-$J$ lines ($J'' \le 64$) with $K = 10$ and 11 did not change parameter 
values much. Adding those with $K = 12$ to 15, the last one up to $J'' = 60$ and the 
remaining high-$K$ transitions with low values of $J$ afforded $\Delta P_{KKJ}$ in the fit.

Trial fits with $G_b(4,7)$ increased changed the $\varv _4 = 1$ distortion parameters 
little and did not improve the quality of the fit. This also applied to trial fits with 
various values of $F(4,8^{2,0})$ in order to test the effect of a purported strong Fermi 
resonance between $\varv _8 = 2^0$ and $\varv _4 = 1$ \cite{FF_Duncan_1978}.

\subsubsection{The $\nu_4$ band}

\begin{figure}
 \begin{center}
  \includegraphics[width=8.5cm]{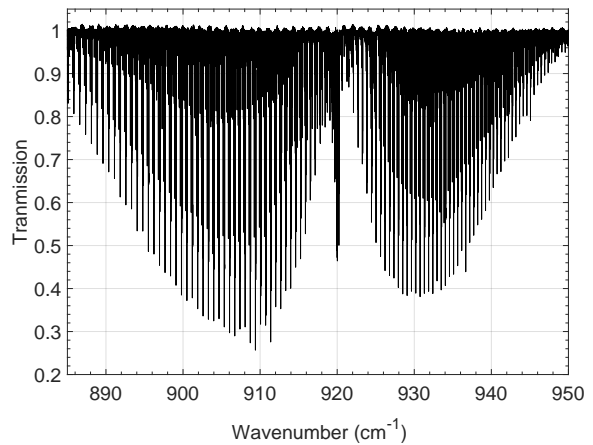}
 \end{center}
  \caption{Overview of the $\nu_4$ band of CH$_3$CN. The strong $\nu_3$ band of OCS becomes 
           prominent at lower wavenumbers. A small amount of OCS was added for calibration 
           purpose, see section~\ref{exptl_IR}.}
  \label{nu4_overview}
 \end{figure}

\begin{figure}
 \begin{center}
  \includegraphics[width=8.5cm]{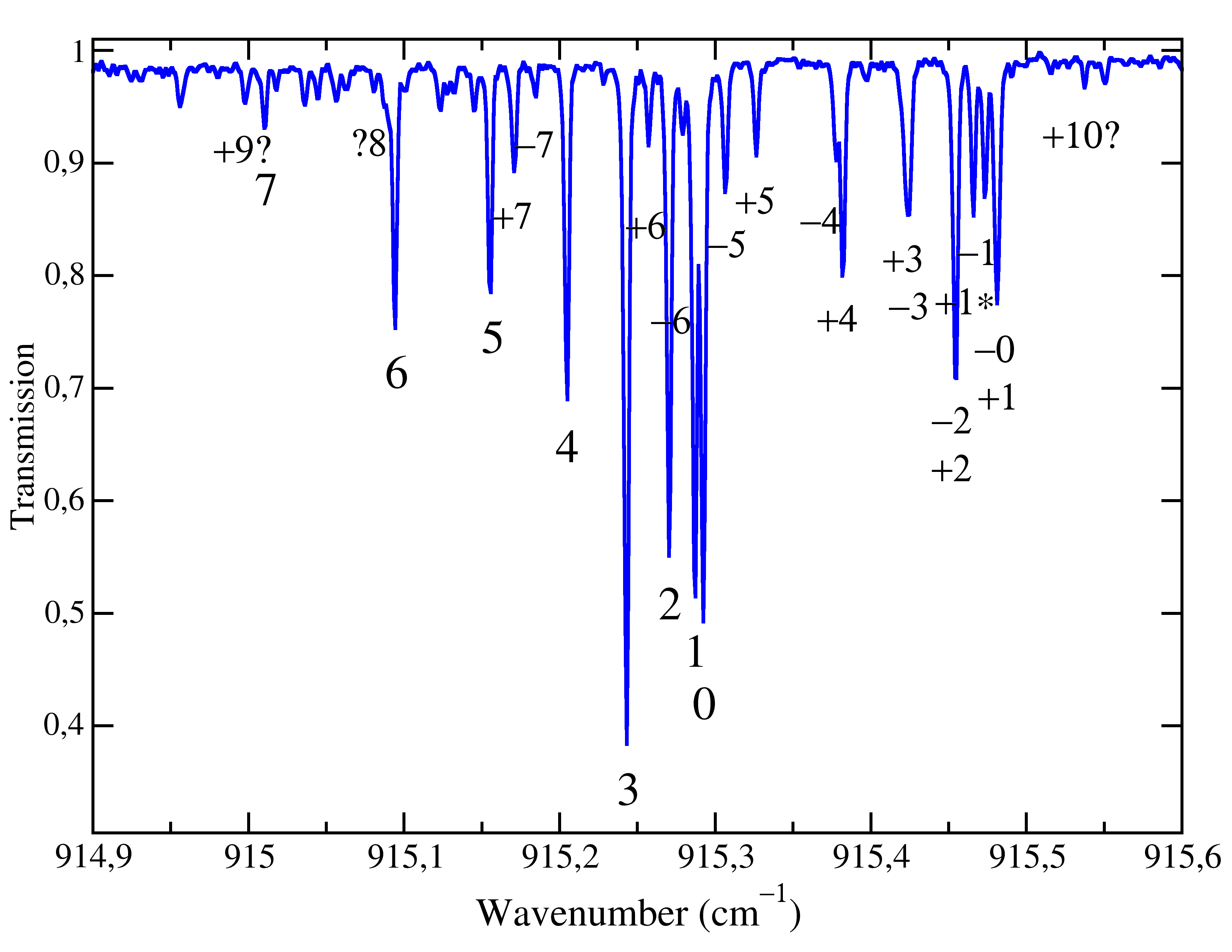}
 \end{center}
  \caption{Section of the $\nu_4$ band of CH$_3$CN showing the $J = 7 - 8$ transitions of $\nu_4$ 
           and $J = 14 - 15$ transitions of $\nu_4 + \nu_8 - \nu_8$. The $K$ quantum numbers 
           are given centered below each line for $\nu_4$; $k = K \times l$ is given in the 
           case of $\nu_4 + \nu_8 - \nu_8$. Question marks indicate tentative assignments 
           for $k = +9$ and +10; for $K = 8$, it indicates that the $K$ value is quite certain, 
           but not if it is $l = +1$ or, more likely, $l = -1$. Please note that $J = 13 - 14$ 
           for the tentative $k = +10$ transition. Unassigned transitions are probably due to 
           $\nu_4 + 2\nu_8 - 2\nu_8$.}
  \label{nu4_detail}
 \end{figure}

No line lists were provided with the earlier combined analysis of $\nu_4$, $\nu_7$, and 
$3\nu_8$ \cite{MeCN_nu4_nu7_3nu8_1993}. Therefore, we relied entirely on our new spectrum 
which is shown in \textbf{Fig.~\ref{nu4_overview}}. A pronounced Herman-Wallis effect 
causes the $P$-branch to be substantially stronger than the $R$-branch. Assignments in 
$P$- and $R$-branches of the $\nu_4$ cold band were straightforward based on the earlier 
analysis. Nevertheless, ground state combination differences were used in the initial 
round to establish the assignments. The assignments were extended to $J$ of 56 and 61 
in the $P$- and $R$-branches, respectively, and up to $K = 13$ in the final rounds, 
taking relative intensities strongly into account. While many transitions with $K = 12$ 
had good or very good S/N, almost all of the $K = 13$ lines were near or below the 
detection limit. Nevertheless, two $P$-branch transitions were assigned unambiguously 
that had sufficient intensities and appeared to be not blended with other lines. 
\textbf{Fig.~\ref{nu4_detail}} shows the $P(8)$ transitions of $\nu _4$ together with 
$P(15)$ transitions of $\nu _4 + \nu _8 - \nu _8$. The assignments in $\nu _4$ cover the 
entire range of the line list which was truncated at the high end because of the weakness 
of the $\nu_4$ lines and the onset of $\nu _7$ and at the low end by the overlap with 
the $\nu_3$ band of OCS. Even though the $Q$-branch was rather compact, many lines up 
to $J = 42$ and $K = 13$ were isolated or occured as sufficiently narrow groups of lines 
to permit inclusion in the fit. Uniform uncertainties of 0.0001~cm$^{-1}$ were attributed 
to the lines, even though this may be somewhat conservative for parts of the stronger 
lines. Statistics to this data set can be found in \textbf{Table~\ref{statistics}}.


\begin{table}
\begin{center}
\caption{Transition dipole moments $\mu$ (D), Herman-Wallis correction $\mu_{\rm HW}$$^a$ (D), and 
         integrated intensities $I_{\rm H}$$^b$ (10$^{-19}$~cm/molecule) of IR bands of CH$_3$CN 
         calculated using the present Hamiltonian model together with the extrapolated and 
         experitentally measured total integrated intensities from previous studies.}
\label{mu-IR_intensities}
{\footnotesize
\begin{tabular}[t]{lcclccccc}
\hline
        &  & & &  \multicolumn{5}{c}{Integrated intensity} \\
\cline{5-9}
band  & $\mu$ & $\mu_{\rm HW}$ & & $I_{\rm H}$ & \cite{MeCN-int_2005} & 
\cite{MeCN-int_1995} & \cite{MeCN-int_1985} & \cite{MeCN-int_1984} \\
\hline
$\nu _4$$^c$  & 0.023 & $-$0.0022 & & 2.028 & 1.95 & 1.93 & 2.30 & 2.06 \\
$2\nu _8$$^d$ & 0.030 &           & & 2.580 & 2.63 & 2.50 & 3.43 &      \\
$\nu _8$$^d$  & 0.043 &           & & 2.285 &      & 1.77 & 1.81 & 2.79 \\
\hline \hline
\end{tabular}\\[2pt]
}
\end{center}
{\footnotesize
$^a$ First-order Herman-Wallis correction; coefficient of $i(\{\phi_c,J_b\} - \{\phi_b,J_c\})/2$.\\
$^b$ Estimated from the cold bands, $\varv _4 = 1 - 0$, $\varv _8 = 2 - 0$ and $1 - 0$, by multiplication 
     with the vibrational partition factor 1.5006 at 296~K. An increase of 2.677\,\% from other isotopic 
     species was considered.\\
$^c$ $\mu$, $\mu_{\rm HW}$, and $I_{\rm H}$ from this work.\\ 
$^d$ $\mu$ and $I_{\rm H}$ from Ref.~\cite{MeCN_v8le2_2015}; $\mu_{\rm HW}$ was not needed. Please note 
     that the transition dipole moments in Ref.~\cite{MeCN_v8le2_2015} were given correctly in the text, 
     but were a factor of 10 too large in Table~10. 
}
\end{table}


The amount of CH$_3$CN in the ground vibrational state can be evaluated from the partial 
pressure of methyl cyanide by subtracting off the contributions of other isotopic species 
and taking into account the vibrational factor of CH$_3$CN at 291~K. However, a Herman-Wallis 
correction had to be evaluated also because the $P$-branch is considerably stronger than the 
$R$-branch. A transition dipole moment of 0.023~D combined with $-$0.0022~D as the coefficient 
of $i(\{\phi_c,J_b\} - \{\phi_b,J_c\})/2$ as first-order Herman-Wallis correction reproduced 
the band shape sufficiently well. These values and the resulting integrated band strength 
determined in this study is given in \textbf{Table~\ref{mu-IR_intensities}} together with 
our earlier values for $\nu_8$ and $2\nu_8$ \cite{MeCN_v8le2_2015} and band strength data from 
previous studies. The line intensities are modeled very well around the $P$- and $R$-branch 
maxima. The intensities are slightly small at low $J$ in the $R$-branch and at high $J$ 
in the $P$-branch, whereas they are slightly high at low $J$ in the $P$-branch and at 
high $J$ in the $R$-branch, or, in other words, the present Herman-Wallis correction is 
slightly too small at low $J$ and slightly too large at high $J$. There is no appropriate 
correction available in SPCAT to the best of our knowledge to reduce these small, albeit 
systematic deviations. The $\nu_4$ intensities did not show any significant intensity 
deviations with $K$. We point out that the same dipole parameters reproduce the 
$\nu_4 + \nu_8 - \nu_8$ intensities well, as is usually the case.


\subsection{The $\varv _8 = 2$ state}
\label{results_v8=2}

\subsubsection{Rotational data and analysis}
\label{results_v8=2_rot}

The $\varv _8 = 2$ rotational data in the present study were taken to a large extent 
from our previous investigation \cite{MeCN_v8le2_2015}. These data comprise parts of 
earlier data, $J'' = 0$, 4, and 5 from Ref.~\cite{MeCN-v8=2_1969} and $J'' = 19$ from 
Ref.~\cite{MeCN-vib_le_v4_J=19_1988}.

In the course of recording transitions pertaining to $\varv _4 = 1$ at 2 and 1~mm, 
we also covered $\varv _8 = 2$ transitions in this region to test the impact in the fit. 
This was partially based on the fact that the presently calculated uncertainties of 
transitions of $\varv _8 = 2^{-2}$, $K = 2$ and of $\varv _8 = 2^{0}$, $K = 4$ had 
larger uncertainties than those of most other transitions in $\varv _8 = 2$, and these 
uncertainties were large enough to warrant measurement. These larger uncertainties 
were caused by the $q_{22}$ interaction between these two $K$ levels with a crossing 
at $J = 40$. Interestingly, transitions connecting these two different $K$ ladders 
had sufficient intensities between $J'' = 8$ and 12 that they could be recorded 
with fairly good S/N. These cross-ladder transitions were comparatively strong 
far away from the resonance because the mixing coefficients for these $J$ levels 
change much more than for two successive $J$ levels closer to the resonance, 
where the mixing coefficients are almost constantly close to 0.5. These transition 
frequencies together with the remaining lines recorded in the 2 and 1~mm regions 
accounted for a considerable part of the improved $\varv _8 = 2$ parameters, in 
particular those of lower order.

Further measurements in the 770$-$1086~GHz region were mostly carried out to improve 
the coverage of interactions with $\varv _8 = 1$ and 3 and with $\varv _4 = 1$ 
through new or more accurate measurements. The interactions with $\varv _4 = 1$ 
were already described in Section~\ref{results_v4=1_rot}.

There is a Fermi resonance between $\varv _8 = 1^{-1}$ and $\varv _8 = 2^{+2}$ with 
the largest perturbations in $K = 14$ and a crossing between $J = 91$ and 92.  
Perturbations from this resonance are smaller at $K = 13$ and even smaller at other $K$. 
We added several new or remeasured transitions with $41 \le J'' \le 58$ and $K = 13$ 
to 15 in both vibrational substates. The highest $J''$ value for these $K$ was 
64 from our previous study \cite{MeCN_v8le2_2015}.

There are, however, additional resonances, which perturb the $K = 13$ levels of both 
vibrational substates, and which are described by $\Delta \varv_8 = \pm 1$, 
$\Delta K = \mp 2$, $\Delta l = \pm 1$. $K = 13$ of $\varv _8 = 1^{-1}$ interacts 
with $K = 11$ of $\varv _8 = 2^{0}$ with a crossing near $J = 60$. 
All four strong transitions were observed earlier \cite{MeCN_v8le2_2015}; 
the transitions with $J'' = 59$ are between the two vibrational states. 
Several further $K = 11$ lines of $\varv _8 = 2^{0}$ were recorded in addition. 
$K = 13$ of $\varv _8 = 2^{+2}$ interacts with $K = 15$ of $\varv _8 = 1^{+1}$; 
a crossing occurs between $J = 52$ and 53. Of the 8 relatively strong transitions 
perturbed most, the transition with $J'' = 51$ to 53 within each vibrational substate 
and the fairly strong cross-ladder transitions with $J'' = 52$, only the $J'' = 51$ 
with $K = 15$ of $\varv _8 = 1^{+1}$ is not in our line list. Three transitions were 
recorded newly and two were remeasured.

Further resonances exist between $\varv _8 = 2$ on one hand and $\varv _8 = 3$ or 
$\varv _7 = 1$ on the other hand. There is a Fermi resonance between $\varv _8 = 2^{0}$ 
and $\varv _8 = 3^{+3}$ at $K = 14$ and, more pronounced, at $K = 15$ with a crossing 
currently calculated between $J = 73$ and 74. The previous lines in $\varv _8 = 2^{0}$ 
extended to $K = 13$, and only one line each was added to $K = 12$ and 13.
Additional Fermi resonances are found between $\varv _8 = 2^{-2}$ and $\varv _8 = 3^{+1}$ 
at $K = 12$ and 13 and between $\varv _8 = 2^{-2}$ and $\varv _7 = 1^{+1}$ at $K = 13$ 
and 14. Transitions in $\varv _8 = 2^{-2}$ extend to $K = 11$, and four lines were added 
to this $K$ among others. Several $\Delta K = \mp 2$, $\Delta l = \pm 1$ interactions were 
noticed, but their effects were not yet strong enough to warrant treatment in the fit.

The change in $\varv _8 = 2$ parameter values in the course of the present investigation 
and in particular caused by the lines around the $\varv _8 = 2$/$\varv _4 = 1$ resonances 
resulted in additional parameter constraints and unconstraining another pair of parameters. 
Since $\Delta D_K$ of $\varv _8 = 2^0$ and $\varv _8 = 2^2$ were the same within three 
times the comparatively large uncertainty of the latter, the two parameters were 
constrained to be the same, essentially without deterioration of the quality of the fit. 
The resulting $\Delta D_K(\varv _8 = 2)$ was identical to two times 
$\Delta D_K(\varv _8 = 1)$ well within the larger uncertainty of the latter. Therefore, 
the ratio of $2 : 1$ was constrained. Subsequently, the values of $\eta _K$ and $q_K$ of 
$\varv _8 = 2$ were contrained to be identical to the respective $\varv _8 = 1$ values with 
only a marginal deterioration of the fit. A slight improvement was achieved by decoupling 
$\eta _{JK}$ of $\varv _8 = 1$ and 2. The mutually constrained parameters 
$F_{2,J}(8^{\pm1},8^{2,0})$ and $F_{2,J}(8^{\pm1},8^{2,\pm2})$ were introduced 
to the fit as a result of the improved data situation.

In the course of these analyses, two very weak $2\nu _8$ lines, $^oR_9(38)$ and $^oR_9(43)$, 
were found to deviate from their calculated positions by $\sim$11 times the attributed 
uncertainties of 0.0003~cm$^{-1}$ and were omitted. Also omitted was the $^oR_7(41)$ line 
with a residual slightly more than three times the uncertainty. Two high-$K$ rotational lines 
were omitted which were already weighted out because of their large residuals, $J'' = 23$, 
$K = 13$ of $\varv _8 = 2^0$ and $J'' = 55$, $K = 15$ of $\varv _8 = 2^{+2}$. Two more 
lines, $J'' = 55$, $K = 14$ and $J'' = 53$, $K = 15$ of $\varv _8 = 2^{+2}$, with 
somewhat large residuals were omitted after inspection of their intensities and their 
line shapes. Statistics to the $\varv _8 = 2$ rotational data sets are given in 
\textbf{Table~\ref{statistics}}. Please note that the maximum $K$ values of 
$\varv _8 = 2^0$ and $\varv _8 = 2^{+2}$ were exchanged erroneously in our previous 
study \cite{MeCN_v8le2_2015}.

\subsubsection{The $2\nu_8$ band}

The $2\nu _8$ data were initially taken from our previous study \cite{MeCN_v8le2_2015}, 
but with three lines omitted as described in Section~\ref{results_v8=2_rot}. 
We scrutinized the $2\nu_8$ line positions and corrected about half of the $2\nu_8^0$ 
position because these were from the initial calibration and assignments and were on 
average $\sim$0.0002~cm$^{-1}$ lower compared to the final calibration. A small number 
of transition frequencies with large residuals were corrected or were omitted because 
of blending with unknown transitions. The uncertainties of parts of the $2\nu_8^0$ 
$Q$-branch and $2\nu_8^{\pm2}$ lines were reevaluated and additional assignments were 
made which, most notably, included 5 $2\nu_8^0$ $Q$-branch transitions with $K = 13$. 
These changes affected the parameter values and uncertainties fairly little, but reduced 
the rms and rms errors of the $2\nu_8$ data sets. Isolated lines with sufficiently good 
S/N were given uncertainties of 0.0002~cm$^{-1}$ as previously. Noisier lines or lines 
in more crowded regions, such as in the $Q$-branch, were often given larger 
uncertainties of 0.0003~cm$^{-1}$ to 0.0010~cm$^{-1}$. The $J$ and $K$ ranges and 
other information on this data set are given in \textbf{Table~\ref{statistics}}. 
The quality of the data in our present fit is better than that in our previous fit, 
but still slightly worse than in the original study, in which 0.12, 0.18, and 
$0.18 \times 10^{-3}$~cm$^{-1}$ were reported for $2\nu _8^{0}$, $2\nu _8^{+2}$, 
and $2\nu _8^{-2}$, respectively from the IR + MW fit \cite{MeCN_2nu8_1993}.

\subsection{The $\varv _8 = 1$ state}
\label{results_v8=1}

\subsubsection{Rotational data and analysis}
\label{results_v8=1_rot}

The $\varv _8 = 1$ rotational data in the present study were taken for the most part 
from our previous investigation \cite{MeCN_v8le2_2015}. These data comprise parts of 
earlier data, $J'' = 3$ to 7 from Ref.~\cite{MeCN-v8=1_1969} and $k \times l = +1$ 
direct-$l$-type transitions from Refs.~\cite{MeCN-v8=1-l-type_1968} and 
\cite{MeCN-v8=1-l-type_etc_1991}.

New transition frequencies involving the $\varv _8 = 1$ and 2 interaction were mentioned 
in Section~\ref{results_v8=2_rot}. The impact of transition frequencies with $J'' = 9$ 
and $K \le 5$ was marginal in spite of uncertainties of mostly 3 or 5~kHz. We also 
determined new transition frequencies involving the $\varv _8 = 0$ and 1 interaction. 
These were remeasurements of the $J'' = 42$ and 43 transitions of $\varv _8 = 0$, 
$K = 14$ and $\varv _8 = 1^{+1}$, $K = 12$, measurement of one of the two $J'' = 42$ 
cross-ladder transitions and remeasurement of one of the two $J'' = 43$ cross-ladder 
transitions. Statistics to the $\varv _8 = 1$ rotational data sets are given in 
\textbf{Table~\ref{statistics}}. Please note that the maximum $K$ value of 
$\varv _8 = 1^{+1}$ was increased from 19 to 20 with respect to our previous study 
\cite{MeCN_v8le2_2015} because we overlooked the previously recorded $J'' = 60$ 
transition in the course of analyzing the fit statistics.


\begin{table*}
\begin{center}
\caption{Present spectroscopic parameters or differences $\Delta$ thereof $^{a,b}$ (cm$^{-1}$, MHz)$^c$ of 
         methyl cyanide within vibrational states $\varv = 0$ and $\varv_4 = 1$ in comparison to previous values.}
\label{parameter_v0_v4}
{\footnotesize
\begin{tabular}[t]{lr@{}lr@{}lrr@{}lr@{}l}
\hline 
 & \multicolumn{4}{c}{$\varv = 0$} & & \multicolumn{4}{c}{$\varv_4 = 1$} \\
\cline{2-5} \cline{7-10} 
Parameter $X$ & \multicolumn{2}{c}{Present} & \multicolumn{2}{c}{Ref.~\cite{MeCN_v8le2_2015}} & & \multicolumn{2}{c}{Present} & \multicolumn{2}{c}{Ref.~\cite{MeCN_nu4_nu7_3nu8_1993}} \\
\hline
$E_{\rm{vib}}$$^c$                          &      0&.0             &      0&.0           & &    920&.290464~(5) &    920&.290284~(13)\\
$(\Delta)^b\ (A - B)$                       & 148900&.011~(38)      & 148900&.103~(66)    & & $-$165&.941~(14)   & $-$166&.205~(33)   \\
$(\Delta)^b\ B$                             &   9198&.899163~(10)   &   9198&.899167~(11) & &  $-$46&.14784~(5)  &  $-$46&.14822~(66) \\
$(\Delta)^b\ D_K \times 10^3$               &   2827&.9~(6)         &   2830&.6~(18)      & &  $-$25&.13~(26)    &  $-$31&.63~(63)    \\
$(\Delta)^b\ D_{JK} \times 10^3$            &    177&.40780~(25)    &    177&.40787~(25)  & &      7&.130~(5)    &      7&.165~(17)   \\
$(\Delta)^b\ D_J \times 10^6$               &   3807&.572~(7)       &   3807&.576~(8)     & &   $-$5&.510~(24)   &   $-$5&.702~(99)   \\
$(\Delta)^b\ H_K \times 10^6$               &    156&.2~(20)        &    164&.6~(66)      & &     15&.5~(12)     &  $-$22&.2~(28)     \\
$(\Delta)^b\ H_{KJ} \times 10^6$            &      6&.0618~(14)     &      6&.0620~(14)   & &  $-$11&.705~(95)   &  $-$13&.895~(105)  \\
$(\Delta)^b\ H_{JK} \times 10^9$            &   1025&.62~(14)       &   1025&.69~(15)     & &    244&.7~(13)     &    315&.7~(33)     \\
$(\Delta)^b\ H_J \times 10^{12}$            & $-$238&.7~(20)        & $-$237&.4~(21)      & &    150&.3~(29)     &       &            \\
$(\Delta)^b\ L_{KKJ} \times 10^{9}$         &   $-$0&.4441~(25)     &   $-$0&.4443~(25)   & &  $-$23&.85~(69)    &       &            \\
$(\Delta)^b\ L_{JK} \times 10^{12}$         &  $-$52&.65~(51)       &  $-$52&.75~(51)     & &     78&.7~(15)     &       &            \\
$(\Delta)^b\ L_{JJK} \times 10^{12}$        &   $-$7&.876~(31)      &   $-$7&.901~(32)    & &   $-$3&.51~(16)    &       &            \\
$(\Delta)^b\ L_J \times 10^{15}$            &   $-$3&.00~(16)       &   $-$3&.10~(17)     & &       &            &       &            \\
$(\Delta)^b\ P_{KKJ} \times 10^{12}$        &       &               &       &             & &  $-$24&.3~(16)     &       &            \\
$(\Delta)^b\ P_{KJ} \times 10^{12}$         &       &               &       &             & &      2&.482~(43)   &       &            \\
$(\Delta)^b\ P_{JK} \times 10^{15}$         &      0&.507~(67)      &      0&.552~(68)    & &  $-$50&.1~(14)     &       &            \\
$(\Delta)^b\ P_{JJK} \times 10^{18}$        &     53&.9~(21)        &     55&.3~(22)      & &       &            &       &            \\
$(\Delta)^b\ eQq$                           &   $-$4&.22297~(103)   &   $-$4&.22308~(107) & &       &            &       &            \\
$(\Delta)^b\ C_{bb} \times 10^3$            &      1&.840~(89)      &      1&.845~(90)    & &       &            &       &            \\
$(\Delta)^b\ (C_{aa} - C_{bb}) \times 10^3$ &   $-$1&.17~(30)       &   $-$1&.15~(31)     & &       &            &       &            \\
\hline \hline
\end{tabular}\\[2pt]
}
\end{center}
{\footnotesize
$^a$ Numbers in parentheses are one standard deviation in units of the least significant figures. 
     Empty entries indicate parameters not applicable or not used in the fit. See Ref.~\cite{MeCN_v8le2_2015} 
     for sign and value considerations.\\
$^b$ Parameter $X$ given for $\varv = 0$; $\Delta X = X_i - X_0$, with $i$ representing an excited 
     vibrational state.\\
$^c$ All parameters given in units of megahertz, except for $E_{\rm{vib}}$, which is given in 
     units of inverse centimeters.
}
\end{table*}

\subsubsection{The $\nu_8$ band}

Transition frequencies of $\nu _8$ were taken from Ref.~\cite{MeCN_nu8_1992} without 
any change with respect to our previous study \cite{MeCN_v8le2_2015}. Briefly, almost 
all fully weighted lines were assigned uncertainties of 0.0002~cm$^{-1}$, 0.0004~cm$^{-1}$ 
for a small number of weak lines and 0.0006~cm$^{-1}$ for lines with weight 0.1. 
The $J$ and $K$ ranges and other information on this data set are given in 
\textbf{Table~\ref{statistics}}. The quality of the data in our present and previous 
fits are about the same and only slightly worse than in the original study in which 0.17 
and $0.16 \times 10^{-3}$~cm$^{-1}$ were reported for $\nu _8^{+1}$ and $\nu _8^{-1}$, 
respectively from the IR + MW fit.

\subsection{The $\varv = 0$ state}
\label{results_v=0}

The $\varv = 0$ data are identical to those of our previous investigation \cite{MeCN_v8le2_2015} 
with the exception of the new transition frequencies involving the $\varv _8 = 0$ and 1 interaction 
mentioned in Section~\ref{results_v8=1_rot}. The previous data consist of Lamb-dip measurements 
covering parts of $J'' = 4$ to 42 from 91 to 780~GHz \cite{MeCN_rot_2006}, terahertz data 
with $59 \le J'' \le 81$ between 1.09 and 1.51~THz \cite{MeCN_v8le2_2015} and with $J'' = 86$ 
to 88 near 1.6~THz \cite{MeCN_rot_2009}. Also included were perturbed $K = 14$ transitions 
mostly from Ref.~\cite{MeCN_v8le2_2015} and from Ref.~\cite{MeCN_rot_2004} as well as 
hyperfine split transitions with $J'' = 0$ \cite{MeCN_1-0}, $J'' = 1$ \cite{MeCN-12-13b_2-1}, 
and $J'' = 2$ and 7 \cite{MeCN-Lille_1977}. Direct information on the axial parameters 
come from five ground state combination loops with $K = 3 - 0$ to $7 - 4$ which were 
averaged over several different $J$ \cite{MeCN_DeltaK=3_1993}. Trial fits with these 
parameters floated showed that the they still differed somewhat from our previous values 
\cite{MeCN_v8le2_2015}. But these differences were deemed to be small enough to keep 
these parameters floated, even more so, as their uncertainties improved substantially 
with respect to those in \cite{MeCN_v8le2_2015}. 
Fit statistics are again provided in \textbf{Table~\ref{statistics}}.


\begin{table}
\begin{center}
\caption{Spectroscopic parameters or differences $\Delta$ thereof $^{a,b}$ (cm$^{-1}$, MHz)$^c$ of 
         methyl cyanide within vibrational states $\varv_7 = 1$ and $\varv_8 = 3$ mostly taken 
         from Ref.~\cite{MeCN_nu4_nu7_3nu8_1993}.}
\label{parameter_v7_v8x3}
{\footnotesize
\begin{tabular}[t]{lr@{}lr@{}lr@{}l}
\hline 
Parameter $X$ & \multicolumn{2}{c}{$\varv_7 = 1$} & \multicolumn{2}{c}{$\varv_8 = 3^1$} & \multicolumn{2}{c}{$\varv_8 = 3^3$}  \\
\hline
$E_{\rm{vib}}$$^c$           &   1041&.85471       &   1077&.7863        &   1122&.15          \\
$E_{\rm{vib}}$$^{c,d}$       &       &             &   1077&.7919~(2)    &   1122&.3489~(1)    \\
$\Delta (A - B)$             &    889&.440         & $-$302&.030         & $-$347&.76          \\
$\Delta (A - B)$$^d$         &       &             &       &             & $-$442&.16~(6)      \\
$\Delta B$                   &   $-$5&.73413       &     80&.29485       &     81&.289         \\
$\Delta B$$^d$               &       &             &       &             &     80&.6130~(13)   \\
$\Delta D_K \times 10^3$     &    149&.333         &  $-$22&.8$^e$       &  $-$22&.8$^e$       \\
$\Delta D_K \times 10^3$$^d$ &       &             &  $-$30&.6$^e$       &  $-$30&.6$^e$       \\
$\Delta D_{JK} \times 10^3$  &      0&.9731        &      3&.5046$^e$    &      3&.5046$^e$    \\
$\Delta D_J \times 10^6$     &     14&.001         &    462&.$^e$        &    462&.$^e$        \\
$\Delta H_{JK} \times 10^9$  & $-$148&.1           &       &             &       &             \\
$A\zeta$                     &  66663&.668         & 138665&.87          & 138527&.8           \\
$A\zeta$$^d$                 &       &             & 138656&.0           & 138654&.7           \\
$\eta_K$                     &      7&.2248        &     11&.013$^e$     &     11&.013$^e$     \\
$\eta_J$                     &      0&.078153      &      0&.40154$^e$   &      0&.40154$^e$   \\
$\eta_{KK} \times 10^6$      &     64&.8           &       &             &       &             \\
$\eta_{JK} \times 10^6$      &  $-$50&.34          &       &             &       &             \\
$\eta_{JJ} \times 10^6$      &      2&.385         &      7&.285$^e$     &      7&.285$^e$     \\
$q$                          &      4&.7634        &     17&.683$^{e,f}$ &     17&.683$^{e,f}$ \\
$q_{J} \times 10^6$          &  $-$10&.85          &  $-$75&.79$^{e,f}$  &  $-$75&.79$^{e,f}$  \\
\hline \hline
\end{tabular}\\[2pt]
}
\end{center}
{\footnotesize
$^a$ Empty entries indicate parameters not applicable or not used in the fit. 
     Signs of $q$ and $q_J$ altered, see Ref.~\cite{MeCN_v8le2_2015}.\\
$^b$ Parameter difference $\Delta (X) = X_i - X_0$, given for rotational and distortion parameters  
     ($i$ represents an excited vibrational state).\\
$^c$ All parameters given in units of megahertz, except for $E_{\rm{vib}}$, which is given in 
     units of inverse centimeters.\\
$^d$ Value determined or estimated in the present work, see also Section~\ref{results_v4=1_rot}.\\
$^e$ Ratio kept fixed in the fit for parameter $X$ or $\Delta X$.\\
$^f$ The parameter $q$ and its distortion corrections connect levels with $\Delta K = \Delta l = 2$, 
     see subsection~\ref{intro-spec} and Ref.~\cite{MeCN_v8le2_2015}.
}
\end{table}


\begin{table}
\begin{center}
\caption{Present and previous interaction parameters$^{a}$ (MHz) between low-lying vibrational states of methyl cyanide.}
\label{interation-parameter}
{\footnotesize
 \renewcommand{\arraystretch}{1.10}
\begin{tabular}[t]{lr@{}lr@{}l}
\hline 
Parameter      & \multicolumn{2}{c}{Present} & \multicolumn{2}{c}{Ref.~\cite{MeCN_v8le2_2015}} \\
\hline
$F_2(0,8^1) \times 10^3$                     &   $-$70&.9033~(44)    &   $-$70&.897~(27)       \\
$F(8^{\pm1},8^{2,\mp2})$                     &   53137&.8~(19)       &   53157&.7~(33)         \\
$F_K(8^{\pm1},8^{2,\mp2})$                   &    $-$6&.$^{b,c}$     &    $-$6&.$^{b,c}$       \\
$F_J(8^{\pm1},8^{2,\mp2}) \times 10^3$       &  $-$370&.42~(30)$^b$  &  $-$369&.89~(44)$^b$    \\
$F_{JJ}(8^{\pm1},8^{2,\mp2}) \times 10^6$    &       1&.389~(63)$^b$ &       1&.681~(87)$^b$   \\
$F_2(8^{\pm1},8^{2,0}) \times 10^3$          &   $-$62&.32~(31)$^d$  &   $-$65&.491~(24)$^d$   \\
$F_2(8^{\pm1},8^{2,\pm2}) \times 10^3$       &  $-$124&.64~(63)$^d$  &  $-$130&.982~(48)$^d$   \\
$F_{2,J}(8^{\pm1},8^{2,0}) \times 10^6$      &    $-$1&.11~(10)$^e$  &        &                \\
$F_{2,J}(8^{\pm1},8^{2,\pm2}) \times 10^6$   &    $-$2&.21~(21)$^e$  &        &                \\
$F_{ac}(8^{2,\pm2},4^1)$ $[2w_{488}]$        &       8&.76557~(13)   &       8&.7362~(21)      \\
$F_{2ac}(8^{2,0},4^1) \times 10^6$           &    $-$6&.941~(9)      &    $-$7&.98~(16)        \\
$F(8^{2,\pm2},7^{\mp1})$ $[W_{788}]$           &   45170&.8$^f$        &   45170&.8$^f$          \\
$F(8^{2,\pm2},8^{3,\mp1})$                   &   75985&.~(66)        &   77208&.~(93)          \\
$F(8^{2,0},8^{3,3})$                         &   90876&.~(47)        &   91509&.~(131)         \\
$G_b(4,7)$ $[2W_{47}]$                       &     909&.$^f$         &     909&.~(2)$^{f,g}$   \\
$F_{bc}(4,7)$ $[2w_{47}]$                    &    $-$1&.9325~(22)    &    $-$1&.84~(2)$^{f,g}$ \\
$F(4,8^{3,\pm3})$ $[W_{4888}]$               &   11381&.1~(16)       &   11430&.~(6)$^{f,g}$   \\
$F_J(4,8^{3,\pm3}) \times 10^3$              &      56&.85~(38)      &        &                \\
$F_2(4,8^{3,\pm1}) \times 10^3$              &     116&.8~(16)       &        &                \\
$F_{2J}(4,8^{3,\pm1}) \times 10^6$           &    $-$7&.50~(91)      &        &                \\
$F_4(4,8^{3,\mp1}) \times 10^6$              &    $-$2&.53~(13)      &        &                \\
$F(7^{\pm1},8^{3,\pm1})$ $[W_{7888}]$        &   50129&.2$^f$        &   50129&.2$^f$          \\
$G_a(7^{\pm1},8^{3,\pm1})$ $[2W^k_{7888}]$   & $-$2239&.1$^f$        & $-$2239&.1$^f$          \\
\hline \hline 
\end{tabular}\\[2pt]
}
\end{center}
{\footnotesize
$^a$ Alternative designations from Ref.~\cite{MeCN_nu4_nu7_3nu8_1993} given in brackets. 
     Numbers in parentheses after the interaction parameter designate the vibrational 
     states separated by a comma, see also Ref.~\cite{MeCN_v8le2_2015}. Numbers in 
     parentheses after the values are one standard deviation in units of the least 
     significant figures.\\
$^b$ $J$ and $K$ distortion corrections to $F(8^{2,\pm2},8^{3,\mp1})$ and $F(8^{2,0},8^{3,3})$ 
     kept fixed to $\sqrt2$ and $\sqrt3$, respectively, times the corresponding 
     $F(8^{\pm1},8^{2,\mp2})$ value; see, e.g., Ref.~\cite{CH3CCH_v39_2004}.\\
$^c$ Estimated assuming $F_K/F_J \approx A/B$, see Ref.~\cite{MeCN_v8le2_2015}.\\
$^{d,e}$ Ratios constrained, see Ref.~\cite{MeCN_v8le2_2015}.\\
$^f$ Kept fixed to values from Ref.~\cite{MeCN_nu4_nu7_3nu8_1993}.\\
$^g$ Uncertainties reported in Ref.~\cite{MeCN_nu4_nu7_3nu8_1993}.
}
\end{table}


\subsection{The global fit up to $\varv _4 = 1$}
\label{global_fit}

Our global fit of low-lying vibrational states of methyl cyanide up to $\varv _8 = 2$ 
was extended to include $\varv _4 = 1$, as described in Section~\ref{results_v4=1}, 
supplemented by additional data pertaining to lower vibrational states and minor omissions 
as detailed in Sections~\ref{results_v8=2}, \ref{results_v8=1}, and \ref{results_v=0}. 
As in our previous work \cite{MeCN_v8le2_2015}, our fit takes into account states up to 
$\varv _8 = 3$. This truncation will have some effect on the parameter values of 
$\varv _8 = 2$ and $\varv _4 = 1$; $\varv _8 = 0$ and 1 should be affected to a much 
lesser extent. Transition frequencies with large residuals were retained in the fit 
if they may be caused by insufficiently accounted perturbations, but were usually 
weighted out. The rms error of the global fit is 0.853, slightly below 1.0. 
Details on separate data sets along with additional fit statistics are given in 
\textbf{Table~\ref{statistics}}. We also state the rms values, which are often given 
as only information on the quality of the fit. These values are useful if all or 
most of the uncertainties are the same and the spread of the uncertainties is not 
more than a factor of a few. This applies to most of the separate IR data sets, 
but not to the rotational data sets as in these cases the uncertainties differ 
by around a factor of 100. The rms is then dominated by transitions with 
relatively large residuals.

Similar to our previous work, we provide the spectroscopic parameters of our global fit in 
several tables. In all instances, these are compared to previous values. These were taken 
from Ref.~\cite{MeCN_v8le2_2015} for $\varv _8 \le 2$ and Ref.~\cite{MeCN_nu4_nu7_3nu8_1993} 
for $\varv _4 = 1$. Parameters for $\varv _7 = 1$ and $\varv _8 = 3$ were taken from the 
latter work as well; the small number of $\varv _8 = 3$ parameters determined or adjusted 
in the present study are given in separate rows. Previous interaction parameters were taken 
from one of the two references. The ground state and $\varv _4 = 1$ parameters are listed 
in \textbf{Table~\ref{parameter_v0_v4}}, $\varv _7 = 1$ and $\varv _8 = 3$ parameters in 
\textbf{Table~\ref{parameter_v7_v8x3}}, interaction parameters in 
\textbf{Table~\ref{interation-parameter}}, and $\varv _8 = 1$ and 2 parameters are given 
in \textbf{Table~\ref{parameter_v8x1_v8x2}}. 
The line, parameter, and fit files are available as supplementary material. 
They will also be provided in the Cologne Spectroscopy Data section of the 
CDMS\footnote{https://cdms.astro.uni-koeln.de/classic/predictions/daten/CH3CN/CH3CN/}.
Calculations of the rotational ($\varv _4 = 1$) and rovibrational ($\nu_4$) spectra will be 
deposited in the catalog section of the CDMS\footnote{https://cdms.astro.uni-koeln.de/}.


\begin{table*}
\begin{center}
\caption{Spectroscopic parameters or differences $\Delta$ thereof $^{a,b}$ (cm$^{-1}$, MHz)$^c$ of 
         methyl cyanide in its $\varv_8 = 1$ and 2 states in comparison to previous values.}
\label{parameter_v8x1_v8x2}
{\footnotesize
\begin{tabular}[t]{lr@{}lr@{}llr@{}lr@{}llr@{}lr@{}l}
\hline 
              &  \multicolumn{4}{c}{$\varv_8 = 1$} & & \multicolumn{4}{c}{$\varv_8 = 2^0$} & & \multicolumn{4}{c}{$\varv_8 = 2^2$}  \\
\cline{2-5} \cline{7-10} \cline{12-15} 
Parameter $X$ & \multicolumn{2}{c}{Present} & \multicolumn{2}{c}{Ref.~\cite{MeCN_v8le2_2015}} & & \multicolumn{2}{c}{Present} 
& \multicolumn{2}{c}{Ref.~\cite{MeCN_v8le2_2015}} & & \multicolumn{2}{c}{Present} & \multicolumn{2}{c}{Ref.~\cite{MeCN_v8le2_2015}}  \\
\hline
$E_{\rm{vib}}$$^c$          &    365&.024349~(7)    &    365&.024365~(9)    & &    716&.749852~(49)   &    716&.75042~(13)    & &    739&.147650~(34)       &    739&.148225~(56)     \\
$\Delta (A - B)$            & $-$115&.871~(10)      & $-$115&.930~(26)      & & $-$187&.539~(16)      & $-$187&.404~(18)      & & $-$260&.109~(26)          & $-$259&.956~(122)       \\
$\Delta B$                  &     27&.53032~(4)     &     27&.53028~(5)     & &     54&.05757~(6)     &     54&.05732~(11)    & &     54&.50276~(4)         &     54&.50273~(7)       \\
$\Delta D_K \times 10^3$    &  $-$10&.21~(9)$^d$    &  $-$11&.46~(48)       & &  $-$20&.42~(17)$^d$   &  $-$20&.2~(3)         & &  $-$20&.42~(17)$^d$       &   $-$7&.5~(16)          \\
$\Delta D_{JK} \times 10^3$ &      0&.9892~(5)      &      0&.9875~(6)      & &      1&.6657~(11)     &      1&.6755~(25)     & &      1&.8036~(10)         &      1&.8088~(15)       \\
$\Delta D_J \times 10^6$    &     95&.999~(13)      &     95&.599~(17)      & &    216&.372~(27)      &    216&.319~(37)      & &    189&.151~(22)          &    189&.162~(31)        \\
$\Delta H_K \times 10^6$    &     20&.8~(5)         &     14&.9~(22)        & &       &               &       &               & &       &                   &       &                 \\
$\Delta H_{KJ} \times 10^9$ &     43&.1~(19)$^d$    &     34&.~(2)          & &     86&.2~(37)$^d$    &    150&.~(24)         & &     86&.2~(37)$^d$        &     25&.~(6)            \\
$\Delta H_{JK} \times 10^9$ &      2&.58~(6)        &      2&.59~(6)        & &     14&.09~(18)       &     17&.71~(34)       & &      1&.93~(12)           &   $-$0&.37~(21)         \\
$\Delta H_J \times 10^{12}$ &    317&.8~(25)        &    315&.3~(30)        & &    208&.1~(51)        &    200&.7~(63)        & &    631&.3~(48)            &    627&.8~(59)          \\
$\Delta L_J \times 10^{15}$ &   $-$2&.82~(17)$^d$   &   $-$2&.64~(20)$^d$   & &   $-$5&.64~(35)$^d$   &   $-$5&.28~(40)$^d$   & &   $-$5&.64~(35)$^d$       &   $-$5&.28~(40)$^d$     \\
$\Delta(eQq)$$^d$           &   $-$0&.0391~(19)$^d$ &   $-$0&.0387~(19)$^d$ & &   $-$0&.0782~(38)$^d$ &   $-$0&.0774~(38)$^d$ & &   $-$0&.0782~(38)$^d$     &   $-$0&.0774~(38)$^d$   \\
$eQq\eta$                   &      0&.1519\,(113)   &      0&.1519\,(113)   & &       &               &       &               & &       &                   &       &                 \\
$A\zeta \times 10^{-3}$     &    138&.65600~(5)     &    138&.65620~(7)     & &       &               &       &               & &    138&.65535~(6)         &    138&.65604~(10)      \\
$\eta_K$                    &     10&.311~(3)$^d$   &     10&.333~(7)       & &       &               &       &               & &     10&.311~(3)$^d$       &     10&.405~(10)        \\
$\eta_J$                    &      0&.390457~(5)    &      0&.390469~(7)    & &       &               &       &               & &      0&.394523~(4)        &      0&.394512~(6)      \\
$\eta_{KK} \times 10^6$     & $-$677&.~(18)$^d$     & $-$834&.~(41)$^d$     & &       &               &       &               & & $-$677&.~(18)$^d$         & $-$834&.~(41)$^d$       \\
$\eta_{JK} \times 10^6$     &  $-$33&.91~(6)        &  $-$34&.06~(6)$^d$    & &       &               &       &               & &  $-$34&.66~(6)            &  $-$34&.06~(6)$^d$      \\
$\eta_{JJ} \times 10^6$     &   $-$2&.3668~(12)$^d$ &   $-$2&.3595~(24)$^d$ & &       &               &       &               & &   $-$2&.3668~(12)$^d$     &   $-$2&.3595~(24)$^d$   \\
$\eta_{JKK} \times 10^9$    &      2&.20~(17)$^d$   &      2&.59~(17)$^d$   & &       &               &       &               & &      2&.20~(17)$^d$       &      2&.59~(17)$^d$     \\
$\eta_{JJK} \times 10^9$    &      0&.511~(5)$^d$   &      0&.509~(6)$^d$   & &       &               &       &               & &      0&.511~(5)$^d$       &      0&.509~(6)$^d$     \\
$q$                         &     17&.798485~(24)   &     17&.798438~(23)   & &       &$^e$           &       &$^e$           & &     17&.73001~(8)$^e$     &     17&.72986~(14)$^e$  \\
$q_{K} \times 10^3$         &   $-$2&.6290~(50)$^d$ &   $-$2&.6645~(111)    & &       &$^e$           &       &$^e$           & &   $-$2&.6290~(50)$^{d,e}$ &   $-$2&.6153~(89)$^{e}$ \\
$q_{J} \times 10^6$         &  $-$63&.861~(14)      &  $-$63&.842~(14)      & &       &$^e$           &       &$^e$           & &  $-$68&.719~(22)$^e$      &  $-$68&.668~(31)$^e$    \\
$q_{JK} \times 10^9$        &     93&.35~(45)$^d$   &     93&.19~(53)$^d$   & &       &$^e$           &       &$^e$           & &     93&.35~(45)$^{d,e}$   &     93&.19~(53)$^{d,e}$ \\
$q_{JJ} \times 10^{12}$     &    309&.9~(14)        &    311&.5~(15)        & &       &$^e$           &       &$^e$           & &    197&.1~(21)$^e$        &    191&.9~(26)$^e$      \\
\hline \hline
\end{tabular}\\[2pt]
}
\end{center}
{\footnotesize
$^a$ Numbers in parentheses are one standard deviation in units of the least significant figures. 
     Empty entries indicate parameters not applicable or not used in the fit. See Ref.~\cite{MeCN_v8le2_2015} for sign and value considerations.\\
$^b$ Parameter difference $\Delta (X) = X_i - X_0$ given for rotational and distortion parameters ($i$ represents an excited vibrational state).\\
$^c$ All parameters given in units of megahertz, except for $E_{\rm{vib}}$, which is given in 
     units of inverse centimeters.\\
$^d$ Ratio kept fixed in the fit for respective $X$ or $\Delta X$.\\
$^e$ The parameter $q$ and its distortion corrections connect levels with $\Delta K = \Delta l = 2$, 
     see subsection~\ref{intro-spec} and Ref.~\cite{MeCN_v8le2_2015}.
}
\end{table*}

\subsection{Single state analysis of $\varv _4 = \varv _8 = 1$}
\label{analysis_v4=v8=1}

The $\nu _4$ IR spectrum contained numerous unassigned features which displayed fairly regular 
patterns for which the assignment to the $\nu _4 + \nu _8 - \nu _8$ hot band suggested itself. 
Transitions originating in $\varv _8 = 1$ are exactly a factor of 6 weaker than corresponding 
transitions in $\varv = 0$ at 293~K. Combining the $\varv _4 = 1$ vibrational changes 
with the complete set of $\varv _8 = 1$ parameters, which are the $\varv = 0$ parameters, 
the $\varv _8 = 1$ vibrational changes and intravibrational Coriolis- and $q_{22}$-derived 
parameters, and all appropriate hyperfine parameters, yielded good estimates of the 
$\varv _4 = \varv _8 = 1$ parameters. From these, numerous transitions could be assigned 
tentatively or with certainty in the $\nu _4 + \nu _8 - \nu _8$ hot band. 
Uncertainties of 0.0002 or 0.0004~cm$^{-1}$ were assigned to most hot band transitions 
for isolated lines with good S/N and other lines, respectively, compared to the smaller 
0.0001~cm$^{-1}$ in the $\nu_4$ cold band, because of many perturbations, in particular 
at higher $K$, and because lines are frequently close to other lines, as can be seen in 
\textbf{Fig.~\ref{nu4_detail}}. Transition frequencies with large residuals and tentative 
assignments were weighted out. Ultimately, assignments in the $R$- and $P$-branch 
extended to $J' = 43$ and 49, respectively. A small number of assignments were also made 
in the $Q$-branch, notably transitions with $k = K \times l = +7$ and $J = 7$ to 10.

Transitions with $K \le 4$ and $k = +5$ appear to be largely unperturbed and could be fit 
within the currently assigned uncertainties. However, some $R$-branch transitions with 
$k = -2$, $-3$, $-4$, and $+4$ had small, but systematic deviations for a few $J$. 
They were weighted out if the deviations were considerably larger than the uncertainties. 
All other $k$ values appear to be perturbed by different degrees. An additional feature of 
the $\nu _4 + \nu _8 - \nu _8$ hot band is, that transitions having the same $K$ and 
opposite $l$ occur frequently close to each other; in the $P$-branch, e.g., $K = 2$, 3, 
and 4, see \textbf{Fig.~\ref{nu4_detail}}. The $k = -5$ lines are all nearly constantly 
about 0.015~cm$^{-1}$ lower than calculated, whereas the $k = -6$ lines are shifted 
in the opposite direction by around 0.03~cm$^{-1}$. These perturbations are caused, at least 
in part, by anharmonic resonances with $\varv _7 = \varv _8 = 1^{+2}$ and $\varv _8 = 4^{+2}$, 
as was observed for CH$_3$C$^{15}$N \cite{Pentade_15N_1984}. The latter resonances corresponds to the 
$\varv _4 = 1$/$\varv _8 = 3$ resonance discussed in Section~\ref{results_v4=1_rot}, but was not 
considered in the analysis in Ref.~\cite{pentade_1994}, most likely caused by the lack of data in 
$\varv _4 = \varv _8 = 1$ and in $\varv _8 = 4^2$. The resonance with $\varv _7 = \varv _8 = 1^{+2}$, 
however, was included in their model. A pronounced change in the $k = -6$ residuals may be caused 
in addition by an interaction with the higher lying $K = 4$ levels of $\varv _7 = \varv _8 = 1^{0}$. 
The origin of the residuals in $k = -7$ are less clear. They may still be caused by the two anharmonic 
resonances affecting $k = -5$ and $-6$. They may also be caused by $\Delta K = -1$, $\Delta l = +2$ 
interaction with $\varv _6 = 1$, which was included in the analysis of Ref.~\cite{pentade_1994}, for 
which the $K$ levels are closest at $K = 10$ and 9, respectively. The $k = -7$ lines appear to be 
blended in the $P$-branch with the stronger $k = +7$ lines around $J'' = 35$ and are shifted 
increasingly up with decreasing $J$ noticeably starting at $J'' = 32$. The shift is 0.011~cm$^{-1}$ 
at $J'' = 15$ in \textbf{Fig.~\ref{nu4_detail}}. The transitions with $k = +6$ and $+7$ are shifted 
down by about 0.0015 and 0.008~cm$^{-1}$, respectively, with essentially no change in $J$, almost 
certainly caused by an anharmonic resonance with $\varv _8 = 4^{+4}$, as observed for 
CH$_3$C$^{15}$N \cite{Pentade_15N_1984} and again equivalent to the $\varv _4 = 1$/$\varv _8 = 3$ 
resonance discussed in Section~\ref{results_v4=1_rot}. This resonance should push up levels with 
$k = +8$, as in the case of CH$_3$C$^{15}$N. We identified transitions with the correct intensity 
for $k = +8$ or $k = -8$; these transitions should have the same intensity. They are, however, 
lower in frequency than calculated, thus possibly favoring the assignment to $k = -8$. 
At any rate, one series of $K = 8$ transitions appears to be shifted by amounts which make it 
difficult to identify these transitions at present. Some very tentative assignments were made 
for $k = +9$ and +10. Even transitions having $k = -11$ and possibly $-9$ and $-10$ may be 
assignable eventually. It is not only the presumably large perturbations in these levels which 
make assignments difficult so far, but also the amount of unassigned transitions with similar 
or even higher intensities, as can be noted in \textbf{Fig.~\ref{nu4_detail}}. 
These transitions belong most likely to the $\nu_4 + 2\nu_8 - 2\nu_8$ hot band, which is 
a factor of 36 weaker than $\nu_4$ and a factor of six weaker than $\nu_4 + \nu_8 - \nu_8$. 
Transitions of $2\nu_4 - \nu_4$ or of the $\nu_4$ bands of the two methyl cyanide isotopologs 
with one $^{13}$C could complicate the situation further as these are less than a factor of 
100 weaker than $\nu_4$ of the main isotopolog.

\begin{figure}
 \begin{center}
  \includegraphics[width=8.5cm]{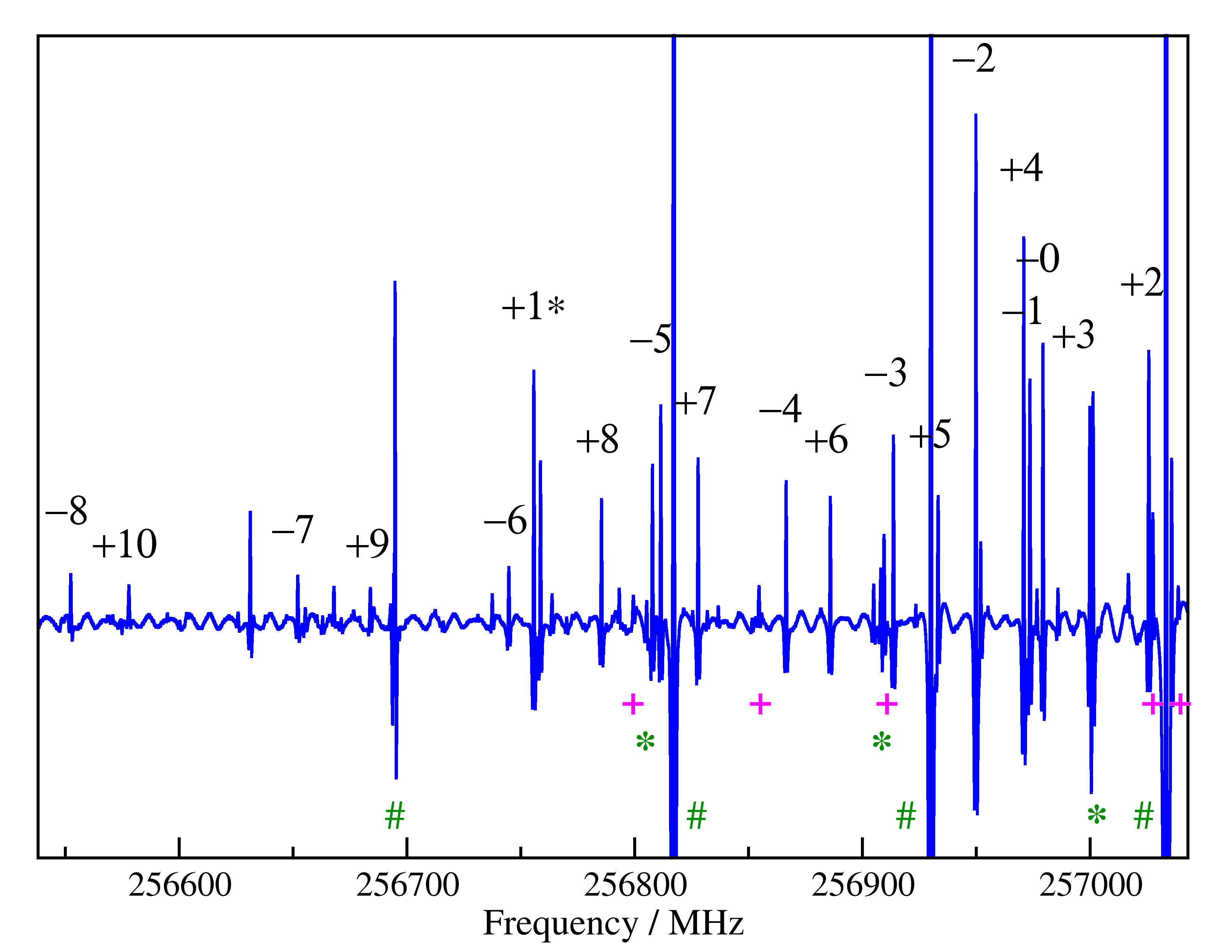}
 \end{center}
  \caption{Detail of the millimeter spectrum of CH$_3$CN in the region of the $J = 14 - 13$ 
           transition of $\varv_4 = \varv_8 = 1$. The $k = K \times l$ quantum numbers are 
           given for this state. The lower frequency transition with $k = +1$ is marked with 
           an asterisk; the higher frequency transition is more than 200~MHz higher than 
           the $k = +2$ line. Assigned, sufficiently strong transitions are marked by a 
           pound and plus signs below the zero level for $\varv = 0$ and $\varv _7 = 1$ 
           of CH$_3$CN, respectively, asterisks below the zero level indicate transitions 
           of $\varv = 0$ of CH$_3$$^{13}$CN.}
  \label{v4=v8=1_J13}
 \end{figure}

From these IR transition frequencies, some small corrections $\Delta \Delta X := X_{4,8} 
- \Delta X_{4} - \Delta X_{8}$ could be determined; $\Delta X_{4} := 0$ if the parameter $X$ 
is not defined for $\varv _4 = 0$ and 1. Rotational transitions above 440~GHz could be assigned 
subsequently up to 1.2~THz. Lower frequency transitions were measured later. 
\textbf{Fig.~\ref{v4=v8=1_J13}} displays a large part of the $J = 14 - 13$ transitions. 
The assignments were straightforward and are considered to be secure for $K$ series that 
appeared to be unperturbed or only slightly perturbed. 
More strongly perturbed transitions could be assigned with confidence quite often taking 
into account relative intensities and trends in the deviations, however it may be that 
blending of some lines escaped notice. Caution is advised if the deviations display complex 
patterns, in particular if they are quite different for a small number of transitions. 
We mention in the following only interactions that did not show up in the IR data because 
of their small effects or because of the absence of data.

Transitions in $\varv _4 = \varv _8 = 1^{+1}$ appear to be not or only slightly perturbed, 
less than $\sim$0.3~MHz for $1 \le K \le 5$, and the coverage is good with $\sim$30 
transitions up to mostly $J'' = 64$. The coverage is also good for $K = 6$ to 8. 
$K = 6$ is only slightly perturbed, less than 1~MHz up to $J'' = 53$ and slightly more 
than 3~MHz for $J'' = 64$. Transitions with $K = 7$ are shifted slightly down at low $J$, 
then up for $J'' = 18$ to 44 or 55, then decreasing again for still higher $J$. 
An assignment of a line 67~MHz higher than calculated for $J'' = 53$ is uncertain. 
Transitions with $K = 8$ are shifted decreasingly up for $J'' \le 16$ and then down by 
up to $\sim$130~MHz for $J'' = 49$. The assignment of $J'' = 60$ to a line close to 
the prediction is very tentative and appears to require that the $J'' = 61$ and 63 lines 
are blended with the respective lines of $\varv _4 = \varv _8 = 1^{-1}$ with $K = 3$ and 
2, respectively, or shifted so much that they could not be assigned with confidence. 
The perturbations in $K = 6$ to 8 are caused at least in part by an anharmonic resonance 
with $\varv _8 = 4^{+4}$, as mentioned further up in this section. 
Additional perturbations in $K = 7$ may be caused by $K = 8$ of $\varv _8 = 3^{-1}$, 
which is only $\sim$0.7~cm$^{-1}$ higher at low $J$, and by $K = 6$ of $\varv _8 = 3^{-3}$, 
which crosses $K = 7$ between $J = 20$ and 21 according to current calculations. 
These resonances are analogous to $\varv _4 = 1$/$\varv _8 = 2^{-1}$ interactions mentioned 
in Section~\ref{results_v4=1_rot}. Further perturbations in $K = 8$ may originate in an 
interaction with $K = 6$ of $\varv _8 = 4^{+2}$ directly or through $q_{22}$ interaction 
with $K = 6$ of $\varv _4 = \varv _8 = 1^{-1}$ as judged from the energy level diagram in 
Ref.~\cite{pentade_1994}. The residuals in $K = 9$ are below 1~MHz for $J'' = 9$ and 
13$-$18; assignments for $J'' = 60$ and 61 are tentative; the magnitudes of the residuals 
are slightly larger for these two transitions. Assignments for $K = 10$ cover $J'' = 13$ 
to 18 and 60 and 61. The measured frequencies are between 18 and 26~MHz lower than the 
calculated ones. A likely explanation is a perturbation with $K = 11$ of $\varv _7 = 1^{-1}$, 
which is $\sim$1~cm$^{-1}$ higher at low $J$, increasing at higher $J$. The assignment of 
$J'' = 61$ in $K = 11$ to a line close to the calculated position is very tentative.


\begin{table}
\begin{center}
\caption{Spectroscopic parameters $X$ or differences $\Delta \Delta X$ thereof $^{a,b}$ 
         (cm$^{-1}$, MHz)$^c$ of methyl cyanide in its $\varv_4 = \varv_8 = 1$ state.}
\label{parameter_v4=v8=1}
{\footnotesize
\begin{tabular}[t]{lr@{}l}
\hline 
Parameter $X$                       & \multicolumn{2}{c}{Value} \\
\hline
$E_{\rm{vib}}$$^c$                        &   1290&.052130~(12) \\
$\Delta \Delta (A - B)$                   &  $-$26&.44~(4)      \\
$\Delta \Delta B$                         &   $-$0&.13917~(11)  \\
$\Delta \Delta D_{JK} \times 10^3$        &      2&.582~(16)    \\
$\Delta \Delta D_J \times 10^6$           &  $-$25&.69~(6)      \\
$\Delta \Delta H_{KJ} \times 10^6$        &     29&.8~(7)       \\
$\Delta \Delta H_{JK} \times 10^9$        &  $-$97&.0~(14)      \\
$\Delta \Delta H_J \times 10^{12}$        &     86&.6~(98)      \\
$\Delta \Delta A\zeta$                    &  $-$21&.93~(6)      \\
$\Delta \Delta \eta_J \times 10^3$        &     25&.58~(9)      \\
$\Delta \Delta \eta_{JK} \times 10^6$     & $-$106&.~(5)        \\
$\Delta \Delta q$                         &   $-$0&.23155~(20)  \\
$\Delta \Delta q_{J} \times 10^6$         &  $-$14&.98~(16)     \\
$\Delta \Delta q_{JJ} \times 10^{12}$     & $-$348&.~(27)       \\
\hline \hline
\end{tabular}\\[2pt]
}
\end{center}
{\footnotesize
$^a$ Numbers in parentheses are one standard deviation in units of the least significant figures. 
     See Ref.~\cite{MeCN_v8le2_2015} for sign and value considerations.\\
$^b$ Parameter difference $\Delta \Delta X = X - \Delta X_4 - \Delta X_8$; $\Delta X_4 := 0$  
     if $X_4$ is not defined.\\
$^c$ All parameters given in units of megahertz, except for $E_{\rm{vib}}$, which is given in 
     units of inverse centimeters.
}
\end{table}


Transitions in $\varv _4 = \varv _8 = 1^{-1}$ extend to $K = 8$, and the coverage is 
good up to $K = 5$. There appear to be no perturbations in $K = 0$, and residuals in 
$K = 1$ to 3 are below 1.5~MHz. $K = 4$ is also only slightly perturbed for $J'' \le 33$, 
less than 1.3~MHz. However, five transitions with $43 \le J'' \le 63$ display rapidly 
increasing residuals from more than 6~MHz to $\sim$162~MHz. An interaction with $K = 2$ 
of $\varv _3 = 1$ is a probable explanation. This interaction is in the model of 
Ref.~\cite{MeCN_nu4_nu7_3nu8_1993}. The anharmonic resonances with 
$\varv _7 = \varv _8 = 1^{+2}$ and $\varv _8 = 4^{+2}$ mentioned earlier in this section 
may cause the small deviations in $K = 5$ and slightly larger ones in $K = 6$ of 
$\varv _4 = \varv _8 = 1^{-1}$. Transitions with $J'' = 9$ and 13$-$18 were assigned quite 
confidently for $K = 6$. The assignment of a somewhat more perturbed line to $J'' = 61$ 
is tentative. Residuals in $K = 7$ increase in magnitude from $-8.5$~MHz to $-15.9$~MHz 
at $J'' = 9$ and 13$-$18 and decrease somewhat for $J'' = 60$ and 61. The residuals in 
$K = 8$ are larger, increasing in magnitude from $-16.9$~MHz to $-34.4$~MHz at $J'' = 9$ 
and 13$-$18. The model in Ref.~\cite{pentade_1994} includes a $\Delta K = -1$, 
$\Delta l = +2$ interaction with $\varv _6 = 1^{+1}$, which may be responsible 
for the perturbations.

Perturbed rotational transitions were weighted out by adding sufficiently large megahertz 
values to the uncertainties that they do not contribute adversely to the rms error. 
The resulting vibrational energy and the spectroscopic parameters $\Delta \Delta X$ are 
given in \textbf{Table~\ref{parameter_v4=v8=1}}. There are 267 rotational transitions with 
256 different frequencies not weighted out with an rms error of 0.984 and an rms of 58.4~kHz. 
As usual in line lists with greatly varying uncertainties, the rms is dominated by the data 
with relatively large residuals. 
The 626 IR transitions (562 different frequencies) with 0.0002~cm$^{-1}$ uncertainty and 
the 120 IR transitions (111 different frequencies) with mostly 0.0004~cm$^{-1}$ uncertainty 
display rms values of 0.00016 and 0.00034~cm$^{-1}$, respectively, possibly judged somewhat 
conservatively considering that deperturbations may improve the rms values.  
The line, parameter, and fit files are available as supplementary material. 
Additional files will also provided in the CDMS, as in the case of the global fit up to 
$\varv _4 = 1$.

\section{Astronomical results}
\label{s:astro}

We use the spectroscopic results obtained in Sect.~\ref{results} to investigate 
the methyl cyanide emission toward the main hot molecular core of the Sgr~B2(N) 
star-forming region. We employ data acquired in the course of the imaging spectral 
line survey ReMoCA carried out toward Sgr~B2(N) with ALMA in the 3~mm 
atmospheric window. A detailed description of the observations and
data analysis was reported in \cite{Belloche19}. In short, the survey was
performed with five different frequency tunings, which we call Setups 1 to 5 
(S1--S5), covering the full frequency range from 84.1 to 114.4~GHz with a 
spectral resolution of 488~kHz that corresponds to a velocity resolution of 1.7 
to 1.3~km~s$^{-1}$. The phase center of the interferometric observations was set
at ($\alpha, \delta$)$_{\rm J2000}$= 
($17^{\rm h}47^{\rm m}19.87^{\rm s}$, $-28^\circ22'16.0''$). This position is 
located half-way between the hot cores Sgr~B2(N1) and Sgr~B2(N2) which are 
separated by 4.9$''$ or $\sim$0.2~pc at the distance of Sgr~B2 ($\sim$8.2~kpc,
\cite{Reid19}). Following the same strategy as in \cite{Belloche19}, we analyzed 
the spectrum at the offset position Sgr~B2(N1S) at which the optical depth of the 
continuum emission is lower than toward the peak of Sgr B2(N1), where it is partially 
optically thick, obscuring the molecular line emission arising from compact regions. 
This offset position is located at ($\alpha, \delta$)$_{\rm J2000}$= 
($17^{\rm h}47^{\rm m}19.870^{\rm s}$, $-28^\circ22'19.48''$), about 1$''$ to 
the south of Sgr~B2(N1). Depending on the Setup, the angular resolution (HPBW) 
varies between $\sim$0.3$''$ and $\sim$0.8$''$, with a median value of 0.6$''$. 
The rms sensitivity ranges from 0.35~mJy~beam$^{-1}$ to 1.1~mJy~beam$^{-1}$, 
with a median value of 0.8~mJy~beam$^{-1}$. The continuum and line contributions 
were separated as described in \cite{Belloche19}, with an improvement performed 
later as reported in \cite{Melosso20}.

The density of the gas probed in emission with ReMoCA toward Sgr~B2(N)'s hot
cores is higher than 10$^7$~cm$^{-3}$ (\cite{Bonfand19}) which allows us to 
assume local thermodynamic equilibrium (LTE) to produce synthetic spectra of 
complex organic species such as methyl cyanide. For this, we used the Weeds 
software (\cite{Maret11}). We modeled each identified species with a set of 
five parameters: size of the emitting region ($\theta_{\rm s}$), 
column density ($N$), temperature ($T_{\rm rot}$), linewidth ($\Delta V$), and 
velocity offset ($V_{\rm off}$) with respect to the assumed systemic velocity 
of the source ($V_{\rm sys}=62$~km~s$^{-1}$). We derived a best-fit model for
each species and added the contributions of all identified species together.

We followed this strategy to model the rotational emission of methyl cyanide, 
its singly-substituted $^{13}$C and $^{15}$N isotopologs, and its 
doubly-substituted $^{13}$C isotopolog toward Sgr~B2(N1S). Apart from the 
predictions obtained in this work for CH$_3$CN $\varv_4=1$ and 
$\varv_4=\varv_8=1$, we used the spectroscopic predictions available in the 
Cologne Database for Molecular Spectroscopy (CDMS, \cite{CDMS_2005}), more 
precisely version 2 of the entry with hyperfine structure 41505 for 
CH$_3$CN $\varv=0$, version 1 of the entries with hyperfine structure 41509 
for CH$_3$CN $\varv_8=1$, 41510 for CH$_3$CN $\varv_8=2$, 42508 for 
$^{13}$CH$_3$CN $\varv=0$, 42513 for $^{13}$CH$_3$CN $\varv_8=1$, 42509 for 
CH$_3$$^{13}$CN $\varv=0$, 42514 for CH$_3$$^{13}$CN $\varv_8=1$, 43513 for 
$^{13}$CH$_3$$^{13}$CN $\varv=0$, and version 1 of the entries 42510 for 
CH$_3$C$^{15}$N $\varv=0$ and 42515 for CH$_3$C$^{15}$N $\varv_8=1$. 
The CH$_3$CN entries with $\varv _8 \le 2$ are based on 
Ref.~\cite{MeCN_v8le2_2015} with considerable ground state contributions, 
also in the range of our study, from Ref.~\cite{MeCN_rot_2006}. 
The isotopic ground state entries are based on \cite{MeCN_rot_2009} with 
data from that study and in part from previous reports; CH$_3$$^{13}$CN 
and $^{13}$CH$_3$CN transitions in the range of our study were taken from 
Ref.~\cite{MeCN_div-isos_rot_1979}; transitions of CH$_3$C$^{15}$N from 
Ref.~\cite{summary_CH3CN-15_rot_1975}. The isotopic entries involving 
$\varv _8 = 1$ are based on \cite{MeCN_isos_v8_rot_2016}; the lower 
frequency data of CH$_3$C$^{15}$N, encompassing the range of our survey, 
are from Ref.~\cite{MeCN-v8=1_1969}.

\begin{figure*}
 \begin{center}
  \includegraphics[width=0.75\hsize]{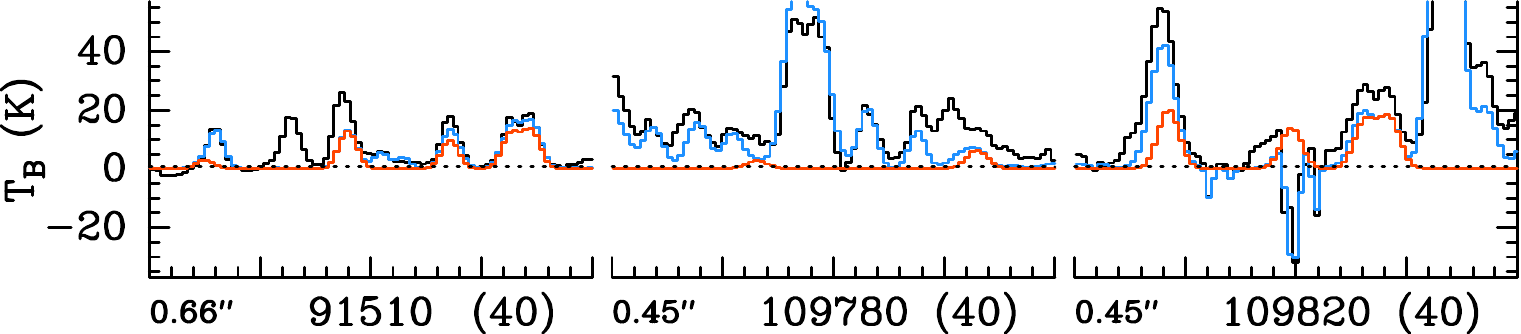}
 \end{center}
 \caption{Transitions of CH$_3$CN, $\varv_4 = 1$ covered by our ALMA 
survey. The best-fit LTE synthetic spectrum of CH$_3$CN, $\varv_4 = 1$ 
is displayed in red and overlaid on the observed spectrum of Sgr~B2(N1S) 
shown in black. The blue synthetic spectrum contains the contributions of all 
molecules identified in our survey so far, including the species shown in red. 
The central frequency and width (in parenthesis) are indicated in MHz below 
each panel. The angular resolution (HPBW) is also indicated. The y-axis is 
labeled in brightness temperature units (K). The dotted line indicates the 
$3\sigma$ noise level.}
 \label{f:spec_ch3cn_v4e1}
\end{figure*}

\begin{figure*}
 \begin{center}
  \includegraphics[width=1.0\hsize]{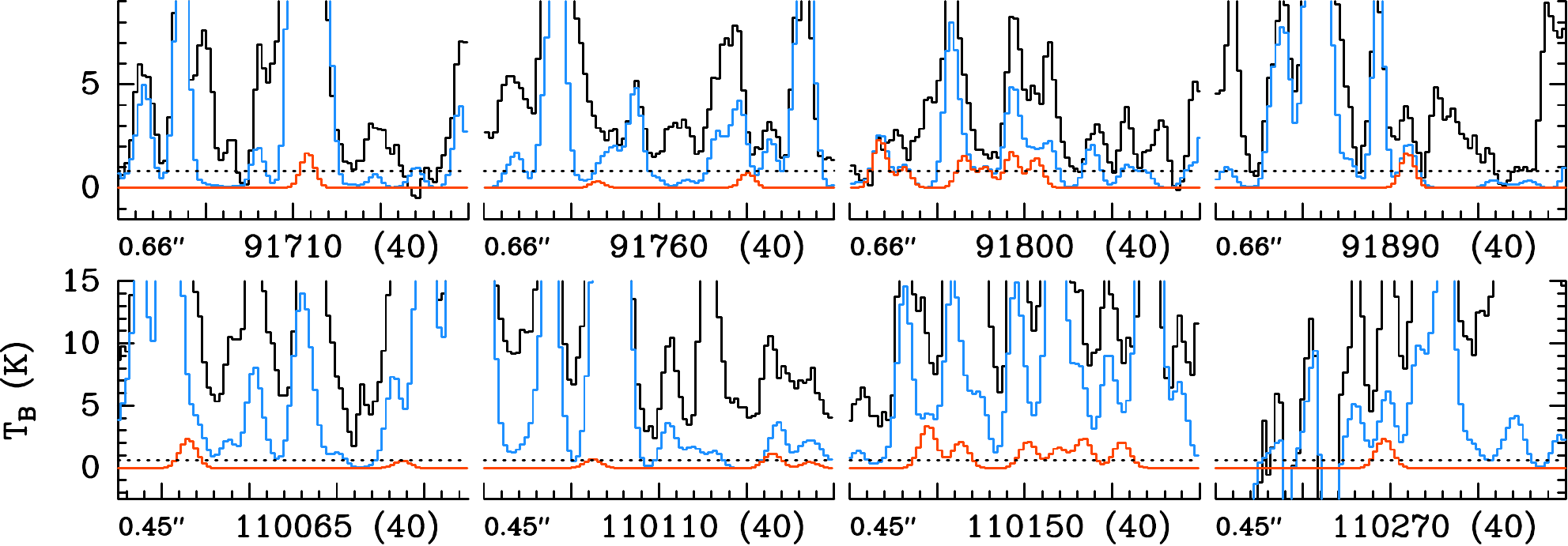}
 \end{center}
 \caption{Same as Fig.~\ref{f:spec_ch3cn_v4e1} but for CH$_3$CN, 
$\varv_4=\varv_8=1$.}
 \label{f:spec_ch3cn_v4e1v8e1}
\end{figure*}

The best-fit spectra obtained for the various vibrational states of methyl 
cyanide and its isotopologs are shown in \textbf{Figs.~\ref{f:spec_ch3cn_v4e1}}, 
\textbf{\ref{f:spec_ch3cn_v4e1v8e1}}, and 
\textbf{\ref{f:spec_ch3cn_ve0}--\ref{f:spec_ch3cn_15n_v8e1}}. These spectra were 
obtained in the following way. On the basis of our complete model that includes 
all molecules identified in the spectrum of Sgr~B2(N1S) so far, we selected 
the transitions of methyl cyanide and its isotopologs that are not 
significantly contaminated by the emission of other species. We produced 
integrated intensity maps of these transitions to measure the size of the 
methyl cyanide emission. These maps are shown in 
\textbf{Figs.~\ref{f:map_ch3cn}--\ref{f:map_ch3cn_13c2}}, and the sizes 
derived from two-dimensional Gaussian fits to these maps are plotted in 
\textbf{Figs.~\ref{f:size_ch3cn}--\ref{f:size_ch3cn_13c2}}. We see a clear 
dependence of the emission size with the upper-level energy of the transitions, 
varying from $\sim$2.5$''$ at low energy to $\sim$1.5$''$ at high energy, 
which is still at least a factor two larger than the angular resolution of the 
ReMoCA survey. A similar behavior is also seen for methanol in the same data 
set (see Fig.~4 of \cite{Motiyenko20}). 
To produce the synthetic spectra of methyl cyanide and its isotopologs, we 
assumed an emission size of 1.6$''$, close to the emission size of the 
high-energy transitions.

We selected the transitions of methyl cyanide and its isotopologs that are 
not too much contaminated by emission from other species and that have a peak 
optical depth lower than 2.5 to produce population diagrams. Following the 
strategy of \cite{EMoCA_2016}, we corrected these diagrams for the line optical 
depths and contamination from other species. These population diagrams are shown 
in \textbf{Figs.~\ref{f:popdiag_ch3cn}--\ref{f:popdiag_ch3cn_13c2}}. 
A fit to these diagrams yields an estimate of the rotational temperature. 
The fit results are listed in \textbf{Table~\ref{t:popfit}}. We obtain rotational 
temperatures on the order of 300~K. However, as explained in \cite{EMoCA_2016} 
and \cite{Belloche19}, these estimates are affected by several systematic 
uncertainties such as the varying level of continuum emission and, more 
importantly, the residual contamination of still unidentified species. In 
addition, our simple Weeds modeling procedure, that assumes a uniform emission, 
cannot account for density and temperature gradients along the line of sight. 
Optically thick lines trace more external layers of the hot core which have 
lower gas temperatures, hence they saturate at lower intensities compared to 
the synthetic spectra. This is particularly obvious for the highly optically 
thick $\varv=0$ lines of CH$_3$CN shown in \textbf{Fig.~\ref{f:spec_ch3cn_ve0}}, 
but it also affects the lower-energy lines of $\varv_8=1$ and $\varv_8=2$ of 
CH$_3$CN as well as $\varv=0$ of the singly-substituted $^{13}$C isotopologs. 
These optically thick lines at low energy, tracing colder material and thus being 
weaker, bias the fits of the population diagrams toward higher temperatures. 
To produce the synthetic spectra, we decided to use a lower temperature of 
260~K. The LTE parameters  of the best-fit models are listed in 
\textbf{Table~\ref{t:coldens}}. Another constraint that we took into account 
to obtain the best-fit model is the $^{12}$C/$^{13}$C isotopic ratio that 
is known to be in the range 20--25 for a number of complex organic molecules 
in Sgr~B2(N) (see, e.g., \cite{Mueller08}, \cite{EMoCA_2016}, \cite{Mueller16}).

\input{tab_ch3cn_popfit.tex}
\input{tab_ch3cn_weedsmodel.tex}

On the basis of the best-fit synthetic spectra obtained above, we get a clear 
detection of methyl cyanide transitions from within $\varv_4=1$ toward 
Sgr~B2(N1S) (\textbf{Fig.~\ref{f:spec_ch3cn_v4e1}}). Because of blends with other 
species, the detection of $\varv_4=\varv_8=1$ is more tentative, but the 
excellent agreement we obtain between the synthetic and observed spectra at 
91784~MHz gives us confidence in the identification of this state in the 
ReMoCA data \textbf{(Fig.~\ref{f:spec_ch3cn_v4e1v8e1})}.

The doubly-substituted $^{13}$C isotopolog of methyl cyanide is only tentatively 
detected, on the basis of one line at 107106~MHz which is partly contaminated 
by emission from CH$_3$COOH $\varv_{\rm t}=1$ 
(\textbf{Fig.~\ref{f:spec_ch3cn_13c13c_ve0}}). The synthetic spectra that we obtained 
for CH$_3$C$^{15}$N $\varv=0$ and $\varv_8=1$ are consistent with the observed 
spectra, but there are too many blends with other species to claim a detection. 
The column density listed in \textbf{Table~\ref{t:coldens}} for this isotopolog 
should rather be seen as an upper limit.


\section{Discussion of spectroscopic results}
\label{spec-discussion}

Our model of low-lying vibrational states of CH$_3$CN up to $\varv _8 = 2$ 
\cite{MeCN_v8le2_2015} was extended to include $\varv _4 = 1$. Additional or improved 
data were obtained in particular for $\varv _8 = 2$. The ground state spectroscopic 
parameter values and uncertainties in \textbf{Table~\ref{parameter_v0_v4}} were 
largely unaffected, at least with respect to the uncertainties. Notable exceptions 
are the purely axial parameters $A - B$, $D_K$, and $H_K$, whose uncertainties were 
lowered mostly by the $\varv _4 = 1$ resonances with $\varv _8 = 2$ and 3. Their values, 
however, appear to change too much as the rms error of the $\Delta K = 3$ ground state 
combination loops roughly doubles to about 1.3, see \textbf{Table~\ref{statistics}}, 
a value that is somewhat too high for a fully satisfactory fit. Possible explanations 
may be unaccounted perturbations in the present data set or a too optimistic judgment 
of the $\Delta K = 3$ ground state combination loops \cite{MeCN_DeltaK=3_1993}. 
Unsurprisingly, the rms and the rms error hardly changed for the $\varv = 0$ and 
$\varv _8 = 1$ rotational data and the $\nu_8$ data with respect to our previous 
study \cite{MeCN_v8le2_2015}. Small to moderate changes occured for the $\varv _8 = 2$ 
rotational and rovibrational data. The present $\nu_4$ data fit to 0.00009~cm$^{-1}$, 
compared with 0.00013~cm$^{-1}$ previously \cite{MeCN_nu4_nu7_3nu8_1993}. Part of 
the somewhat large rms error of the $\varv _4 = 1$ rotational data is caused by 
residuals from transitions involving $J = 56$ and $K = 7$, pointing at the need 
to improve the treatment of the interaction with $\varv _8 = 3^{-1}$.

The $\varv _4 = 1$ parameters from the present investigation and those from the previous 
study \cite{MeCN_nu4_nu7_3nu8_1993} are in a complex relationship, as can be seen in 
\textbf{Table~\ref{parameter_v0_v4}}. First, the extensive amount of rotational data 
in the present analysis required many more spectroscopic parameters, up to tenth order. 
$E_{\rm{vib}}$, $\Delta B$, $\Delta D_{JK}$, and $\Delta D_J$ are quite similar. 
The values of $\Delta (A - B)$, $\Delta D_K$, and in particular $\Delta H_K$ differ 
considerably with respect to the uncertainties. The value previously employed for $H_K$ 
in the ground vibrational state, $\sim$156~Hz \cite{MeCN_nu4_nu7_3nu8_1993}, is identical 
to our present value, therefore not suited to explain the difference. One explanation may 
be that our IR data extend to $K = 13$, whereas those in Ref.~\cite{MeCN_nu4_nu7_3nu8_1993} 
extend only to $K = 12$. While this is not a large difference, it may be enough to explain 
the differences in $\Delta H_K$, especially taking into account the relatively large 
uncertainties, and the concomitant changes in $\Delta (A - B)$ and $\Delta D_K$. Another 
explanation may be associated with the differences in observed and treated perturbations.

The magnitudes of $\Delta X$ compared to those of the ground state $X$ are noteworthy for the 
parameters of sixth order and higher in our present parameter set. The magnitudes of $\Delta X$ 
are much larger in particular for the highly $K$ dependent parameters $\Delta H_{KJ}$, 
$\Delta L_{KKJ}$, and $\Delta P_{KKJ}$, and all these vibrational changes are negative. 
This could be an indication of a perturbation. However, there is no indication of a 
correspondence in the $\varv _8 = 2^0$ parameters caused by a purported strong Fermi resonance 
\cite{FF_Duncan_1978}. In addition, trial fits with various values of $F(4,8^{2,0})$ did not 
affect more than one or two of the $\Delta X$ substantially. Literature values do not provide 
evidence that these peculiar parameter values may be caused by Coriolis interaction with 
$\varv _7 = 1$ or $\varv _6 = 1$. The $\Delta H_{KJ}$ value in Ref.~\cite{MeCN_nu4_nu7_3nu8_1993} 
is similar in value to ours, and their IR and rotational data extend to $K = 12$. In contrast, 
their data in $\nu _7$ and the $\nu _6$ data in Ref.~\cite{pentade_1994} extend to $K = 15$, and 
$\Delta H_{KJ}$ was constrained to zero for either vibrational state. Trial fits with $G_b(4,7)$ 
increased pointed in the same direction. It may thus be that these vibrational changes are 
intrinsic to this vibrational mode which is the stretching mode of a long and fairly weak C$-$C 
single bond.

Our integrated band strength of $\nu_4$ agrees well with previously reported values from low- or 
medium-resolution spectroscopic studies, as summarized in \textbf{Table~\ref{mu-IR_intensities}}; 
the agreement is slightly worse for $2\nu_8$ and more so for $\nu_8$, possibly reflecting the 
challenges of far-IR measurements.

Some $\varv _8 = 3$ parameters in \textbf{Table~\ref{parameter_v7_v8x3}} differ with respect 
to the entirely fixed values used in our previous study. The value of $A\zeta$ in 
$\varv _8 = 3^3$ is expected to be closer to the value presently assumed than the one from 
Ref.~\cite{MeCN_nu4_nu7_3nu8_1993}, as can be seen by comparison with the $\varv _8 = 1$ 
and 2 values in \textbf{Table~\ref{parameter_v8x1_v8x2}}. 
Our $3\nu_8^3$ origin of 1122.35~cm$^{-1}$ is only slightly different from 1122.15~cm$^{-1}$ 
in Ref.~\cite{MeCN_nu4_nu7_3nu8_1993} and in very good agreement with the extrapolated value 
of 1122.34~cm$^{-1}$ in Ref.~\cite{MeCN_nu4_nu7_3nu8_1993}. In addition, our $\varv _8 = 3$ 
$\Delta (A-B)$ value of $-442.06 \pm 0.06$~MHz agrees quite well with the extrapolated 
$-459.6 \pm 4.5$~MHz derived from Ref.~\cite{MeCN_nu4_nu7_3nu8_1993}. This demonstrates that 
the rotational transitions in $\varv _4 = 1$ contain information on the band origins of 
$\varv _8 = 3^3$ and $3^1$ through perturbations between $\varv _8 = 3$ and $\varv _4 = 1$, 
notably the proximity of $J = 42$ of $K = 7$ in $\varv _4 = 1$ and $K = 1$ in 
$\varv _8 = 3^{-3}$. An improved treatment of the $\varv _8 = 3$ interactions with 
$\varv _8 = 2$ and $\varv _4 = 1$ requires $\varv _8 = 3$ and $\varv _7 = 1$ data to be 
added to the line list. The $\varv _7 = 1$ data are necessary because of the widespread 
interaction between $\varv _7 = 1$ and $\varv _8 = 3$. The $\varv _8 = 3$ and $\varv _7 = 1$ 
data need to be added with caution because both states are perturbed by higher lying states 
such as $\varv _4 = \varv _8 = 1$ or $\varv _8 = 4$. A more firm determination of the purely 
axial $\varv _8 = 3^3$ parameters may necessitate assignments in $3\nu _8^3 - 2\nu _8^2$ 
because $3\nu _8^3 - \nu _8^1$ is forbidden to first order and acquires intensity only 
through $q_{22}$ interaction with the strongly allowed $3\nu _8^1 - \nu _8^1$ hot band.

Several of the interaction parameters in \textbf{Table~\ref{interation-parameter}} have now 
smaller uncertainties; $F_2$ of the $\varv _8 = 1$ and 2 interaction is an exception because 
of $F_{2,J}$ in the present fit. The $\varv _8 = 2$ and 3 Fermi parameters not only have 
smaller uncertainties, but $F(8^{2,\pm2},8^{3,\mp1}) \approx 75967$~MHz is much closer 
to the theoretical $\sqrt2 \times F(8^{\pm1},8^{2,\mp2}) = 75148$~MHz than $\sim$77208~MHz 
from our previous work \cite{MeCN_v8le2_2015}. The value 90876~MHz of $F(8^{2,0},8^{3,3})$, 
on the other hand, differs slightly more from $\sqrt3 \times F(8^{\pm1},8^{2,\mp2})$ or 
92037~MHz, see also, e.g., Ref.~\cite{CH3CCH_v39_2004}. Such deviations are common if 
a resonance is approached only from one side and for one of the interacting (sub-) states. 
Since the final interaction parameter values should be closer to the ideal values, we tried 
a fit constraining these parameters to the ideal ratios with $F(8^{\pm1},8^{2,\mp2})$. 
The rms error of the fit deteriorated moderately from 0.853 to 0.914, but the largest part 
of this was caused by a deterioration of the $\varv _8 = 2^0$ rotational data from 0.965 
to 1.403, which is not surprising, given that the current $F(8^{2,0},8^{3,3})$ value 
deviated much more from the ideal ratio than $F(8^{2,\pm2},8^{3,\mp1})$. All attempts 
to improve the quality of the fit were unsatisfactory, and we discarded this fit. 
Possible explanations for the deterioration may be, for example, unfavorable 
$\varv _8 = 3$ parameters or the truncation of the Hamiltonian at $\varv _8 = 3$. 
The values describing the resonances between $\varv _4 = 1$ and $\varv _8 = 3$ look 
quite reasonable, but the number and choice of parameters as well as their values may 
change when extensive $\varv _8 = 3$ data will have been added to the line list or when 
truncation of the Hamiltonian, in particular in $\varv _8$, is less of an issue.

The $\varv _8 = 1$ and 2 parameters in \textbf{Table~\ref{parameter_v8x1_v8x2}} display some 
changes in values and uncertainties which are caused in complex and different ways by introducing 
several constraints, additional transition frequencies for these states, and further energy 
constraints through the interactions of $\varv _4 = 1$ with $\varv _8 = 2$ and 3. 
The changes are most pronounced in $E_{\rm{vib}}$, $\Delta (A - B)$, $\Delta D_K$, and in 
the one $\Delta H_K$ used for $\varv _8 = 1$. Trial fits with the two $\varv _8 = 2$ 
$\Delta H_K$ values constrained to two times the $\varv _8 = 1$ value increased the rms error 
of the entire fit and of the $\Delta K = 3$ ground state combination loops somewhat more, 
therefore, these constraints were omitted for now.

A set of spectroscopic parameters of $\varv _4 = \varv _8 = 1$ was determined. 
The vibrational changes are often not particularly small, but this may be caused, at least 
in part, by not treating any of the perturbations. One of the important states in this 
regard is $\varv _8 = 4$, whose estimation of spectroscopic parameters will likely benefit 
from improved $\varv _8 = 3$ parameters. It will be very important that these estimates 
are very good in order to be able to account properly for the perturbations in the 
$\varv _4 = \varv _8 = 1$ rotational data because of the strong $\Delta K = \pm2$, 
$\Delta l = \pm2$ interactions and possibly $\Delta K = \pm4$, $\Delta l = \pm4$ 
interactions. We point out that interacrtions up to $\Delta K = \pm8$, $\Delta l = \pm8$ 
are allowed in $\varv _8 = 4$.

Quantum-chemical calculations may be useful to disentangle the plethora of perturbations 
in low-lying vibrational states of CH$_3$CN. To the best of our knowledge, there are no 
published results available on, e.g., interaction parameters. However, numerous calculations 
are available on calculations of the anharmonic vibrations of CH$_3$CN, e.g. 
Refs.~\cite{MeCN_anh_calc_2005,MeCN_anh_calc_2011,MeCN_anh_calc_2015}. A very sensitive probe 
into the quality of such calculations is the energy splitting of the $l$ components of 
$\varv _8 = n$. Presently available experimental numbers are summarized in 
\textbf{Table~\ref{l-splitting}}. The splitting is determined to first order by $g_{ll}l^2$; 
a value of $g_{ll} = 5.6$~cm$^{-1}$ accounts well for the experimental CH$_3$CN values 
in Table~\ref{l-splitting}.


\begin{table}
\begin{center}
\caption{Experimental vibrational energy $E_{\rm vib}$ of $\varv _8 = 1$ of CH$_3$CN (cm$^{-1}$) 
         and splitting $\Delta _{n, l1 - l2}$ (cm$^{-1}$) of the $l$ components of $\varv _8 = n$ 
         in comparison to experimental values of related molecules and in comparison to 
         quantum-chemical calculations.}
\label{l-splitting}
{\footnotesize
\begin{tabular}[t]{lccclcc}
\hline 
                   & \multicolumn{3}{c}{Experiment} & &      \multicolumn{2}{c}{Quantum-chemistry} \\
\cline{2-4} \cline{6-7} 
                     & CH$_3$CN$^a$ & CH$_3$CCH$^b$ & CH$_3$NC$^c$ & & CH$_3$CN$^d$ & CH$_3$CN$^e$ \\
\hline
$E_{\rm vib}$        &   365.0      &   330.9       &     267.3    & &    364       &   360.991    \\
$\Delta _{2, 2 - 0}$ &   22.397     &   20.416      &     20.662   & &     6        &   $-$0.646   \\
$\Delta _{3, 3 - 1}$ &   44.56      &   40.872      &              & &     7        &   $-$1.222   \\
$\Delta _{4, 2 - 0}$ &    22        &               &              & &              &   $-$0.576   \\
$\Delta _{4, 4 - 2}$ &    66        &               &              & &              &   $-$1.726   \\
\hline \hline
\end{tabular}\\[2pt]
}
\end{center}
{\footnotesize
$^a$ Ref.~\cite{MeCN_v8le2_2015} and this work.\\
$^b$ Refs.~\cite{MeCCH_nu10+Dyade_2002} and \cite{CH3CCH_v39_2004}; except $\varv _b = 3$ data from Ref.~\cite{MeCCH_10mue_2009}.\\
$^c$ Ref.~\cite{MeNC_v8le2_2011}.\\
$^d$ Ref.~\cite{MeCN_anh_calc_2005}.\\
$^e$ Ref.~\cite{MeCN_anh_calc_2011}.\\
}
\end{table}


Even though the $\varv _8 = 4$ values were estimated, there is little doubt on the origin 
of $\varv _8 = 4^4$; its value can be evaluated from various information available for CH$_3$CN 
and CH$_3$C$^{15}$N. Combining the origin of $4\nu _8^4 - \nu _8$ of CH$_3$C$^{15}$N, 
$1140.626 \pm 0.032$~cm$^{-1}$ \cite{Pentade_15N_1984}, with the estimate of the $\nu _8$ band 
origin, 362.41~cm$^{-1}$ \cite{MeCN_isos_v8_rot_2016,FF_Duncan_1978}, yields 1503.04~cm$^{-1}$ 
as estimate of the $4\nu _8^4$ band origin. This value may be scaled with the $^{14}$N/$^{15}$N 
ratios of the $\nu _8$ \cite{MeCN_isos_v8_rot_2016}, $3\nu _8^3$ 
\cite{Triade_15N_1984,MeCN_nu4_nu7_3nu8_1993}, and $4\nu _8^2$ band origins 
\cite{Pentade_15N_1984,MeCN_isos_v8_rot_2016,pentade_1994} to arrive at an estimate for 
the $4\nu _8^4$ band origin of CH$_3$CN. The respective values are 1513.9, 1514.6, and 
1514.3~cm$^{-1}$, in excellent agreement with our estimate of $\sim$1514~cm$^{-1}$ 
\cite{MeCN_v8le2_2015}, see also \textbf{Table~\ref{vib_energies}}.

Experimental values of the related molecules propyne and methyl isocyanide in 
\textbf{Table~\ref{l-splitting}} are in line with those of methyl cyanide as far as 
they are available. In contrast, quantum-chemically derived values of CH$_3$CN agree poorly 
with the experimental ones. The authors of Ref.~\cite{MeCN_anh_calc_2005} did not discuss 
their poor value for $\varv _8 = 2$ and alleged that the experimental $\varv _8 = 3$ 
value must be incorrect because the band centers were only ''estimated experimental 
transitions''. The $\varv _8 = 3^1$ origin, however, was determined directly through 
assignments in the perturbation allowed $3\nu_8^1$ band and through the fully allowed 
$3\nu_8^1 - \nu_8^1$ band \cite{MeCN_nu4_nu7_3nu8_1993}. The $\varv _8 = 3^3$ origin 
was derived through perturbations, but its value is quite close to the extrapolated 
one of 1122.339~cm$^{-1}$; our origin is actually still closer to this value. 
Assignments in the $3\nu_8^3 - 2\nu_8^2$ hot band would yield this origin directly, 
along with a value for $\Delta (A - B)$ (or $\Delta A$). Our $\varv _8 = 3^1$ origin 
is quite precise, however, it depends somewhat on the $\Delta (A - B)$ value used 
in the fit which is marginally smaller in magnitude than the extrapolated value derived 
from $\Delta A$ in Ref~\cite{MeCN_nu4_nu7_3nu8_1993}. The data in Ref.~\cite{MeCN_anh_calc_2011} 
are somewhat better represented by one $g_{ll}$ value of about $-0.6$~cm$^{-1}$, but additional 
correction would be needed for quantitative agreement. In addition, the values disagree in 
sign and in magnitude and disagree more with the experimental values than those from 
Ref.~\cite{MeCN_anh_calc_2011}. There is only one value for 
$\Delta _{2, 2 - 0} = 0.925$~cm$^{-1}$ in Ref.~\cite{MeCN_anh_calc_2011}. In order to be 
helpful for disentangling the perturbations in low-lying states of methyl cyanide, it is 
necessary that future quantum-chemical calculations reduce these deviations substantially.

\section{Conclusions and outlook}
\label{Conclusions}

Data of $\varv _4 = 1$ have been added to our model of low-lying vibrational states of 
CH$_3$CN up to $\varv _8 = 2$ \cite{MeCN_v8le2_2015}. The analysis revealed new 
rovibrational interactions and improved several interactions treated already earlier. 
One important outcome is that now all four $l$-components of $\varv _8 = 3$ are linked 
in energy to $\varv _4 = 1$. The analysis of interacting states of methyl cyanide up to 
$\varv _4 = 1$ has probably been developed to the extent that is possible without inclusion 
of extensive $\varv _7 = 1$ and $\varv _8 = 3$ data. Earlier, perturbations in $\varv _4 = 1$ 
that were not treated likely caused the problems we encountered in adding this state 
to our global model. Adding $\varv _7 = 1$ and $\varv _8 = 3$ data needs to be done with 
caution because these two states are not only perturbed by each other and the lower lying 
states $\varv _4 = 1$ and $\varv _8 = 2$, but also by higher lying states such as 
$\varv _4 = \varv _8 = 1$ or $\varv _8 = 4$, among others. Our earlier problems with 
including $\varv _7 = 1$ and $\varv _8 = 3$ probably arose from underestimating 
the amount of perturbations in these states. Recently, spectra 
in the region of $\nu _8$ of CH$_3$CN were recorded at the Canadian Light Source. 
These may be very beneficial for the analyses of $\varv _8 = 3$ and possibly even 
$\varv _8 = 4$. Not only transitions of the $\nu _8$ cold band were easily identified, 
but also some of the $2\nu _8 - \nu _8$ hot band. The $3\nu _8 - 2\nu _8$ hot band 
should also be identifiable, potentially even $4\nu _8 - 3\nu _8$. Rotational transitions 
up to $\varv _8 = 4$ have sufficient intensity at room temperature to study them 
quite extensively. Investigations of states up to $\varv _8 = 5$ should be possible, 
but this may be challenging.

Assignments in $\nu _4 + \nu _8 - \nu _8$ and in $\varv _4 = \varv _8 = 1$ were made, 
and a simple model was developed to account for a fair fraction of these data. 
This model should be useful for the upcoming reanalysis of $\varv _7 = 1$ and 
$\varv _8 = 3$ data even if the $\varv _4 = \varv _8 = 1$ data and their perturbations 
will be left untouched because of the plethora of interactions with higher lying states.

Rotational transitions of methyl cyanide from within all vibrational states up to $\varv _4 = 1$ 
were detected with ALMA in the hot molecular core of Sgr~B2(N1). The tentative detection of 
$\varv _4 = \varv _8 = 1$ bodes well for the future detection of the slightly lower-energy 
vibrational states $\varv _7 = 1$ and $\varv _8 = 3$.


\section*{Acknowledgements}

It is our pleasure to dedicate this article to Stephan Schlemmer. We thank the reviewers 
for their questions and suggestions which helped to clarify some aspects of the manuscript. 
We thank Robert L. Sams for recording the $\nu _4$ infrared spectrum, Isabelle Kleiner for 
initial assignments and early modeling efforts, and Linda R. Brown for the final calibration 
of the IR spectrum, the generation of a peak list, and further initial assignments in 
$\nu _4 + \nu _8 - \nu _8$. We also thank John C. Pearson for recording part of the 
submillimeter spectra at JPL. We are grateful for support by the Deutsche 
Forschungsgemeinschaft (DFG) via the collaborative research center SFB~956 (project ID 
184018867), sub-project B3. The portion of this work which was carried out at the Jet 
Propulsion Laboratory, California Institute of Technology was performed under contract 
with the National Aeronautics and Space Administration. 
The infrared spectrum analyzed in the present study was recorded at the W.R. Wiley 
Environmental Molecular Sciences Laboratory, a national scientific user facility 
sponsored by the Department of Energy's Office of Biological and Environmental 
Research located at the Pacific Northwest National Laboratory (PNNL). PNNL is 
operated for the United States Department of Energy by the Battelle Memorial 
Institute under Contract DE-AC05-76RLO1830.
Our research benefited from NASA's Astrophysics Data System (ADS).

\appendix
\section{Supplementary Material}

The following are the Supplementary data to this article: 
The parameter, line, and fit files of the global fit (16u4.$\ast$) and of the isolated 
fit (4plus8QN.$\ast$) along with a readme file are provided.

\section{Additional Figures}

\begin{figure*}
 \begin{center}
  \includegraphics[width=0.75\hsize]{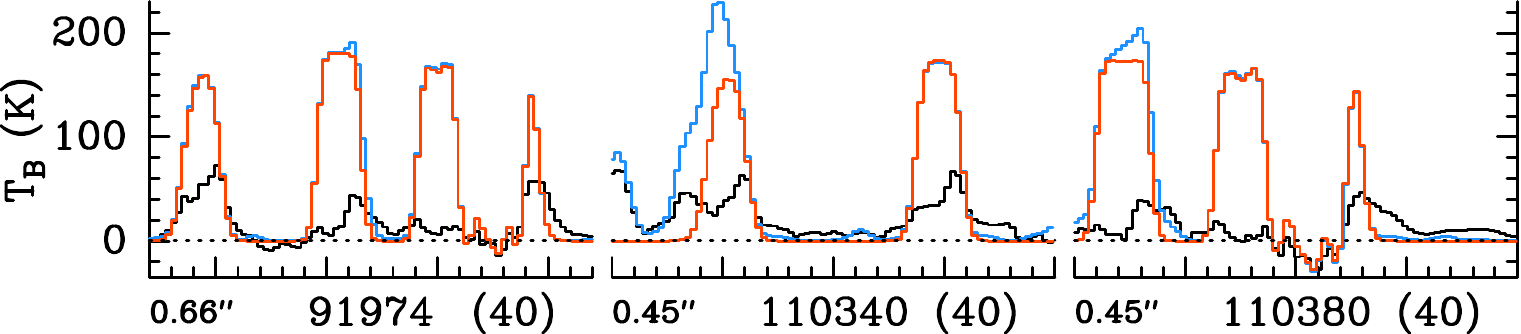}
 \end{center}
 \caption{Transitions of CH$_3$CN, $\varv = 0$ covered by our ALMA 
survey. The best-fit LTE synthetic spectrum of CH$_3$CN, $\varv = 0$ 
is displayed in red and overlaid on the observed spectrum of Sgr~B2(N1S) shown 
in black. The blue synthetic spectrum contains the contributions of all 
molecules identified in our survey so far, including the species shown in red. 
The central frequency and width (in parenthesis) are indicated in MHz below 
each panel. The angular resolution (HPBW) is also indicated. The y-axis is 
labeled in brightness temperature units (K). The dotted line indicates the 
$3\sigma$ noise level.}
 \label{f:spec_ch3cn_ve0}
\end{figure*}

\begin{figure*}
 \begin{center}
  \includegraphics[width=1.0\hsize]{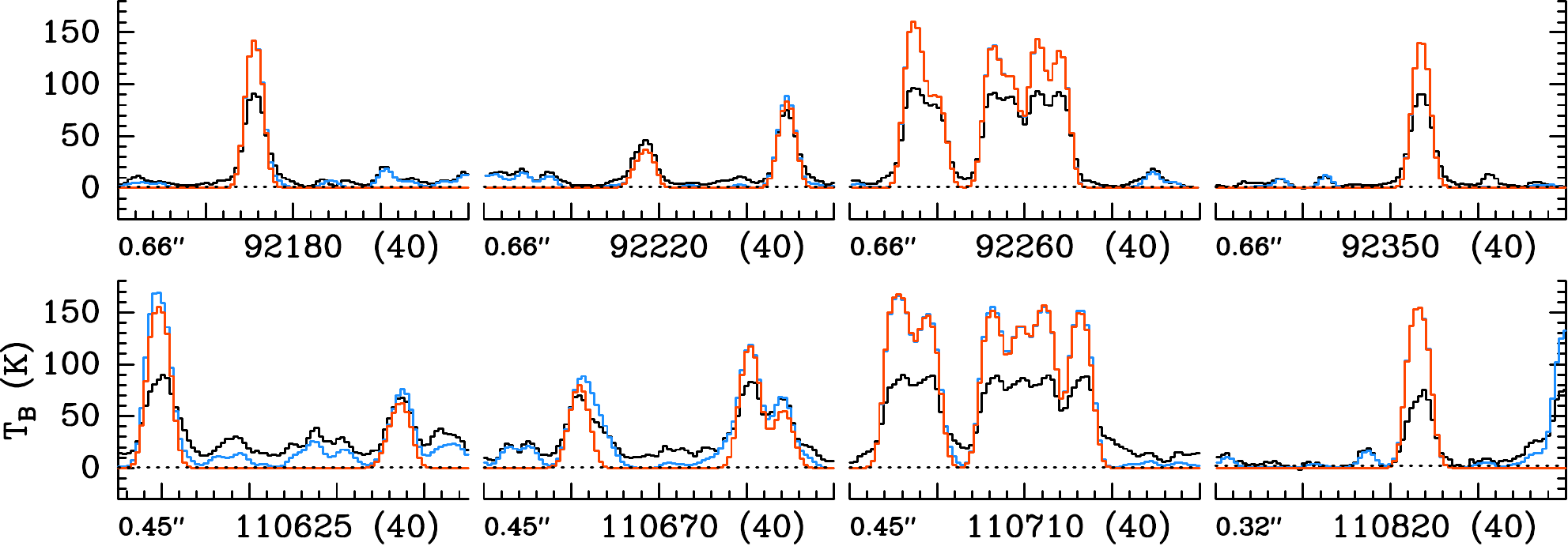}
 \end{center}
 \caption{Same as Fig.~\ref{f:spec_ch3cn_ve0} but for CH$_3$CN, $\varv_8=1$.}
 \label{f:spec_ch3cn_v8e1}
\end{figure*}

\begin{figure*}
 \begin{center}
  \includegraphics[width=1.0\hsize]{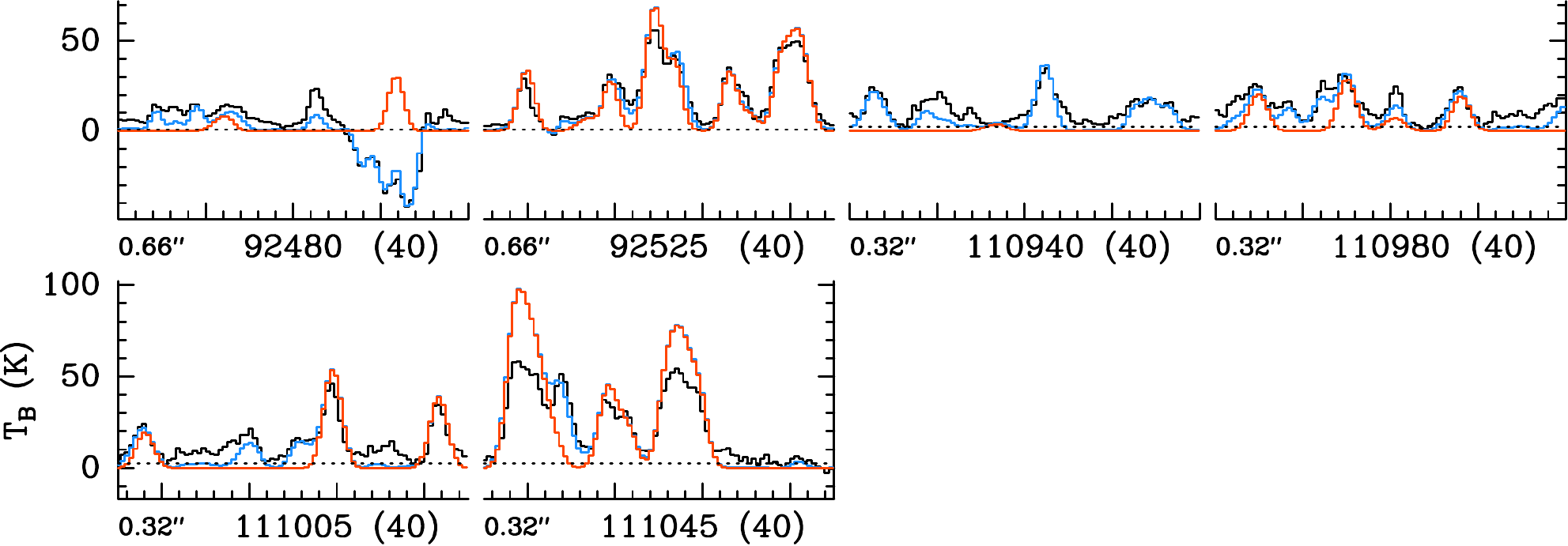}
 \end{center}
 \caption{Same as Fig.~\ref{f:spec_ch3cn_ve0} but for CH$_3$CN, $\varv_8=2$.}
 \label{f:spec_ch3cn_v8e2}
\end{figure*}

\begin{figure*}
 \begin{center}
  \includegraphics[width=0.75\hsize]{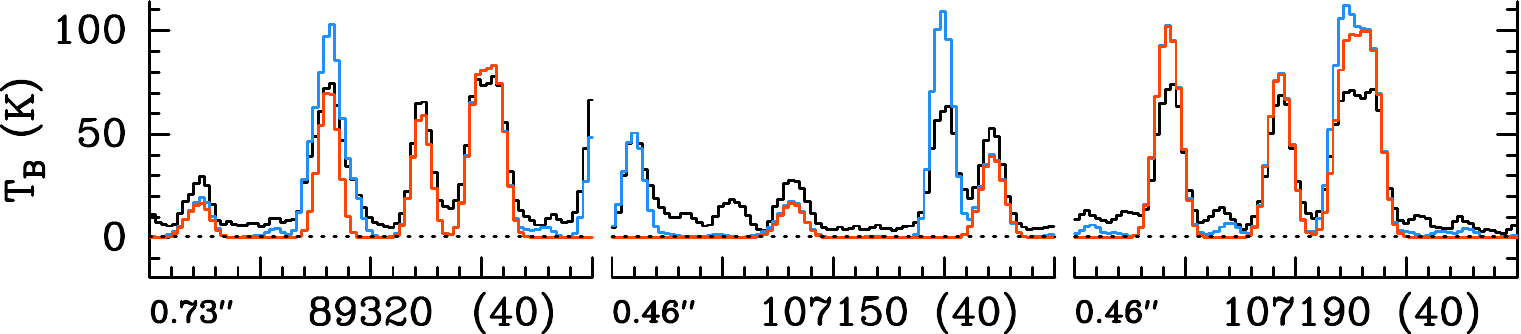}
 \end{center}
 \caption{Same as Fig.~\ref{f:spec_ch3cn_ve0} but for $^{13}$CH$_3$CN, $\varv=0$.}
 \label{f:spec_ch3cn_13c1_ve0}
\end{figure*}

\begin{figure*}
 \begin{center}
  \includegraphics[width=1.0\hsize]{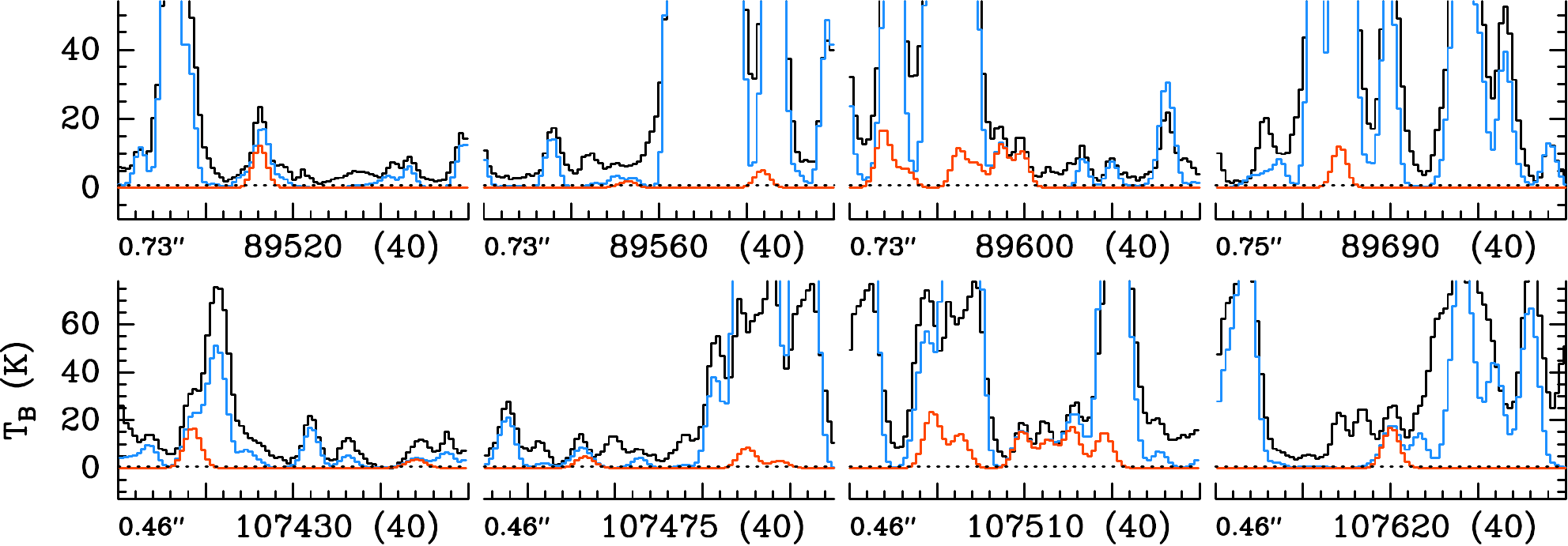}
 \end{center}
 \caption{Same as Fig.~\ref{f:spec_ch3cn_ve0} but for $^{13}$CH$_3$CN, $\varv_8=1$.}
 \label{f:spec_ch3cn_13c1_v8e1}
\end{figure*}

\begin{figure*}
 \begin{center}
  \includegraphics[width=0.75\hsize]{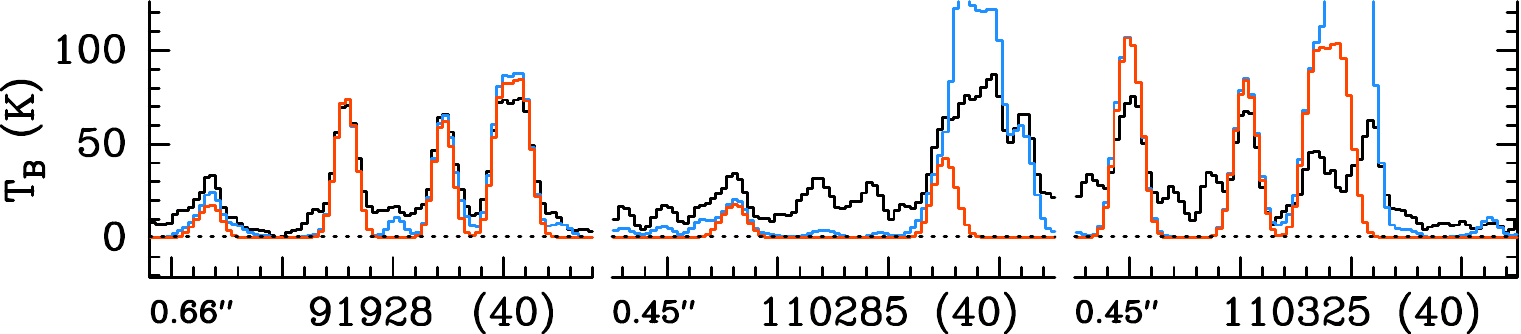}
 \end{center}
 \caption{Same as Fig.~\ref{f:spec_ch3cn_ve0} but for CH$_3$$^{13}$CN, $\varv=0$.}
 \label{f:spec_ch3cn_13c2_ve0}
\end{figure*}

\begin{figure*}
 \begin{center}
  \includegraphics[width=1.0\hsize]{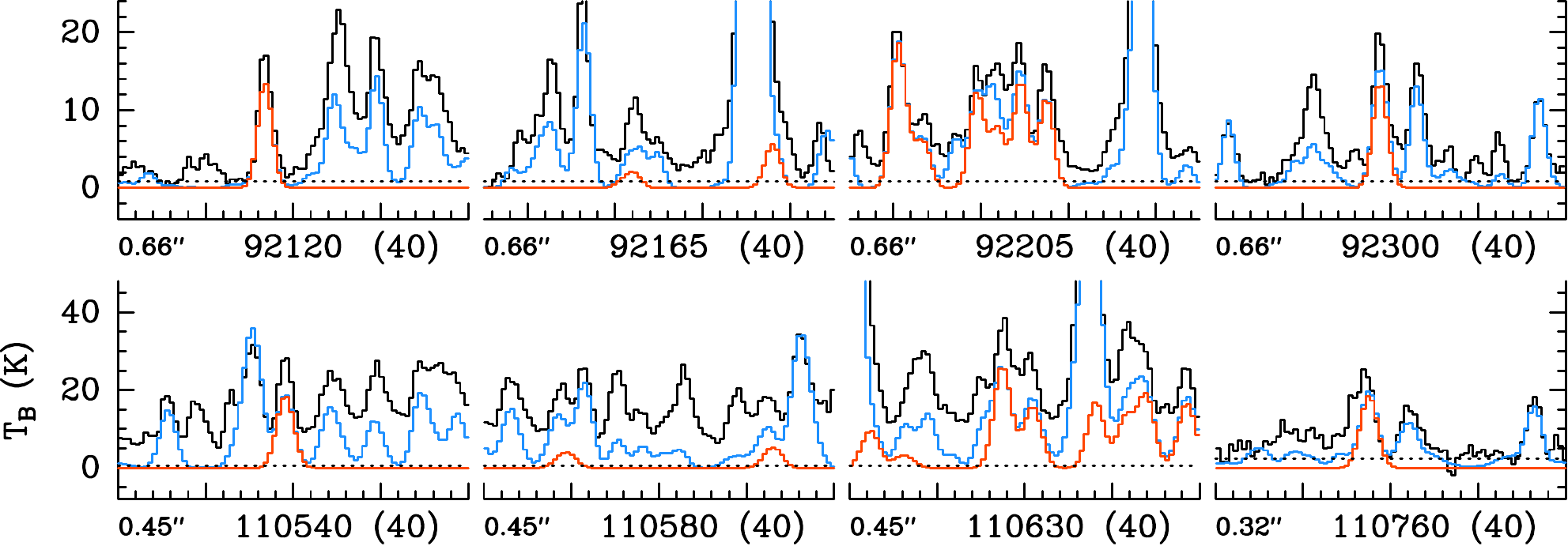}
 \end{center}
 \caption{Same as Fig.~\ref{f:spec_ch3cn_ve0} but for CH$_3$$^{13}$CN, $\varv_8=1$.}
 \label{f:spec_ch3cn_13c2_v8e1}
\end{figure*}

\begin{figure*}
 \begin{center}
  \includegraphics[width=0.75\hsize]{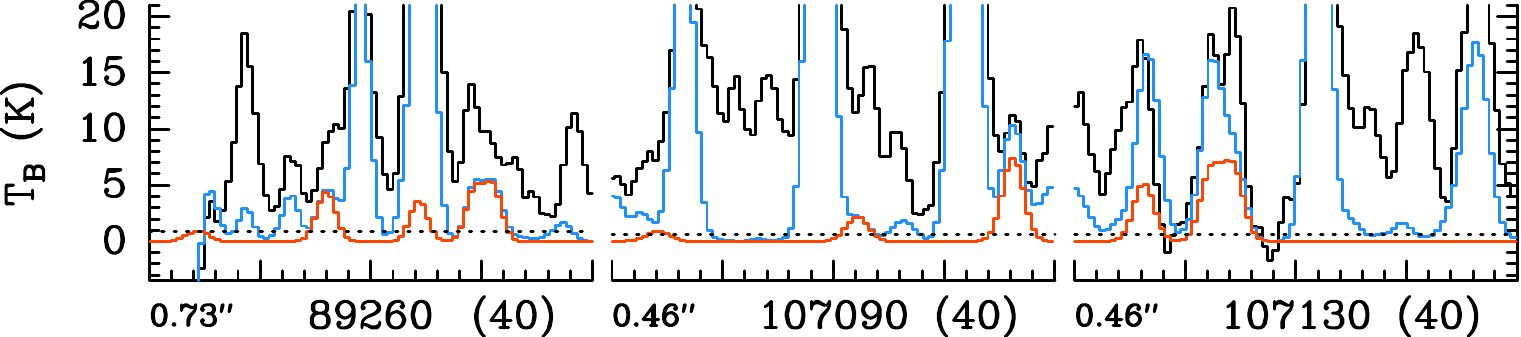}
 \end{center}
 \caption{Same as Fig.~\ref{f:spec_ch3cn_ve0} but for $^{13}$CH$_3$$^{13}$CN, $\varv=0$.}
 \label{f:spec_ch3cn_13c13c_ve0}
\end{figure*}

\begin{figure*}
 \begin{center}
  \includegraphics[width=1.0\hsize]{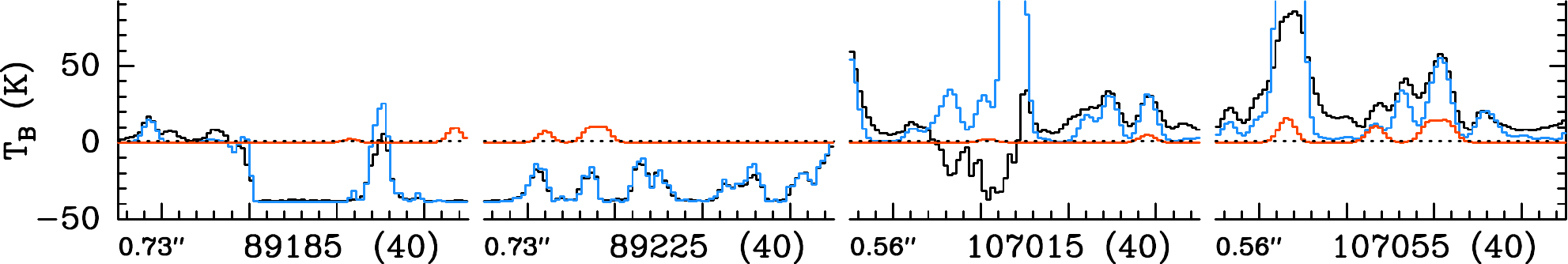}
 \end{center}
 \caption{Same as Fig.~\ref{f:spec_ch3cn_ve0} but for CH$_3$C$^{15}$N, $\varv=0$.}
 \label{f:spec_ch3cn_15n_ve0}
\end{figure*}

\begin{figure*}
 \begin{center}
  \includegraphics[width=1.0\hsize]{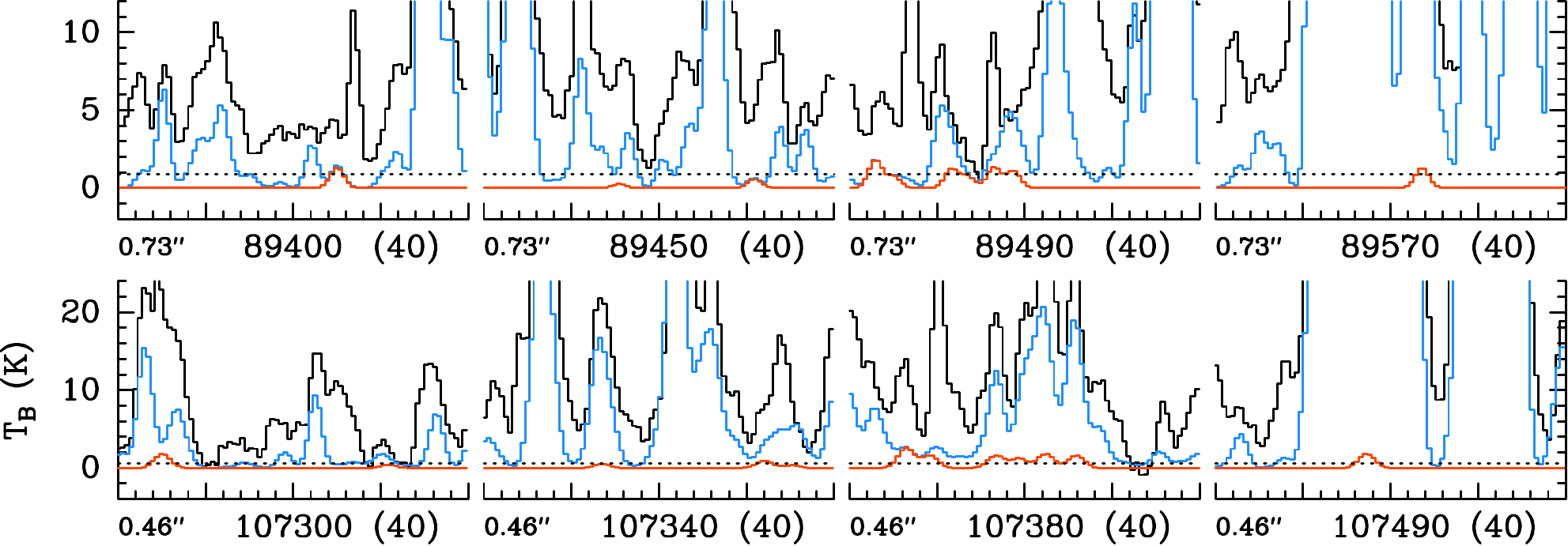}
 \end{center}
 \caption{Same as Fig.~\ref{f:spec_ch3cn_ve0} but for CH$_3$C$^{15}$N, $\varv_8=1$.}
 \label{f:spec_ch3cn_15n_v8e1}
\end{figure*}

\begin{figure*}
 \begin{center}
  \includegraphics[width=0.85\hsize]{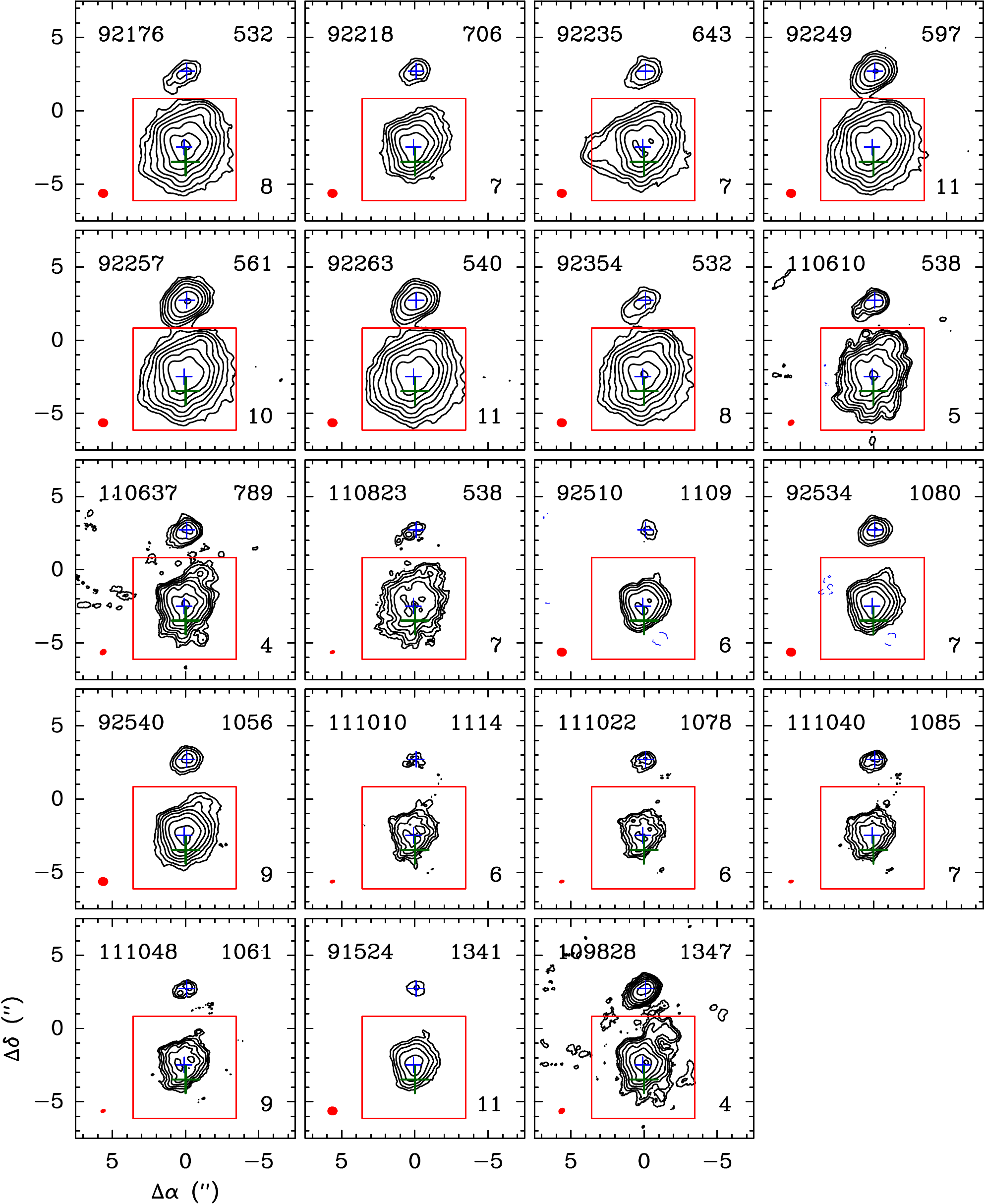}
 \end{center}
 \caption{Integrated intensity maps of transitions or groups of transitions of 
CH$_3$CN that are relatively free of contamination from emission of other 
molecules. From top left to bottom right, the first ten panels show lines from
within $\varv_8=1$, the next seven panels lines from within $\varv_8=2$, and
the last two panels lines from within $\varv_4=1$. In each panel, the line
frequency in MHz is written in the top left corner, the energy of the upper 
level in K is given in the top right corner, the rms noise level $\sigma$ in 
mJy~beam$^{-1}$~km~s$^{-1}$ is written in the bottom right corner, and the
beam (HPBW) is shown in the bottom left corner as a red filled ellipse. The 
black contour levels start at $6\sigma$ and then increase geometrically by a 
factor of two at each step. The blue, dashed contours show the $-6\sigma$ 
level. The bottom and top blue crosses indicate the positions of the hot 
molecular cores Sgr B2(N1) and Sgr B2(N2), respectively. The green cross marks 
the position Sgr~B2(N1S). Because of the variation in systemic velocity across 
the field, the assignment of the detected emission to each line is valid only 
for the region around Sgr B2(N1), highlighted with the red box.}
 \label{f:map_ch3cn}
\end{figure*}

\begin{figure*}
 \begin{center}
  \includegraphics[width=0.6375\hsize]{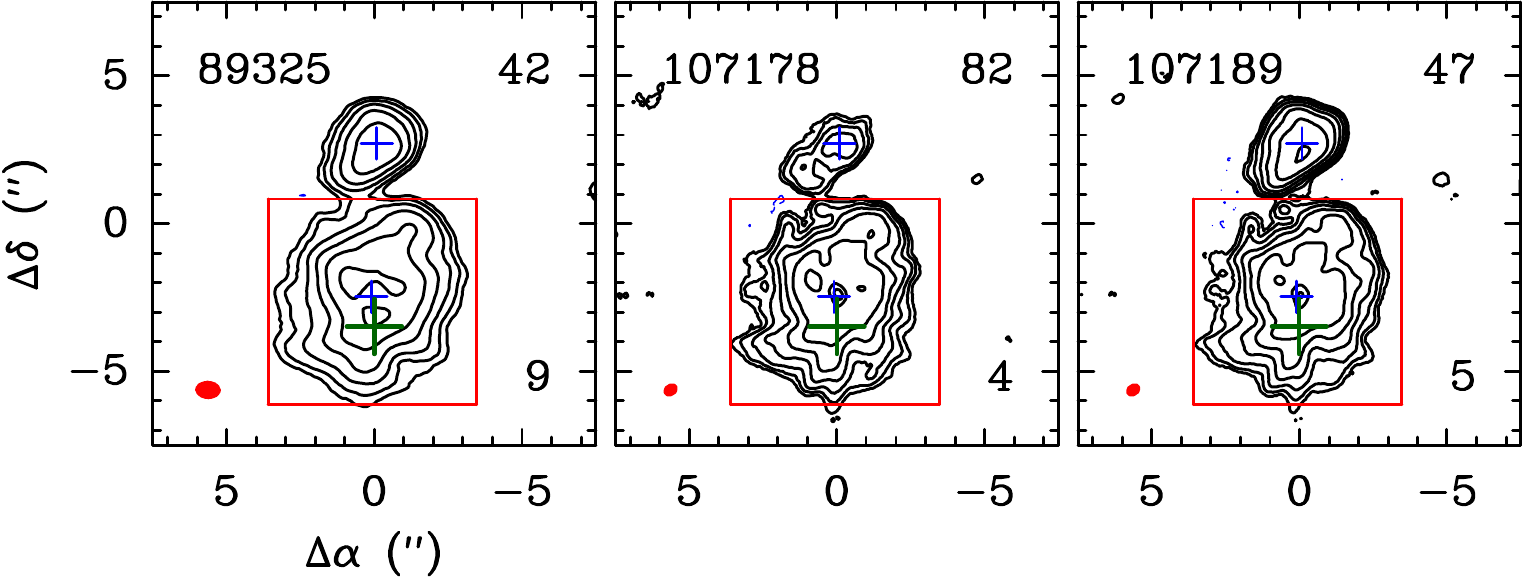}
 \end{center}
 \caption{Same as Fig.~\ref{f:map_ch3cn} but for $^{13}$CH$_3$CN $\varv=0$.}
 \label{f:map_ch3cn_13c1}
\end{figure*}

\begin{figure*}
 \begin{center}
  \includegraphics[width=0.85\hsize]{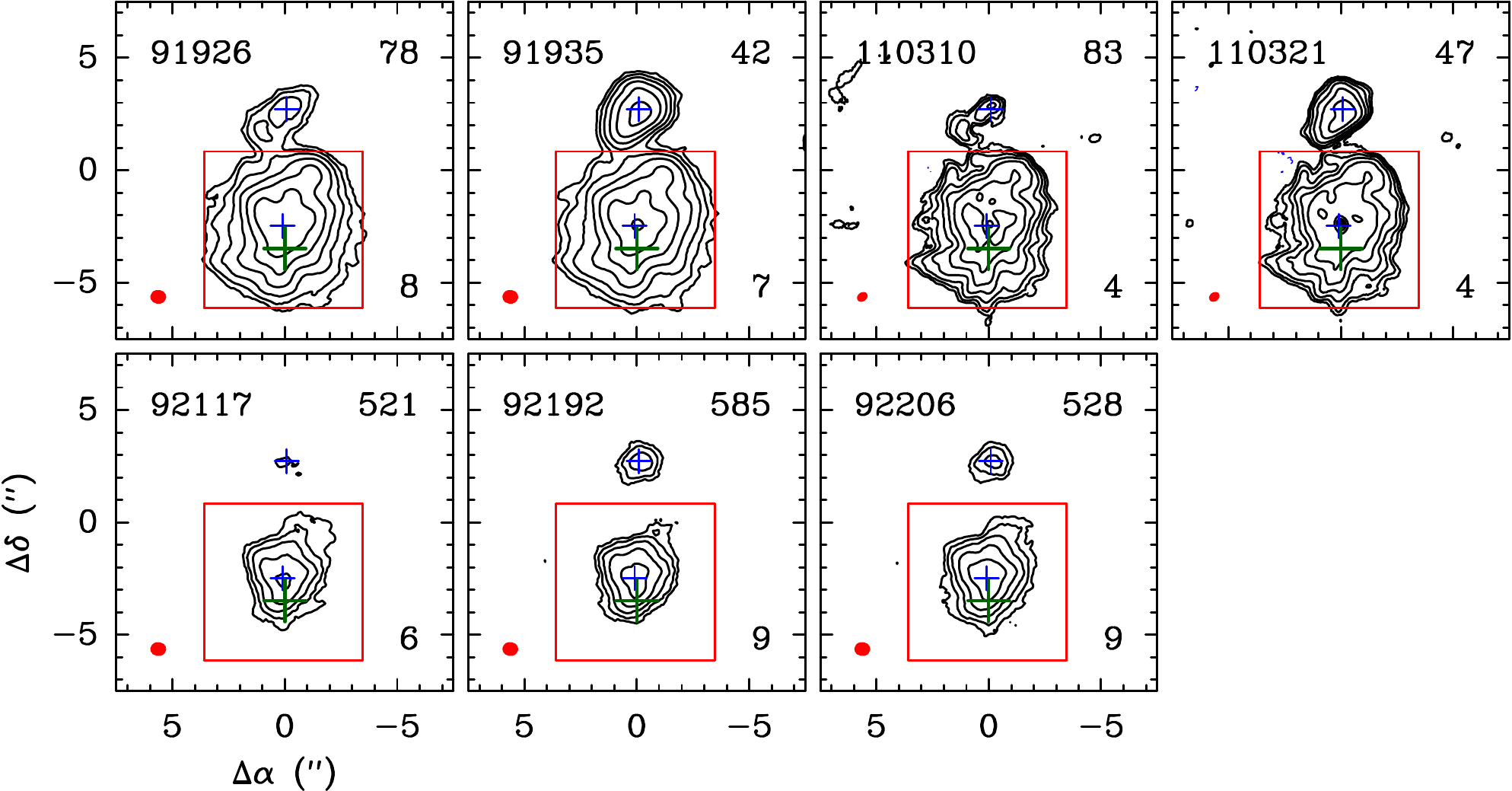}
 \end{center}
 \caption{Same as Fig.~\ref{f:map_ch3cn} but for CH$_3$$^{13}$CN $\varv=0$ 
(top row) and $\varv_8=1$ (bottom row).}
 \label{f:map_ch3cn_13c2}
\end{figure*}

\begin{figure}
 \begin{center}
  \includegraphics[width=1.0\hsize]{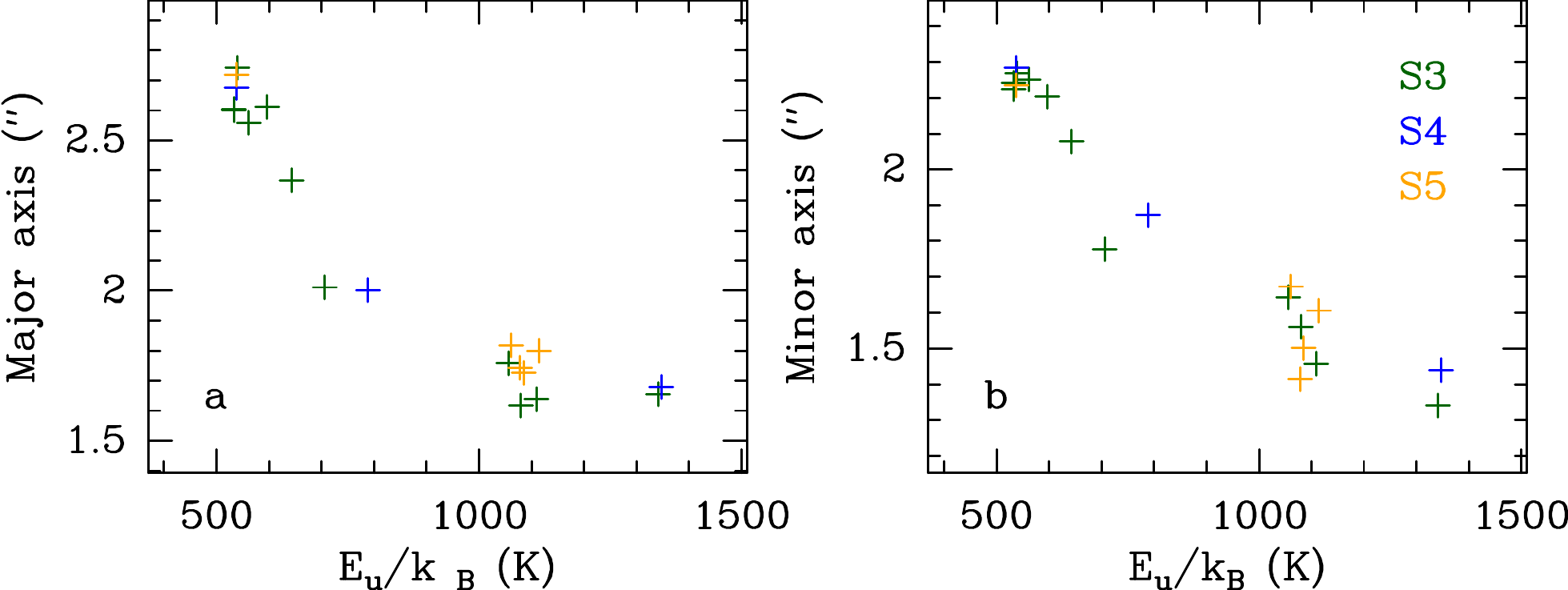}
 \end{center}
 \caption{Deconvolved emission size of uncontaminated CH$_3$CN transitions as 
a function of upper-level energy. The major and minor axes (FWHM) are shown in
panels a and b, respectively. The observational spectral setups with which the 
transitions were measured are color-coded as indicated in panel b.}
 \label{f:size_ch3cn}
\end{figure}

\begin{figure}
 \begin{center}
  \includegraphics[width=1.0\hsize]{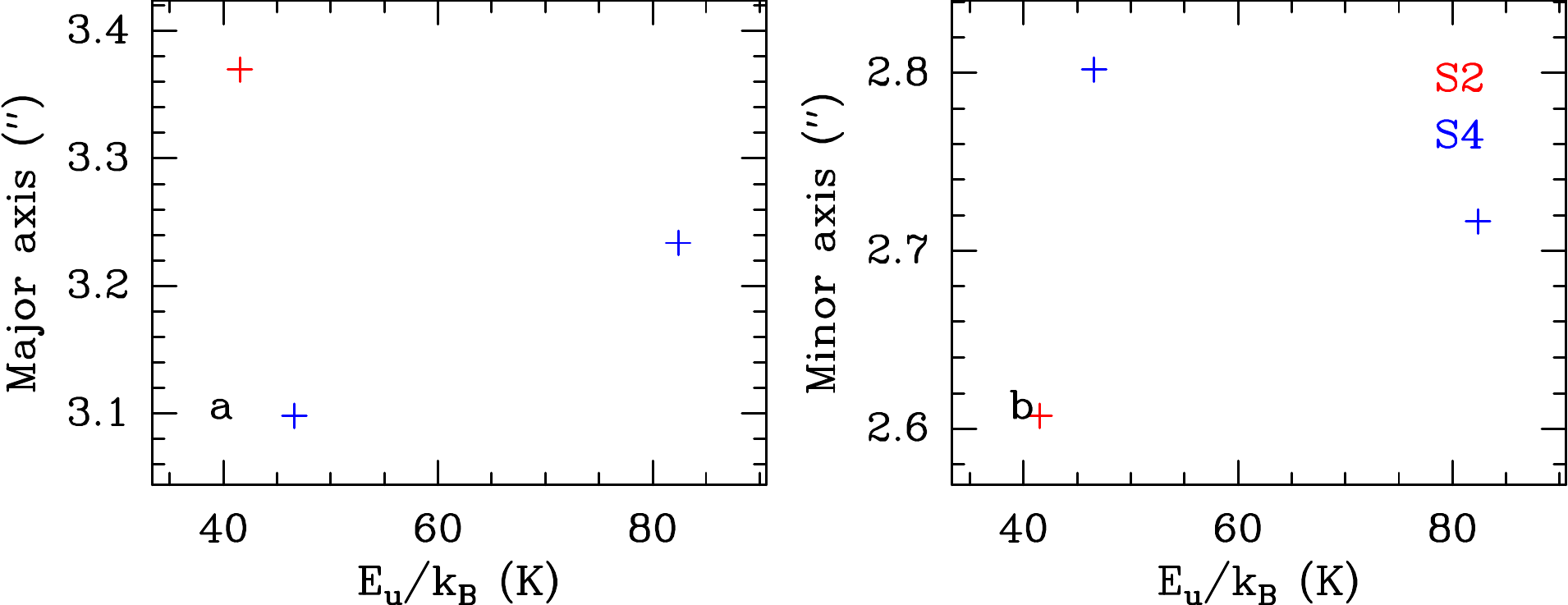}
 \end{center}
 \caption{Same as Fig.~\ref{f:size_ch3cn} but for $^{13}$CH$_3$CN.}
 \label{f:size_ch3cn_13c1}
\end{figure}

\begin{figure}
 \begin{center}
  \includegraphics[width=1.0\hsize]{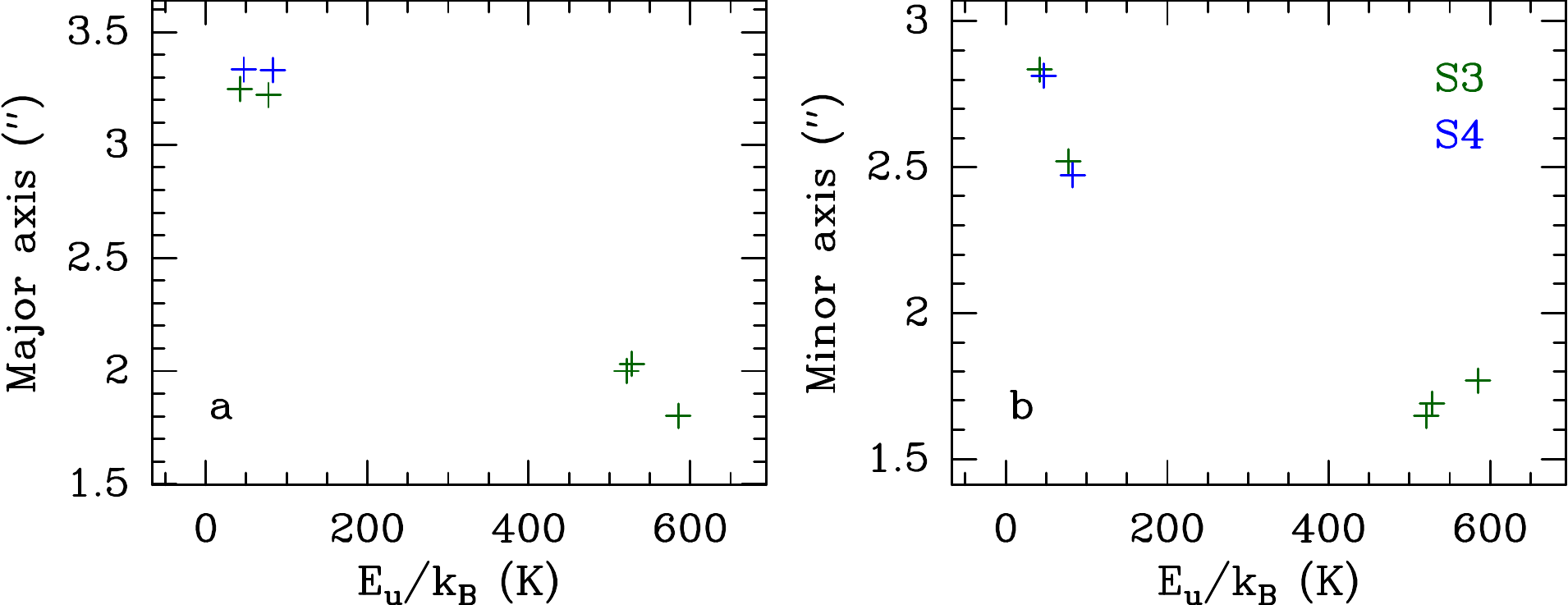}
 \end{center}
 \caption{Same as Fig.~\ref{f:size_ch3cn} but for CH$_3$$^{13}$CN.}
 \label{f:size_ch3cn_13c2}
\end{figure}

\begin{figure}
 \begin{center}
  \includegraphics[width=1.0\hsize]{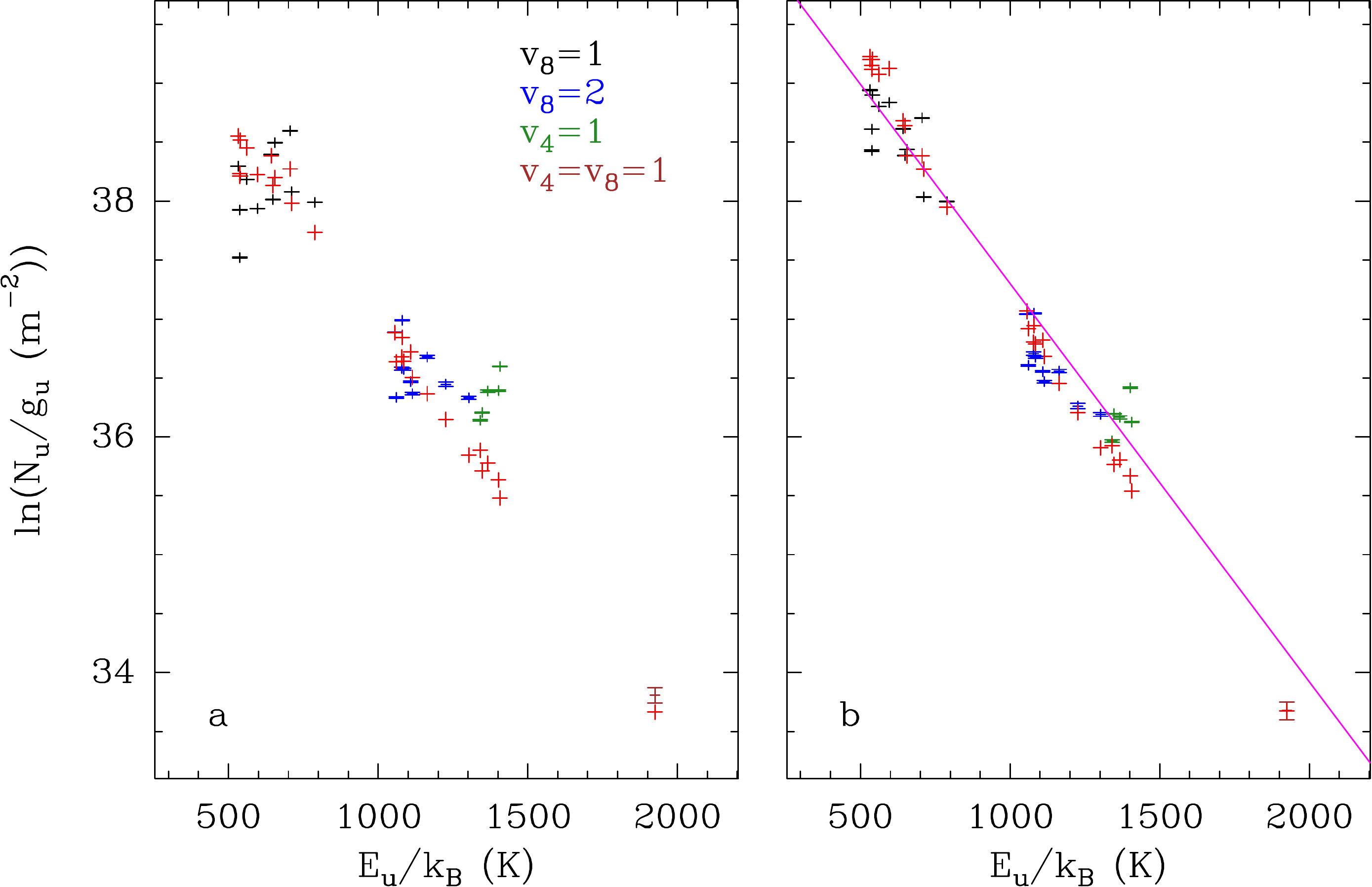}
 \end{center}
 \caption{Population diagram of CH$_3$CN toward Sgr B2(N1S). The observed 
datapoints are shown in various colors (but not red) as indicated in the upper 
right corner of panel a while the synthetic populations are shown in red. No 
correction is applied in panel a. In panel b, the optical depth correction has 
been applied to both the observed and synthetic populations and the 
contamination by all other species included in the full model has been removed 
from the observed datapoints. The purple line is a linear fit to the observed 
populations (in linear-logarithmic space).}
 \label{f:popdiag_ch3cn}
\end{figure}

\begin{figure}
 \begin{center}
  \includegraphics[width=1.0\hsize]{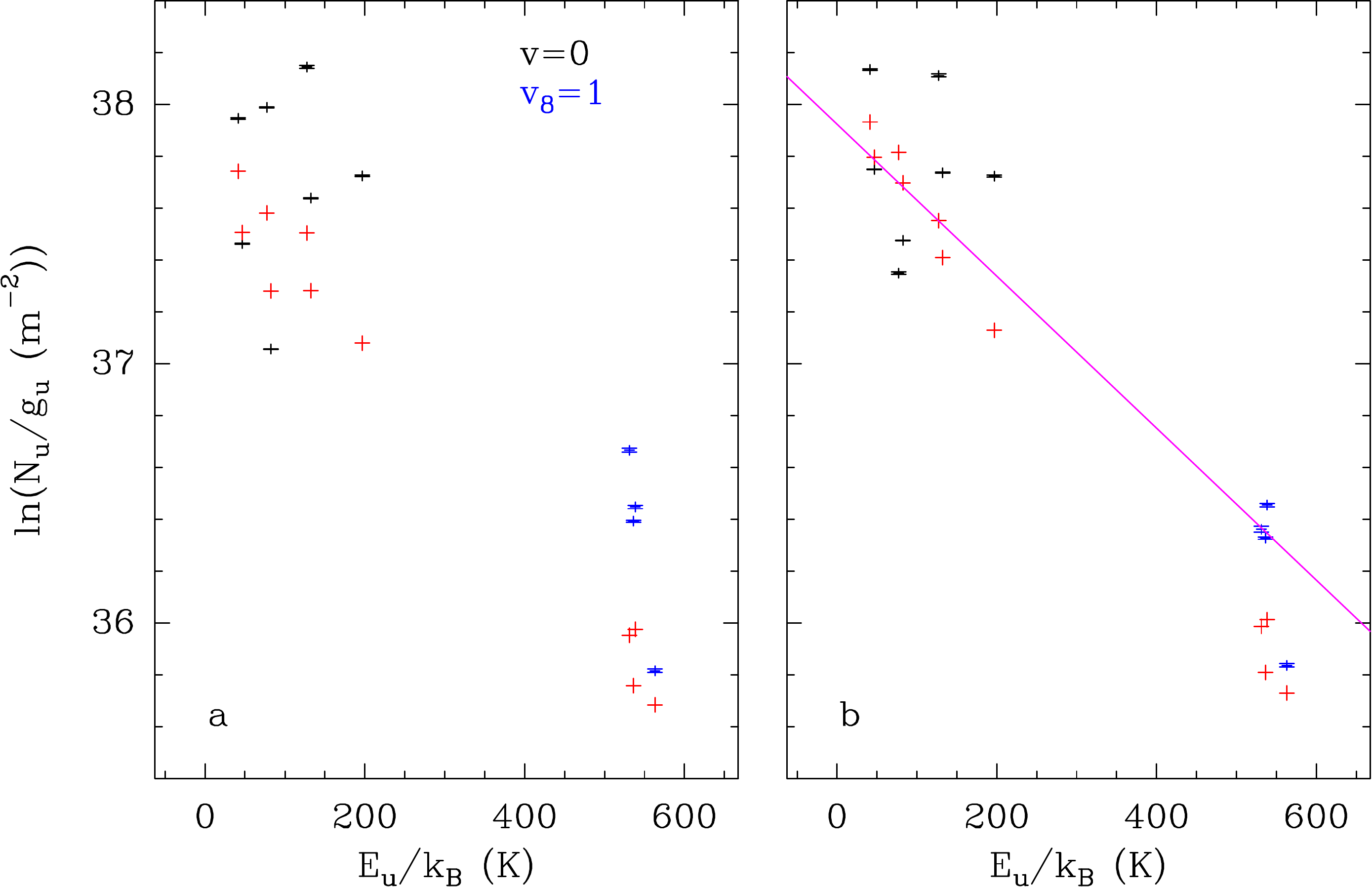}
 \end{center}
 \caption{Same as Fig.~\ref{f:popdiag_ch3cn} but for $^{13}$CH$_3$CN.}
 \label{f:popdiag_ch3cn_13c1}
\end{figure}

\begin{figure}
 \begin{center}
  \includegraphics[width=1.0\hsize]{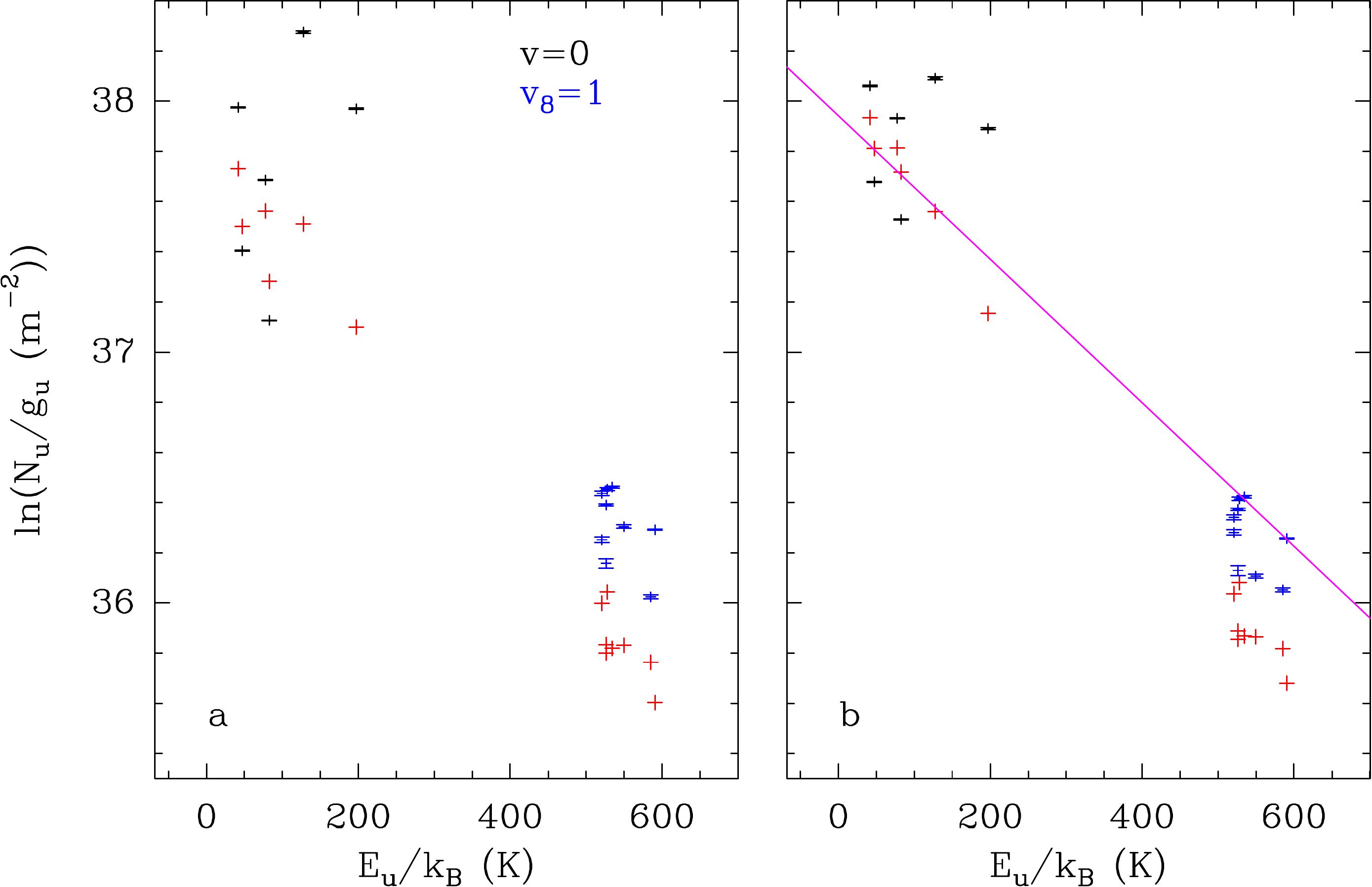}
 \end{center}
 \caption{Same as Fig.~\ref{f:popdiag_ch3cn} but for CH$_3$$^{13}$CN.}
 \label{f:popdiag_ch3cn_13c2}
\end{figure}

Figures~\ref{f:spec_ch3cn_ve0}--\ref{f:spec_ch3cn_15n_v8e1} show spectra of
the transitions of CH$_3$CN $\varv=0$, $\varv_8=1$,  $\varv_8=2$, 
$^{13}$CH$_3$CN $\varv=0$, $\varv_8=1$, 
CH$_3$$^{13}$CN $\varv=0$, $\varv_8=1$, $^{13}$CH$_3$$^{13}$CN $\varv=0$, and 
CH$_3$C$^{15}$N $\varv=0$, $\varv_8=1$, respectively, covered by the ReMoCA 
survey of Sgr~B2(N1S) performed with ALMA. 
Figures~\ref{f:map_ch3cn}--\ref{f:map_ch3cn_13c2} show integrated intensity
maps of uncontaminated transitions of CH$_3$CN, $^{13}$CH$_3$CN, and 
CH$_3$$^{13}$CN, respectively. 
Figures~\ref{f:size_ch3cn}--\ref{f:size_ch3cn_13c2} show the emission sizes
of CH$_3$CN, $^{13}$CH$_3$CN, and CH$_3$$^{13}$CN, respectively, as derived 
from two-dimensional Gaussian fits to these maps. 
Figures~\ref{f:popdiag_ch3cn}--\ref{f:popdiag_ch3cn_13c2} show population
diagrams of CH$_3$CN, $^{13}$CH$_3$CN, and CH$_3$$^{13}$CN, respectively.

\bibliographystyle{elsarticle-num}
\bibliography{MeCN}


\end{document}

%% file: tab_ch3cn_popfit.tex
\begin{table}
 \begin{center}
 \caption{
 Rotational temperatures derived from population diagrams toward Sgr~B2(N1S).
}
 \label{t:popfit}
 \vspace*{1.0ex}
 \begin{tabular}{lll}
 \hline
 \multicolumn{1}{c}{Molecule} & \multicolumn{1}{c}{States$^{\rm a}$} & \multicolumn{1}{c}{$T_{\rm fit}$$^{\rm b}$} \\ 
  & & \multicolumn{1}{c}{\small (K)} \\ 
 \hline
CH$_3$CN & $\varv_8=1$, $\varv_8=2$, $\varv_4=1$, $\varv_4=\varv_8=1$ &   296 (14) \\ 
$^{13}$CH$_3$CN & $\varv=0$, $\varv_8=1$ &   341 (70) \\ 
CH$_3$$^{13}$CN & $\varv=0$, $\varv_8=1$ &   350 (36) \\ 
\hline 
 \hline
 \end{tabular}
 \end{center}
 \vspace*{-1.ex}
 {\footnotesize
 $^{\rm a}$ {Vibrational states that were taken into account to fit the population diagram.}
 $^{\rm b}$ {The standard deviation of the fit is given in parentheses. As explained in Sect.~4.4 of \cite{Belloche19}, this uncertainty is purely statistical and should be viewed with caution. It may be underestimated.}
 }
 \end{table}

%% file: tab_ch3cn_weedsmodel.tex
\begin{table*}[!ht]
 \begin{center}
 \caption{
 Parameters of our best-fit LTE model of methyl cyanide toward Sgr~B2(N1S).
}
 \label{t:coldens}
 \vspace*{1.0ex}
 \begin{tabular}{lcrccccccr}
 \hline
 \multicolumn{1}{c}{Molecule} & \multicolumn{1}{c}{Status$^{\rm a}$} & \multicolumn{1}{c}{$N_{\rm det}$$^{\rm b}$} & \multicolumn{1}{c}{$\theta_{\rm s}$$^{\rm c}$} & \multicolumn{1}{c}{$T_{\mathrm{rot}}$$^{\rm d}$} & \multicolumn{1}{c}{$N$$^{\rm e}$} & \multicolumn{1}{c}{$F_{\rm vib}$$^{\rm f}$} & \multicolumn{1}{c}{$\Delta V$$^{g}$} & \multicolumn{1}{c}{$V_{\mathrm{off}}$$^{\rm h}$} & \multicolumn{1}{c}{$\frac{N_{\rm ref}}{N}$$^{\rm i}$} \\ 
  & & & \multicolumn{1}{c}{\small ($''$)} & \multicolumn{1}{c}{\small (K)} & \multicolumn{1}{c}{\small (cm$^{-2}$)} & & \multicolumn{1}{c}{\small (km~s$^{-1}$)} & \multicolumn{1}{c}{\small (km~s$^{-1}$)} & \\ 
 \hline
 CH$_3$CN, $\varv=0$$^\star$ & d & 9 &  1.6 &  260 &  2.8 (18) & 1.00 & 6.0 & $-0.2$ &       1 \\ 
 \hspace*{7.8ex} $\varv_8=1$ & d & 22 &  1.6 &  260 &  2.8 (18) & 1.00 & 6.0 & $-0.2$ &       1 \\ 
 \hspace*{7.8ex} $\varv_8=2$ & d & 16 &  1.6 &  260 &  2.8 (18) & 1.00 & 6.0 & $-0.2$ &       1 \\ 
 \hspace*{7.8ex} $\varv_4=1$ & d & 4 &  1.6 &  260 &  2.8 (18) & 1.00 & 6.0 & $-0.2$ &       1 \\ 
 \hspace*{7.8ex} $\varv_4=\varv_8=1$ & t & 1 &  1.6 &  260 &  2.8 (18) & 1.00 & 6.0 & $-0.2$ &       1 \\ 
 $^{13}$CH$_3$CN, $\varv=0$ & d & 9 &  1.6 &  260 &  1.4 (17) & 1.35 & 6.0 & $-0.2$ &      21 \\ 
 \hspace*{9.3ex} $\varv_8=1$ & d & 4 &  1.6 &  260 &  1.4 (17) & 1.35 & 6.0 & $-0.2$ &      21 \\ 
 CH$_3$$^{13}$CN, $\varv=0$ & d & 6 &  1.6 &  260 &  1.4 (17) & 1.35 & 6.0 & $-0.2$ &      21 \\ 
 \hspace*{9.3ex} $\varv_8=1$ & d & 10 &  1.6 &  260 &  1.4 (17) & 1.35 & 6.0 & $-0.2$ &      21 \\ 
 $^{13}$CH$_3$$^{13}$CN, $\varv=0$ & t & 1 &  1.6 &  260 &  6.8 (15) & 1.35 & 6.0 & $-0.2$ &     415 \\ 
 CH$_3$C$^{15}$N, $\varv=0$ & n & 0 &  1.6 &  260 & $<$  1.4 (16) & 1.35 & 6.0 & $-0.2$ & $>$     207 \\ 
 \hspace*{9.3ex} $\varv_8=1$ & n & 0 &  1.6 &  260 & $<$  1.4 (16) & 1.35 & 6.0 & $-0.2$ & $>$     207 \\ 
\hline 
 \hline
 \end{tabular}
 \end{center}
 \vspace*{-1.ex}
 {\footnotesize
 $^{\rm a}$ {d: detection, t: tentative detection, n: nondetection.}
 $^{\rm b}$ {Number of detected lines (conservative estimate, see Sect.~3 of \cite{EMoCA_2016}). One line of a given species may mean a group of transitions of that species that are blended together.}
 $^{\rm c}$ {Source diameter (\textit{FWHM}).}
 $^{\rm d}$ {Rotational temperature.}
 $^{\rm e}$ {Total column density of the molecule. $x$ ($y$) means $x \times 10^y$. An identical value for all listed vibrational states of a molecule means that LTE is an adequate description of the vibrational excitation.}
 $^{\rm f}$ {Correction factor that was applied to the column density to account for the contribution of vibrationally excited states, in the cases where this contribution was not included in the partition function of the spectroscopic predictions.}
 $^{\rm g}$ {Linewidth (\textit{FWHM}).}
 $^{\rm h}$ {Velocity offset with respect to the assumed systemic velocity of Sgr~B2(N1S), $V_{\mathrm{sys}} = 62$ km~s$^{-1}$.}
 $^{\rm i}$ {Column density ratio, with $N_{\rm ref}$ the column density of the previous reference species marked with a $\star$.}
 }
 \end{table*}